\documentclass[onecolumn]{aa}
\usepackage{graphicx}
\usepackage{txfonts}

\begin{document}

\title {Sample of minor merger of galaxies: Optical CCD surface photometry and HII region properties.}

\author{D. L. Ferreiro\inst{1} \fnmsep\thanks{Visiting Astronomer at the Cerro Tololo Iter-American
Observatory (CTIO).}
	\and
	M. G. Pastoriza\inst{2}  }

\institute{IATE, Observatorio Astron\'omico, Universidad Nacional de
    C\'ordoba, Laprida 854, 5000, C\'ordoba, Argentina\\
	\email{diegof@mail.oac.uncor.edu}
	\and
	Departamento de Astronomia, IF-UFRGS, C.P. 15051, CEP
    91501-970, Porto Alegre, RS, Brazil\\
	\email{mgp@if.ufrgs.br} }

\date{Received 17/12/2003; accepted 08/07/2004}

\abstract{We present the results of the B, V and I photometry of eleven southern minor mergers. The total apparent B magnitude, integrated B-V and V-I colours were measured. We built B, V, and I equivalent profiles for each galaxy and decomposed them into bulge and disk components when possible. From H$\alpha$+N[II] images we have estimated the basic photometric parameters of the HII regions, such as position, size, B-V and V-I colours, H$\alpha$+[NII] luminosity and $EW(H\alpha+[NII])$ equivalent width. Primary components have blue absolute magnitudes in the range --22 $< M_B <$ --18, with a peak at $M_B$ = -- 22. The magnitudes of the secondary components are in the range --22 $< M_B <$ --16 with a maximum at $M_B$ = -- 19. Most pairs have $\Delta$$M_B$ $\sim$ 2, which means that in luminosity the primary galaxy is on average about 6 times brighter than the secondary. We found a linear correlation between the luminosity ratios of the components and their ratio of major diameters, leading to mass ratios between 0.04 $<$ $\cal M$ $_{secondary}$/$\cal M$ $_{primary}$ $<$ 0.2, suggesting indeed that our sample is formed by minor mergers. On average the galaxies have colours bluer than those of isolated galaxies with the same morphological type. Most of the HII regions and evolved star-forming regions of the sample were formed between 3.6 to 13.7 Myr ago with an average of (6.3$\pm$0.7) Myr. The HII region properties, luminosity, sizes and ages are similar in both components. The HII regions have log(H$\alpha$+[NII]) luminosity between 38.6 and 41.7. The HII region luminosity function for the whole sample fits a power law of index  $\alpha$ = --1.33. The linear correlation between the luminosity $\cal L$(H$\alpha$+[NII]) and the size of the HII regions has slope of 2.12$\pm$0.06. We found that the disk of the primary component is more luminous than those of Lu's sample, while the disk of the secondary is smaller and fainter. A plot of the disk parameters does not change with colour. This indicates that the different stellar populations in the disks were affected in the same way.\footnote{Tables 4 to 6 are only available in electronic form at the CDS via anonymous ftp to cdsarc.u-strasbg.fr (130.79.128.5) or via http://cdsweb.u-strasbg.fr/cgi-bin/qcat?J/A+A/. Tables 2 and 3 and Figures 1 to 11 will only be published in the electronic version of the Journal, that will be available at http://www.edpsciences.org/aa.}\\

 }

\authorrunning {D. Ferreiro \& M. Pastoriza}

\titlerunning {Optical CCD surface photometry and HII region properties}

\maketitle

\section{Introduction}
Previous works have shown that interactions and merger events represent important mechanisms for driving the evolution of galaxies.
Interacting galaxies are more active in the UV (Larson \& Tinsley 1978), in optical emission-line strength (Kennicutt \& Keel 1984), in the
near-IR (Joseph \& Wright 1985) and in radio emission (Hummel 1981).
All these effects have been attributed to induced star-formation as a result of close encounters or tidal interactions (Bushouse 1986; Kennicutt et al. 1987).

From these events, those that involve low-mass satellites are of particular interest since they seem to occur more frequently in the local universe
(Frenk et al. 1988; Carlberg \& Couchman 1989). A good example of these so-called "minor mergers" is the Milky Way, that will form a future minor merger
with the Large Magallanic Cloud ($\cal M$\rm$_{LMC}$/$\cal M$\rm$_{MW}$ $\sim$ 0.1) (Schwarzkopf \& Dettmar 2000,  hereafter SD2000).

Interaction between galaxies may lead to a high star-formation rate (SFR)
in two different ways: by increasing the formation rate per unit mass and by
increasing the concentration of the gas in some regions of the galaxies
(Combes 1993). The efficiency of the star formation process per unit mass
has been found to be higher by a factor of 10 in interacting galaxies than
the average value found in normal galaxies (Young 1993). Moreover, numerical
simulations have shown that interactions result in a redistribution of
substantial quantities of the material into the central regions of the galaxies (Hernquist \& Mihos, 1995). This injection of fresh material (mostly neutral hydrogen) into the potential well of a galaxy is then expected to be followed by
the development of a compression mechanism that could result in a rapid and
efficient formation of $H_2$ molecular clouds and their subsequent collapse
 to form stars.

However, in contrast to previous results, Bergvall et al. (2003) concluded that interacting and merging galaxies do not substantially differ from similar normal isolated galaxies if we consider the global star formation. They also found that there is no significant increased scatter in the colours of Arp galaxies with respect to the normal ones. They only detected a moderate enhancement of the star formation in the centers of interacting galaxies.

What is clear is that little is know about the rate of minor mergers
events, their influence on the structure and kinematics of galactic disks and the efficiency in producing morphological perturbations (SD2000).
However, new works are addressing the influences of a minor merger
on the kinematics, disk scale parameters and Star Formation Rates.
For instance, the numerical simulations of Berentzen et al. (2003) show that the small companion passes through the disk of a larger barred galaxy nearly
perpendicular to its plane. The interactions produce expanding ring structures,
offset bars, spokes and other asymmetries in the stars and gas.

The goal of this paper is to study the effects of the interaction on the integrated photometric properties and star formation activity
in a sample of eleven minor mergers, selected from a larger sample of physical pairs studied by Donzelli \& Pastoriza (1997) (hereafter DP97).
The paper is organized as follows: In Sect.~\ref{obs} we describe the sample
selection and we summarize the observations. Sect.~\ref{red} presents the data reductions. Sect.~\ref{prop} describes the particular properties of the individual galaxies. Sect.~\ref{res} presents results and discussions. The conclusions are given in Sect.~\ref{con}.\\

\section{Sample selection and observations}
\label{obs}

The sample of minor mergers for this study was previously selected from the Catalogue of Arp \& Madore (1987) and their optical spectroscopic properties have been studied by DP97. These pairs of galaxies consist of a main galaxy (component A) and a companion (component B) that has about half or less the diameter of the component A.  The nuclear spectrum of the galaxies in 27 physical pairs were classified by DP97 into four groups according to the emission line spectra and the enhancement of the star formation activity measured in the nuclear region by the equivalent width of the H$\alpha$+[NII] emission lines. For this work we have selected 11 pairs, mainly from groups 2, 3 and 4. The main characteristics of these groups are the following:\\

Group 1 - The spectra of both components are dominated by strong absortion features, typical of early-type galaxy spectra. No emission lines are observed.

Group 2 - The spectra of the A components of this group are dominated by absorption futures e.g., Na 5890 $\AA$, MgI+MgH 5175 $\AA$, G band, CN 4200 $\AA$ and weak H$\alpha$+[NII] emission lines. The $EW(H\alpha+[NII])$ values of these lines are around 4 $\AA$. The B components have HII region emission-line type spectra with moderate excitation.

Group 3 - The main component shows low-excitation HII region type spectra similar to those observed in the bulge of Sa-Sb galaxies. The absorption spectra for these galaxies have important contributions from younger stellar populations. The secondary component has HII region emission-type spectra with a wide range of excitations as inferred from the $EW(H\alpha+[NII])$ values, which range from 6 to 80 $\AA$.

Group 4 - Strong emission lines (H$\alpha$, [NII], [OIII] and H$\beta$) are observed in both components indicating more exited spectra than in group 3. The observed $EW(H\alpha+[NII])$ have values greater than 30 $\AA$ for the main component and the average of 77 $\AA$ for the secondary is almost 3 times higher than that reported for normal galaxies (Liu \& Kennicutt, 1995).\\

Table~\ref{Tsample} summarizes the main characteristics of the sample pairs taken from the NED database: identification, right ascension, declination,
apparent B magnitude and radial velocity for both
members, morphological type, blue extinction and group number according
to DP97 classification.

\begin{table*}[h]
\caption[]{Sample Galaxies}
\label{Tsample}
\begin{center}
\begin{tabular} {lcccccclcc} \hline\hline

\multicolumn{1}{c}{Name}&{R. A.}&{Decl.}&\multicolumn{2}{|c|}{m$_B$}&\multicolumn{2}{|c|}{cz$^{\mathrm(1)}$}&{Morphology}&{A$_b$}&{Group}\\
\cline{4-7}
&{(J2000)}&{(J2000)}&\multicolumn{1}{|c|}{Main}&\multicolumn{1}{|c|}{Secondary}&\multicolumn{1}{|c|}{Main}&\multicolumn{1}{|c|}{Secondary}&&&\\
\hline

AM1256-433 & 12:56:58 & --43:50:11 & 16.02 & 17.04 & 9014 & 9183$^{\mathrm(2)}$ & Interacting   & 0.390 & 4 \\
AM1401-324 & 14:04:15 & --33:01:28 & 14.80 & --- & 10321 & 10426$^{\mathrm(2)}$ & Sb            & 0.355 & 3 \\
AM1448-262 & 14:51:14 & --26:37:49 & 14.14 & 15.61 & 2576 & 2738         & (R?)SB(rs)0/a & 0.676 & 2 \\
AM2030-303 & 20:33:60 & --30:22:24 & 15.18 & 17.05 & 12465$^{\mathrm(2)}$ & 12474    & G Trpl        & 0.299 & 3 \\
AM2058-381 & 21:01:39 & --38:05:00 & 15.16 & 17.34 & 12383$^{\mathrm(2)}$ & 12460$^{\mathrm(2)}$ & S?    & 0.219 & 3 \\
AM2105-332 & 21:08:04 & --33:13:19 & 13.98 & 15.50 & 5449 & 5810         & SBO? pec      & 0.396 & 2 \\
AM2229-735 & 22:33:43 & --73:40:51 & --- & --- & 17535 & 17342       & SO?           & 0.159 & 4 \\
AM2238-575 & 22:41:37 & --57:36:22 & 13.96 & 16.79 & 10838 & 10659       & SAB(rs)bc     & 0.087 & 1 \\
AM2306-721 & 23:09:43 & --72:00:05 & 14.76 & 15.90 & 8916 & 9069         & G             & 0.130 & 4 \\
AM2322-821 & 23:26:29 & --81:54:42 & 13.23 & --- & 3680 & 3424$^{\mathrm{(2)}}$ & SA(r)bc       & 0.782 & 4 \\
AM2330-451 & 23:33:14 & --45:01:38 & 12.34 & 14.65 & 3137 & 3551         & SB(s)0/a? sp  & 0.063 & 2 \\
\hline\hline

\end{tabular}

\begin{list}{}{}
\item[$^{\mathrm{(1)}}$]Km s$^{-1}$
\item[$^{\mathrm{(2)}}$] DP97
\end{list}

\end{center}
\end{table*}

The observations were taken in 1999 July using a Tektronix 2048 $x$ 2048 CCD attached to the Cerro Tololo Inter-American Observatory 0.90 m telescope with a pixel scale of 0.396 arcsec. Seeing conditions were good (1".1 -- 1".4) and nights were photometric. Observations were made using Johnson-Cousins BVI and interference filters of typically 75$\AA$ wide. Table~\ref{Tfilters} list details of the interference filters (for the continuum and emission line images) used for the observation of each galaxy pair. A summary of the log of observations is show in Table~\ref{T log}.\\

\begin{table*}[h]
\caption[]{The Narrow Filters}
\label{Tfilters}
\begin{center}
\begin{tabular}{l|cccc|cccccc} \hline\hline

\multicolumn{1}{c|}{Galaxy}&{Filters}&{Center}&{FWHM}&{Trans}&{Filters}&{Center}&{FWHM}&{Trans}\\
&{Cont.}&{[$\AA$]}&{[$\AA$]}&{[\%]}&{Line}&{[$\AA$]}&{[$\AA$]}&{[\%]} \\
\hline

AM1256-433& 6606/75& 6608& 70& 84.30& 6781/78& 6785& 77& 78.98\\
AM1401-324& 6649/76& 6650& 77& 87.49& 6781/78& 6785& 77& 78.98\\
AM1448-262& 6477/75& 6448& 77& 83.20& 6606/75& 6608& 70& 84.30\\
AM2030-303& 6693/76& 6693& 92& 87.93& 6826/78& 6832& 81& 82.32\\
AM2058-381& 6693/76& 6693& 92& 87.93& 6826/78& 6832& 81& 82.32\\
AM2105-332& 6520/76& 6530& 71& 81.09& 6693/76& 6693& 92& 87.93\\
AM2229-735& 6781/78& 6785& 77& 78.98& 6961/79& 6967& 90& 84.46\\
AM2238-575& 6649/76& 6650& 77& 87.49& 6781/78& 6785& 77& 78.98\\
AM2306-721& 6606/75& 6608& 70& 84.30& 6781/78& 6785& 77& 78.98\\
AM2322-821& 6477/75& 6448& 77& 83.20& 6649/76& 6650& 77& 87.49\\
AM2330-451& 6477/75& 6448& 77& 83.20& 6649/76& 6650& 77& 87.49\\
\hline\hline

\end{tabular}
\end{center}
\end{table*}

\begin{table*}[h]
\caption[]{Photometric Observations}
\label{T log}
\begin{center}
\begin{tabular}{lcccccc} \hline\hline

\multicolumn{1}{c}{Galaxy}&{Date}&{B}&{V}&{I}&{H$\alpha$}&{FWHM}\\
\cline{3-6}
&&\multicolumn{4}{c}{Exposure [s]}&{[arcsec]}\\
\hline

AM1256-433 & 14/7 & 1x1200 & 1x800 & 1x800 & 1x1800 & 1.2\\
AM1401-324 & 30/7 & 1x1000 & 1x700 & 1x700 & 1x1200 & 1.2\\
AM1448-262 & 13/7 & 1x900 & 1x500 & 1x500 & 1x1200 & 1.3\\
AM2030-303 & 13/7 & 1x900 & 1x500 & 1x500 & 1x1200 & 1.1\\
AM2058-381 & 14/7 & 1x1200 & 1x800 & 1x800 & 1x1800 & 1.4\\
AM2105-332 & 30/7 & 1x1200 & 1x900 & 1x900 & 1x1500 & 1.2\\
AM2229-735 & 13/7 & 1x900 & 1x500 & 1x500 & 1x1200 & 1.3\\
AM2238-575 & 31/7 & 1x1000 & 1x900 & 1x900 & 1x1500 & 1.3\\
AM2306-721 & 14/7 & 1x1200 & 1x900 & 1x900 & 1x1800 & 1.1\\
AM2322-821 & 30/7 & 1x1200 & 1x900 & 1x900 & 1x1500 & 1.2\\
AM2330-451 & 30-31/7 & 1x1200 & 1x900 & 1x900 & 1x1500 & 1.4\\
\hline\hline

\end{tabular}
\end{center}
\end{table*}

\section{Data reduction, magnitudes, colours, H$\alpha$ images and luminosity profiles}
\label{red}
\subsection{Image reduction}

All images were reduced following the standard procedures (bias subtraction and flat field normalization). Sky subtraction was performed by fitting the sky value in areas well beyond the limits of the galaxies. Cosmic ray removal, seeing estimation by a Gaussian fit to the field stars, extinction correction and calibration with standard stars were the final reduction steps applied to the images. Most of these reduction steps were done using the IRAF package.

For each night we took 7-18 photometric standards around the celestial equator from Landolt (1992) and at least 3 southern spectrophotometric standards from Stone \& Baldwin(1983). Estimates of the accuracy in the calibrations are $\pm$ 0.04 mag in B, $\pm$ 0.06 mag in (B-V) and $\pm$0.06 mag in (V-I).

For each H$\alpha$ image a suitable H$\alpha$ continuum image was subtracted.
Prior to this subtraction a careful alignment of the images was made.
This alignment was performed using the IMALIGN task of IRAF with at least 7 field stars. The typical accuracy was better than 0.5 pixel. When images had different seeing, the ones with better seeing were convolved with a Gaussian function in order to match the image with the poorer seeing.
Using the same technique, we have also obtained B-I colour images.
Fig.~\ref{fig1} -~\ref{fig11} present the B, B-I and H$\alpha$ images for the galaxies.

\subsection{Integrated Magnitudes and colours}
\label{mag}

We have measured total B, V and I magnitudes using two independent methods.
The first one is the integration of intensity pixels in a series of diaphragms with increasing radius until convergence occurs. The second method consisted of the integration of the luminosity profile (see Sect.~\ref{prof}). The total magnitudes obtained with both methods are in very good agreement. Table~\ref{Tmag} lists total B magnitude and integrated colours B-V and V-I   as well as the absolute $M_B$ magnitude and the major (D) and minor (d) diameters measured inside the 24 mag arcsec$^{-2}$ isophote. All magnitudes are corrected for both galactic and internal extinction. Throughout this paper we adopt a Hubble constant $H_0$ = 75 $km$ $s^{-1} Mpc^{-1}$. The measured B magnitudes are systematically brighter than the B magnitudes listed in the RC3 Catalogue (see Table~\ref{Tsample}). These differences are most probably due to the accuracy of the present observations and data reduction.  \\

\begin{table*}[h]
\caption[]{ B Magnitude, B-V and V-I integrated colours, $M_B$ absolute magnitude,  major (D) and minor (d) diameters}
\label{Tmag}
\begin{center}
\begin{tabular}{lcccccccc} \hline\hline

\multicolumn{1}{c}{Galaxy}&{B}&$E(B-V)_{int}$&$E(B-V)_{gal}$&{B-V}&{V-I}&{$M_B$}&{D["]}&{d["]} \\
\multicolumn{1}{c}{Main} \\

\hline

AM1256-433W &15.04&0.83&0.098&0.47&0.58&--20.76& 83.5 & 17.4\\
AM1401-324N &14.32&--- &0.084&0.43&1.25&--21.90& 58.0 & 56.0\\
AM1448-262NE&14.14&--- &0.169&1.07&1.50&--19.35& 100.8 & 87.2\\
AM2030-303SW&15.25&0.43&0.075&0.73&1.13&--21.27& 43.6 & 40.2\\
AM2058-381N &14.91&0.69&0.055&0.60&0.80&--21.38& 85.7 & 32.2\\
AM2105-332SE&14.22&0.96&0.099&0.96&1.75&--20.55& 65.9 & 41.4\\
AM2229-735E &15.98&0.34&0.040&0.73&1.18&--21.06& 46.0 & 23.3\\
AM2238-575W &14.47&--- &0.022&1.17&1.57&--21.49& 120.2 & 79.9\\
AM2306-721S &14.07&0.32&0.033&0.24&0.55&--21.48& 85.3 & 47.2\\
AM2322-821SE&13.35&0.25&0.196&0.81&0.91&--20.98& 113.8 & 85.9\\
AM2330-451NE&14.47&--- &0.016&1.17&1.57&--18.70& 127.5 & 83.5\\
\hline
\multicolumn{1}{c}{Secondary} \\
AM1256-433E &16.41&--- &0.098&0.61&0.83&--19.48& 18.4 & 16.7\\
AM1401-324S &17.45&--- &0.084&0.77&1.18&--18.69& 15.0 & 12.0\\
AM1448-262SW&15.35&0.35&0.169&0.68&1.26&--18.18& 53.6 & 27.7\\
AM2030-303NE&16.37&--- &0.075&0.71&1.22&--20.15& 33.4 & 27.5\\
AM2058-381S &16.24&--- &0.055&0.40&0.59&--20.07& 33.4 & 18.2\\
AM2105-332NW&16.06&--- &0.099&0.85&1.77&--18.85& 29.3 & 19.2\\
AM2229-735W &17.36&0.58&0.040&0.63&1.02&--19.69& 17.2 & 11.5\\
AM2238-575E &16.77&--- &0.022&1.53&1.77&--19.11& 40.1 & 18.0\\
AM2306-721N &14.47&--- &0.033&0.18&0.44&--21.11& 55.5 & 27.3\\
AM2322-821NW&15.41&0.11&0.196&0.69&0.66&--18.87& 37.7 & 22.3\\
AM2330-451SW&16.77&--- &0.016&1.53&1.77&--16.67& 64.6 & 27.3\\
\hline \hline

\end{tabular}
\end{center}
\end{table*}

\subsection{H$\alpha$ Images}
\label{halpha}

The H$\alpha$+[NII] images show in Fig.~\ref{fig1}C -~\ref{fig11}C were used to identify the HII regions for each galaxy and  measure the X and Y coordinates, relative to the galaxy center. The HII region size was measured using the area ($\cal A$) inside the isophotal level that has an intensity value of 10\% of the central intensity of the HII region. From this area we calculate the equivalent radius given by
 $r_{eq}$ = ($\cal A$/$\pi$)$^{0.5}$.

The H$\alpha$+[NII] flux and luminosity were determined integrating the intensity pixels inside a diaphragm of radius equal to the equivalent radius. The flux calibration was performed through calibrated spectra of some of the observed HII regions. For each H$\alpha$ image a suitable H$\alpha$ continuum image was also observed. Therefore we were able to calculated the equivalent width $EW(H\alpha+[NII])$ for the individual HII regions.
 We have also measured B magnitude and B-V and V-I colours for the HII regions, integrating the intensity pixels inside the equivalent radius. We have estimated the internal reddening for each galaxy from the line intensity ratio H$\alpha$/H$\beta$ observed in their central region by Pastoriza, Donzelli \& Bonatto (1999).
The internal reddenings derived from the H$\alpha$/H$\beta$ ratio of the nuclear spectrum are in the range of  0.25 $< E(B-V) <$ 0.96, lower than those  observed in warm ultraluminous infrared galaxies (Surace et al 1998) of 0.78 $< E(B-V) <$ 1.93, indicating that our galaxies have less dust that the warm ULIG galaxies. On the other hand it is unlikely that the reddening in the  HII regions is larger than the one estimated in the nucleus. As argued by Whitmore \& Schwiser (1995) the dust clearing time is only $10^6$ years for regions of about 100 pc diameter (roughly the size of our regions). Since we have estimated ages from 3 to 10 Myr for our HII regions (see Table~\ref{Tregions}), using as age indicator the equivalent width of H$\alpha$, which is not reddening dependent, possibly few of them still have a significant amount of dust. Therefore, lacking other sources for estimating HII region reddening, we use the reddening measured in the nucleus to represent an upper limit for the knots.
 Colours and fluxes for the HII regions were extinction-corrected assuming a constant E(B-V) value throughout the galaxy, as described above.
Table~\ref{Tregions} list the relative position of the regions with respect to the galaxy nucleus, equivalent radius, total B magnitude, colours, calibrated H$\alpha$+[NII] luminosity inside $\cal A$ and equivalent width $EW(H\alpha+[NII])$. The table also lists the HII region age, estimated in section 5.3\\

\begin{table*}
\caption[]{HII region, position, equivalent radius, colour, (H$\alpha$+[NII]) luminosity , H$\alpha$+[NII] equivalent width and age.}
\label{Tregions}
\begin{center}
\begin{tabular}{cccccccccc} \hline\hline

{Region}&{X}&{Y}&{$r_{eq}$}&{B}&{B-V}&{V-I}&{$\cal L$ (H$\alpha$+[NII])}&{Log(EW)[$\AA$] }&{Age}\\
&{["]}&{["]}&{[pc]}&{[mag]}&&&{[erg s$^{-1}$]}&{$\pm$0.1}&{[10$^6$ yr]} \\
\hline

\multicolumn{10}{c}{AM1256-433W}\\
1&--37&--42&131.52&19.28 $\pm$ 0.27&0.69&	1.26&(4.10 $\pm$ 0.05)$\times 10^{39}$&1.9 &	6.3$\pm$ 0.7\\
2&--30&--28&105.21&19.57 $\pm$ 0.27&0.81&	1.28&(5.65 $\pm$ 0.08)$\times 10^{39}$&2.1 &	6.3$\pm$ 0.7\\
3&--27&--21&105.24&19.30 $\pm$ 0.27&0.76&	1.28&(5.47 $\pm$ 0.07)$\times 10^{39}$&2.0 &	6.0$\pm$ 0.7\\
4&--18&--12&131.52&18.66 $\pm$ 0.27&0.76&	1.31&(7.43 $\pm$ 0.01)$\times 10^{39}$&1.9 &	5.9$\pm$ 0.8\\
5&0  &0  &157.83&18.04 $\pm$ 0.27&0.77&	1.43&(1.79 $\pm$ 0.02)$\times 10^{40}$&2.0 &	6.1$\pm$ 0.7\\
6&39 &31 &236.69&17.86 $\pm$ 0.27&0.81&	1.38&(1.80 $\pm$ 0.02)$\times 10^{40}$&1.9 &	6.4$\pm$ 0.7\\
7&86 &31 &105.20&19.71 $\pm$ 0.27&0.62&	1.05&(3.06 $\pm$ 0.03)$\times 10^{39}$&2.0 &	4.2$\pm$ 0.8\\
\multicolumn{10}{c}{AM1256-433E} \\
b&--9 &56 &268.86&15.64 $\pm$ 0.27&1.61&	1.96&---                              &0.6 &	13.1$\pm$ 1.1\\
a&0  &0  &154.24&18.73 $\pm$ 0.27&1.11&	1.73&---                              &1.4 &	7.0$\pm$ 0.8\\
\hline
\\
\multicolumn{10}{c}{AM1401-324N} \\
N&0&0&210.25	&15.89 $\pm$ 0.30&0.55&	0.74&(7.67 $\pm$ 0.08)$\times 10^{40}$&1.1 &	10.2$\pm$ 1.0\\
1&6&16&120.45	&18.96 $\pm$ 0.30&0.31&	0.66&(2.22 $\pm$ 0.02)$\times 10^{40}$&1.9 &	6.2$\pm$ 0.7\\
2&15&6&210.25	&17.33 $\pm$ 0.30&0.33&	0.70&(1.19 $\pm$ 0.01)$\times 10^{41}$&2.0 &	6.2$\pm$ 0.7\\
3&--11&--6&223.83	&15.91 $\pm$ 0.30&0.47&	0.62&(6.74 $\pm$ 0.07)$\times 10^{40}$&1.1 &	10.1$\pm$ 1.0\\
4&--23&8&178.15	&18.57 $\pm$ 0.30&0.47&	0.72&(1.89 $\pm$ 0.02)$\times 10^{40}$&1.6 &	6.6$\pm$ 0.8\\
5&--43&--13&120.45&21.71 $\pm$ 0.30&0.14&	0.16&(8.30 $\pm$ 0.10)$\times 10^{39}$&2.3 &	5.7$\pm$ 0.7\\
6&--26&38&120.45	&21.01 $\pm$ 0.30&0.35&	0.39&(5.20 $\pm$ 0.20)$\times 10^{39}$&2.1 &	6.0$\pm$ 0.7\\
\multicolumn{10}{c}{AM1401-324S} \\
N&0&0&	152.11	&18.40 $\pm$ 0.30&0.71&	0.92&(6.00 $\pm$ 0.10)$\times 10^{39}$&1.3&	7.5$\pm$ 0.9\\
\hline
\\
\multicolumn{10}{c}{AM1448-262NE} \\
N&0&0&45.09	&16.95 $\pm$ 0.40&1.31&	1.63&	---&	0.5&	13.7$\pm$ 1.2\\
\multicolumn{10}{c}{AM1448-262SW} \\
1&0&0&119.83	&16.29 $\pm$ 0.40&0.76&	1.30&	---&	1.6 &	7.1$\pm$ 0.7\\
2&--3&--2&66.84	&17.36 $\pm$ 0.40&0.85&	1.42&	---&	1.8 &	7.1$\pm$ 1.0\\
3&4&--10&35.73	&18.69 $\pm$ 0.40&0.75&	1.25&	---&	1.7 &	7.0$\pm$ 0.7\\
4&10&--4&23.97	&19.54 $\pm$ 0.40&0.67&	1.11&	---&	1.6 &	7.1$\pm$ 0.8\\
\hline
\\
\multicolumn{10}{c}{AM2030-303SW} \\
1&0&0&314.96	&17.33 $\pm$ 0.40& 0.53&1.05&(8.40 $\pm$1.00)$\times 10^{40}$&	1.9 &	6.5$\pm$ 0.7\\
2&17&--7&109.1	&20.50 $\pm$ 0.40& 0.61&1.02&(8.40 $\pm$0.80)$\times 10^{39}$&	1.9 &	6.5$\pm$ 0.7\\
\multicolumn{10}{c}{AM2030-303NE} \\
1&0&0&272.96	&18.41 $\pm$ 0.40& 0.84&1.30&(1.60 $\pm$0.20)$\times 10^{38}$&	1.7 &	6.4$\pm$ 0.7\\
2&--8&21&345.27	&17.49 $\pm$ 0.40& 0.60&1.15&(1.40 $\pm$8.00)$\times 10^{39}$&	2.1 &	6.0$\pm$ 0.7\\
3&--24&18&252.15	&18.12 $\pm$ 0.40& 0.61&1.17&(1.10 $\pm$0.20)$\times 10^{40}$&	1.7 &	6.4$\pm$ 0.7\\
4&34&19&109.18	&24.36 $\pm$ 0.40& 1.30&0.95&(6.60 $\pm$2.00)$\times 10^{38}$&	2.7 &	4.7$\pm$ 0.5\\
\hline
\\
\multicolumn{10}{c}{AM2058-381N} \\
N&0&0&180.64	&17.91 $\pm$ 0.27&0.99&	1.14&(8.00 $\pm$ 0.10)$\times 10^{39}$&	1.1 &	7.5$\pm$ 1.0\\
1&8&16&144.52	&20.96 $\pm$ 0.27&0.32&	0.67&(2.39 $\pm$ 0.03)$\times 10^{39}$&	2.2 &	5.0$\pm$ 1.0\\
2&12&55&144.52	&20.74 $\pm$ 0.27&0.41&	0.47&(3.13 $\pm$ 0.05)$\times 10^{39}$&	2.4 &	5.6$\pm$ 0.6\\
3&14&47&144.52	&20.69 $\pm$ 0.27&0.31&	0.62&(2.45 $\pm$ 0.05)$\times 10^{39}$&	2.2 &	5.0$\pm$ 1.0\\
4&23&--8&265.49	&17.95 $\pm$ 0.27&0.33&	0.65&(2.26 $\pm$ 0.03)$\times 10^{39}$&	2.0 &	6.2$\pm$ 0.7\\
5&--4&--22&198.71	&18.29 $\pm$ 0.27&0.41&	0.75&(1.41 $\pm$ 0.01)$\times 10^{40}$&	1.9 &	6.3$\pm$ 0.7\\
6&--27&--28&144.52&20.68 $\pm$ 0.27&0.44&	0.50&(2.37 $\pm$ 0.05)$\times 10^{39}$&	2.1 &	5.1$\pm$ 1.0\\
7&22&8&	180.64	&18.97 $\pm$ 0.27&0.44&	0.70&(7.90 $\pm$ 0.10)$\times 10^{39}$&	1.9 &	6.3$\pm$ 0.7\\
8&--15&20&179.74	&18.75 $\pm$ 0.27&0.46&	0.75&(7.10 $\pm$ 0.10)$\times 10^{39}$&	1.7 &	6.5$\pm$ 0.7\\
9&--5&22&179.74	&18.68 $\pm$ 0.27&0.50&	0.77&(7.90 $\pm$ 0.10)$\times 10^{39}$&	1.7 &	3.7$\pm$ 1.1\\
10&2&13&180.64	&18.53 $\pm$ 0.27&0.58&	0.87&(8.30 $\pm$ 0.10)$\times 10^{39}$&	1.6 &	3.6$\pm$ 1.1\\
11&9&5&144.52	&18.99 $\pm$ 0.27&0.58&	0.86&(4.15 $\pm$ 0.06)$\times 10^{39}$&	1.5 &	4.0$\pm$ 1.0\\
\multicolumn{10}{c}{AM2058-381S} \\
N&0&0&345.36	&16.86$\pm$ 0.27&0.27&	0.47&(7.20 $\pm$ 0.10)$\times 10^{40}$&	1.6 &	6.6$\pm$ 0.8\\
\hline\hline

\end{tabular}
\end{center}
\end{table*}

\begin{table*}
\begin{center}
\begin{tabular}{cccccccccc} \hline\hline

{Region}&{X}&{Y}&{$r_{eq}$}&{B}&{B-V}&{V-I}&{$\cal L$ (H$\alpha$+[NII])}&{Log(EW)[$\AA$] }&{Age}\\
&{["]}&{["]}&{[pc]}&{[mag]}&&&{[erg s$^{-1}$]}&{$\pm$0.1}&{[10$^6$ yr]} \\
\hline
\multicolumn{10}{c}{AM2105-332SE} \\
N&0&0&158.98	&15.61$\pm$ 0.30&0.92&	1.65&(5.60 $\pm$0.04)$\times 10^{37}$&	0.6 &	8.4$\pm$ 1.4\\
\multicolumn{10}{c}{AM2105-332NW} \\
N&0&0&202.71	&16.55$\pm$ 0.30&0.63&	1.57&(4.36 $\pm$ 0.06)$\times 10^{38}$&	1.6 &	6.5$\pm$ 0.8\\
\hline
\\
\multicolumn{10}{c}{AM2229-735W} \\
N&0&0&417.2	&17.35$\pm$ 0.40&0.66&	1.21&(2.01 $\pm$ 0.01)$\times 10^{41}$&	1.8 &	7.1$\pm$ 0.8\\

1&4&19&280.22	&19.02$\pm$ 0.40&0.76&	1.28&(5.53 $\pm$ 0.02)$\times 10^{40}$&	1.8 &	6.9$\pm$ 0.9\\
2&--12&14&383.7	&19.53$\pm$ 0.40&0.55&	1.06&(9.84 $\pm$ 0.03)$\times 10^{40}$&	2.4 &	6.5$\pm$ 0.9\\
3&--36&34&204.64	&20.66$\pm$ 0.40&0.71&	1.18&(4.50 $\pm$ 0.10)$\times 10^{39}$&	1.5 &	7.2$\pm$ 0.4\\
4&14&31&306.96	&19.46$\pm$ 0.40&0.70&	1.17&(4.35 $\pm$ 0.02)$\times 10^{40}$&	1.9 &	6.8$\pm$ 0.9\\
5&1&10&260.87	&18.88$\pm$ 0.40&0.72&  1.24&(5.95 $\pm$ 0.03)$\times 10^{40}$&	1.8 &	6.9$\pm$ 0.9\\
6&1&--11&276.69	&18.40$\pm$ 0.40&0.61&  1.16&(5.36 $\pm$ 0.02)$\times 10^{40}$&	1.6 &	7.0$\pm$ 0.8\\
\multicolumn{10}{c}{AM2229-735E} \\
N&0&0&505.97	&18.13$\pm$ 0.40&0.56&	1.12&(1.89 $\pm$ 0.01)$\times 10^{41}$&	2.1 &	6.4$\pm$ 0.7\\
\hline
\\
\multicolumn{10}{c}{AM2238-575W}\\
N&0&0&205.86	&18.87$\pm$ 0.30&1.03&1.91&---&	1.7 &	6.4$\pm$ 0.7\\
1&3&--6&99.86	&20.30$\pm$ 0.30&1.04&1.80&---&	1.7 &	6.5$\pm$ 0.7\\
2&--6&--103&180.02&20.76$\pm$ 0.30&0.82&1.87&---&	1.9 &	6.2$\pm$ 0.7\\
3&--43&--83&315.77&18.38$\pm$ 0.30&0.78&1.70&---&	2.1 &	6.1$\pm$ 0.7\\
4&--64&--63&157.10&20.57$\pm$ 0.30&0.93&1.86&---&	2.1 &	6.1$\pm$ 0.7\\
5&--70&--54&165.59&20.12$\pm$ 0.30&0.78&1.76&---&	1.9 &	6.2$\pm$ 0.7\\
6&--86&34&126.31	&21.06$\pm$ 0.30&0.85&1.43&---&	1.8 &	6.3$\pm$ 0.7\\
7&--61&15&126.31	&20.34$\pm$ 0.30&0.89&1.78&---&	1.9 &	6.2$\pm$ 0.7\\
8&--58&--22&273.47&18.51$\pm$ 0.30&1.03&1.97&---&	1.9 &	6.2$\pm$ 0.7\\
9&--49&--49&157.87&19.49$\pm$ 0.30&0.99&1.80&---&	1.9 &	6.3$\pm$ 0.7\\
10&--39&--53&181.40&19.07$\pm$ 0.30&0.96&1.87&---&	1.8 &	6.3$\pm$ 0.7\\
11&--27&--61&181.40&19.15$\pm$ 0.30&0.85&1.85&---&	1.8 &	6.4$\pm$ 0.7\\
\multicolumn{10}{c}{AM2238-575E}\\
N&0&0&186.59	&17.69$\pm$ 0.30&1.24&1.9&---&0.7 	&	12.8$\pm$ 1.1\\
\hline
\\
\multicolumn{10}{c}{AM2306-721S} \\
N&0&0&	178.75	&17.75$\pm$ 0.27&0.62&	1.28&(3.98 $\pm$ 0.04)$\times 10^{40}$&	1.4 &	7.4$\pm$ 0.9\\
1&--22&23&139.86	&19.14$\pm$ 0.27&0.48&	0.85&(1.77 $\pm$ 0.02)$\times 10^{40}$&	1.9 &	6.5$\pm$ 0.7\\
2&--3&18&76.61	&20.23$\pm$ 0.27&0.50&	0.97&(5.12 $\pm$ 0.08)$\times 10^{39}$&	1.7 &	6.7$\pm$ 0.8\\
3&32&25&76.61	&20.86$\pm$ 0.27&0.43&	0.66&(3.26 $\pm$ 0.05)$\times 10^{39}$&	1.9 &	6.7$\pm$ 0.8\\
4&36&20&76.61	&20.92$\pm$ 0.27&0.43&	0.63&(3.50 $\pm$ 0.06)$\times 10^{39}$&	2.0 &	6.6$\pm$ 0.8\\
5&39&12&76.61	&20.88$\pm$ 0.27&0.44&	0.70&(3.25 $\pm$ 0.04)$\times 10^{39}$&	1.9 &	6.7$\pm$ 0.8\\
6&25&--19&229.82	&18.13$\pm$ 0.27&0.49&	0.93&(4.23 $\pm$ 0.04)$\times 10^{40}$&	1.8 &	6.6$\pm$ 0.8\\
7&11&--20&127.68	&19.12$\pm$ 0.27&0.48&	0.92&(1.37 $\pm$ 0.02)$\times 10^{40}$&	1.7 &	6.6$\pm$ 0.8\\
8&--7&--23&102.14	&19.96$\pm$ 0.27&0.48&	0.91&(9.55 $\pm$ 0.08)$\times 10^{39}$&	1.9 &	6.5$\pm$ 0.8\\
9&--9&--16&102.14	&19.62$\pm$ 0.27&0.48&	0.91&(9.22 $\pm$ 0.08)$\times 10^{39}$&	1.8 &	6.6$\pm$ 0.8\\
10&--28&--5&102.14&19.75$\pm$ 0.27&0.48&	0.88&(5.23 $\pm$ 0.07)$\times 10^{39}$&	1.6 &	6.7$\pm$ 0.8\\
\multicolumn{10}{c}{AM2306-721N} \\
1&0&0&214.96	&16.91$\pm$ 0.27&0.45&	0.83&(4.75 $\pm$ 0.07)$\times 10^{41}$&	1.9 &	6.8$\pm$ 0.8\\
2&9&9&158.76	&17.25$\pm$ 0.27&0.50&	1.00&(2.71 $\pm$ 0.01)$\times 10^{41}$&	1.8 &	6.7$\pm$ 0.8\\
3&19&23&242.51	&16.86$\pm$ 0.27&0.42&	0.59&(5.17 $\pm$ 0.08)$\times 10^{41}$&	2.0 &	6.6$\pm$ 0.8\\
\hline\hline

\end{tabular}
\end{center}
\end{table*}

\begin{table*}
\begin{center}
\begin{tabular}{cccccccccc} \hline\hline

{Region}&{X}&{Y}&{$r_{eq}$}&{B}&{B-V}&{V-I}&{$\cal L$ (H$\alpha$+[NII])}&{Log(EW)[$\AA$] }&{Age}\\
&{["]}&{["]}&{[pc]}&{[mag]}&&&{[erg s$^{-1}$]}&{$\pm$0.1}&{[10$^6$ yr]} \\
\hline
\multicolumn{10}{c}{AM2322-821SE} \\
N&0&0&99.28	&16.46$\pm$ 0.30&1.18&	1.22&(1.61 $\pm$ 0.03)$\times 10^{40}$&	1.5 &	7.3$\pm$ 0.8\\
1&11&8&53.68	&18.71$\pm$ 0.30&1.15&	1.20&(2.25 $\pm$ 0.03)$\times 10^{39}$&	1.6 &	6.1$\pm$ 1.2\\
2&20&9&48.32	&19.21$\pm$ 0.30&1.07&	1.05&(1.86 $\pm$ 0.03)$\times 10^{39}$&	1.8 &	6.1$\pm$ 1.1\\
3&22&3&37.19	&19.84$\pm$ 0.30&1.06&	1.10&(8.90 $\pm$ 0.10)$\times 10^{38}$&	1.7 &	6.6$\pm$ 1.0\\
4&27&--6&59.05	&18.56$\pm$ 0.30&0.92&	0.88&(2.35 $\pm$ 0.04)$\times 10^{39}$&	1.7 &	5.8$\pm$ 1.2\\
5&25&--15&48.32	&19.07$\pm$ 0.30&0.93&	0.89&(1.43 $\pm$ 0.03)$\times 10^{39}$&	1.7 &	6.2$\pm$ 1.0\\
6&19&--22&45.22	&19.57$\pm$ 0.30&0.89& 0.85&(1.28 $\pm$ 0.03)$\times 10^{39}$&	1.7 &	6.1$\pm$ 1.0\\
7&13&--57&43.61	&20.08$\pm$ 0.30&0.86&	0.84&(1.20 $\pm$ 0.01)$\times 10^{39}$&	2.1 &	6.2$\pm$ 0.8\\
8&16&--61&42.61	&20.08$\pm$ 0.30&0.95& 0.84&(1.26 $\pm$ 0.01)$\times 10^{39}$&	2.1 &	6.1$\pm$ 0.9\\
9&4&--63&32.21	&20.76$\pm$ 0.30&0.82&	0.74&(7.50 $\pm$ 0.30)$\times 10^{38}$&	2.2 &	6.2$\pm$ 0.7\\
10&8&--32&59.05	&19.01$\pm$ 0.30&0.97&	0.94&(1.41 $\pm$ 0.03)$\times 10^{39}$&	1.7 &	6.3$\pm$ 1.0\\
11&--24&--29&61.68&19.04$\pm$ 0.30&0.95&	0.89&(1.93 $\pm$ 0.03)$\times 10^{39}$&	1.8 &	6.0$\pm$ 1.0\\
12&--49&--59&58.32&18.87$\pm$ 0.30&0.93& 0.87&(1.43 $\pm$ 0.03)$\times 10^{39}$&	2.0 &	5.7$\pm$ 1.1\\
13&--76&--6&80.35	&18.64$\pm$ 0.30&0.74&	0.71&(3.44 $\pm$ 0.06)$\times 10^{39}$&	2.1 &	5.4$\pm$ 1.2\\
14&--41&4&59.05	&19.06$\pm$ 0.30&0.91&	0.87&(1.61 $\pm$ 0.03)$\times 10^{39}$&	1.8 &	6.1$\pm$ 1.0\\
15&--37&18&48.32	&19.44$\pm$ 0.30&0.93&	0.86&(1.35 $\pm$ 0.01)$\times 10^{39}$&	1.9 &	6.1$\pm$ 0.9\\
16&--34&30&42.95	&20.06$\pm$ 0.30&0.98&	0.99&(8.30 $\pm$ 0.10)$\times 10^{38}$&	1.8 &	6.5$\pm$ 0.9\\
17&--22&52&42.95	&20.26$\pm$ 0.30&0.96&	0.89&(7.60 $\pm$ 0.10)$\times 10^{38}$&	2.0 &	6.4$\pm$ 0.7\\
18&--45&72&42.95	&20.41$\pm$ 0.30&0.84&	0.84&(9.40 $\pm$ 0.10)$\times 10^{38}$&	2.1 &	6.2$\pm$ 0.7\\
19&4&92&45.23	&18.77$\pm$ 0.30&0.87& 0.82&(1.43 $\pm$ 0.03)$\times 10^{39}$&	1.8 &	6.0$\pm$ 1.0\\
20&4&35&72.02	&18.78$\pm$ 0.30&0.99&	1.03&(2.10 $\pm$ 0.03)$\times 10^{39}$&	1.7 &	6.0$\pm$ 1.2\\
21&29&138&58.81	&19.93$\pm$ 0.30&0.73&	0.60&(2.07 $\pm$ 0.03)$\times 10^{39}$&	2.4 &	5.5$\pm$ 1.0\\
22&37&129&48.02	&20.30$\pm$ 0.30&0.71&	0.65&(1.25 $\pm$ 0.01)$\times 10^{39}$&	2.3 &	5.9$\pm$ 0.6\\
23&49&134&59.05	&19.79$\pm$ 0.30&0.63&	0.63&(2.23 $\pm$ 0.04)$\times 10^{39}$&	2.4 &	5.9$\pm$ 1.0\\
24&95&109&72.02	&19.57$\pm$ 0.30&0.68&	0.68&(1.84 $\pm$ 0.03)$\times 10^{39}$&	2.2 &	6.1$\pm$ 1.0\\
25&101&101&42.95&20.47$\pm$ 0.30&0.62&	0.58&(6.00 $\pm$ 0.10)$\times 10^{38}$&	2.1 &	6.4$\pm$ 0.7\\
26&135&--20&64.42&20.22$\pm$ 0.30&0.71&	0.66&(1.40 $\pm$ 0.03)$\times 10^{39}$&	2.3 &	5.8$\pm$ 0.7\\
27&105&--85&64.42&20.62$\pm$ 0.30&0.86&	0.60&(1.51 $\pm$ 0.03)$\times 10^{39}$&	2.4 &	5.7$\pm$ 0.8\\
\multicolumn{10}{c}{AM2322-821NW} \\
1&0&0&81.16	&17.78$\pm$ 0.30&0.67&	0.76&(1.34 $\pm$ 0.01)$\times 10^{40}$&	1.3 &	7.3$\pm$ 0.8\\
2&9&1&56.07	&18.32$\pm$ 0.30&0.87&	0.96&(3.74 $\pm$ 0.04)$\times 10^{39}$&	1.6 &	5.8$\pm$ 1.4\\
3&18&--11&59.94	&18.98$\pm$ 0.30&0.77&	0.81&(2.65 $\pm$ 0.03)$\times 10^{39}$&	1.6 &	6.0$\pm$ 1.2\\
4&15&4&46.86	&18.49$\pm$ 0.30&1.04& 1.08&(1.40 $\pm$ 0.01)$\times 10^{39}$&	1.7 &	6.5$\pm$ 1.0\\
5&41&15&49.95	&19.54$\pm$ 0.30&0.68&	0.77&(1.77 $\pm$ 0.03)$\times 10^{39}$&	1.7 &	6.2$\pm$ 1.0\\
6&31&29&69.21	&18.35$\pm$ 0.30&0.68&	0.66&(2.62 $\pm$ 0.03)$\times 10^{39}$&	2.3 &	5.6$\pm$ 1.0\\
7&20&18&39.96	&18.90$\pm$ 0.30&0.78&	0.85&(1.40 $\pm$ 0.01)$\times 10^{39}$&	2.0 &	6.3$\pm$ 1.0\\
\hline
\multicolumn{10}{c}{AM2330-451NE} \\
N&0&0&66.45	&15.21$\pm$ 0.30&0.80&1.48& ---&0.2 &15.3$\pm$ 1.4\\
\multicolumn{10}{c}{AM2330-451SW} \\
1&0&0&51.80	&19.51$\pm$ 0.30&0.84&1.12&---&1.5 &6.9$\pm$ 0.8 \\
2&18&--17&56.75	&18.92$\pm$ 0.30&0.75&1.05& ---&--- &---\\
3&6&--34&67.14	&18.50$\pm$ 0.30&0.74&1.04&--- &1.5 &6.8$\pm$ 0.8\\
4&--5&--27&1.80	&19.07$\pm$ 0.30&0.77&1.06&--- &--- &---\\

\hline\hline

\end{tabular}
\end{center}
\end{table*}

\subsection{Luminosity profiles}
\label{prof}

Most of the observed galaxies have strong morphological perturbations due to the interaction with the close companion and show very irregular isophotes. For such disturbed galaxies, algorithms that trace azimuthal luminosity profiles (Jedrzejewski 1987) do not converge. In this case it is more convenient to obtain the equivalent profiles as defined by S\'ersic (1982). The equivalent radius $r_{eq}$ is r$_{eq}$ = (S(m)/$\pi$)$^{0.5}$.  S(m) is proportional to the projected area (in arcsec$^2$) for pixels with the intensity I(m) = 10$^{-0.4m}$ satisfying the condition I(m')$>$I(m). These profiles have the advantage that they can be traced for any galaxy morphology.

We have obtained BVI equivalent profiles for each galaxy and decomposed then into bulge and disk components when possible.
The functional form adopted for each of these components is:\\

  \begin{equation}  \label{eq1}
  I(r)  =  I_{\mathrm{e}} \; exp\;\left[ -7.688\;\left( \left( \frac{r} {r_{\mathrm{eff}}} \right)^{0.25}-1\right) \;\right]
  \end{equation}

\noindent

  \begin{equation} \label{eq2}
  I(r)  =  I_{\mathrm{d}} \; exp  \left[ -\; \frac{r} {d_{\mathrm{l}}}\; \right] \\
  \end{equation}

\noindent
for the bulge and the disk, respectively.
In the above equations $I_e$ is the intensity at the effective radius $r_{eff}$, which encloses half of the total luminosity of the bulge, $I_d$ is the central intensity and $d_l$ the scale length of the disk component.

In general terms we followed the method described by Shombert \& Bothum (1987) and we used the NFIT routine implemented in STSDAS in order to obtain the parameters mentioned above. The routine must be provided with appropriate initial parameters in order to apply the fit. Disk parameters can be inferred directly through the profile since it is not seriously contaminated by the bulge in the outermost region of the galaxy. However, bulge parameters were difficult to infer since most of the galaxies show humps in the luminosity profile in the innermost regions. In these cases, we have only fitted a disk profile.
Errors in the parameters were calculated by making small variations to the initial bulge and disk parameters. We verified that the differences remain within 20\%.
The total luminosity for each component was calculated using the expressions 6 and 7 from Boris et al. (2001) with the above parameters.  The equivalent luminosity profiles together with the fitted functions are presented in Fig.~\ref{fig1} -~\ref{fig11}, panels E and F. The photometric bulge and disk parameters for each filter are listed in Table~\ref{Tparameters}. The effective and central surface brightness
$m_e$ = -- 2.5 $log(I_e)$ and $m_0$ = -- 2.5 $log$$(I_0)$, are expressed in $mag$ $arcsec^{-2}$.

\begin{table*}[h]
\caption[]{ Bulge and disk parameters}
\label{Tparameters} 

\begin{center}
\begin{tabular}{l|cccc|cccc|cccc} \hline\hline

\multicolumn{1}{c|}{Galaxy}&{m$_e$}&{r$_{eff}$}&{b$_0$}&{d$_l$}&{m$_e$}&{r$_{eff}$}&{b$_0$}&{d$_l$}&{m$_e$}&{r$_{eff}$}&{b$_0$}&{d$_l$}\\
&\multicolumn{4}{c}{B}&\multicolumn{4}{|c|}{V}&\multicolumn{4}{c}{I}\\

\hline

AM1256-433W & --- & --- & 20.96 & 6.55 & --- & --- & 20.82 & 7.01 & --- & --- & 20.00 & 6.50 \\

AM1256-433E & --- & --- & 21.99 & 5.14 & --- & --- & 21.10 & 4.49 & --- & --- & 20.60 & 4.99 \\

AM1401-324N & 20.65 & 2.86 & 22.65 & 10.90 & 21.20 & 5.60 & 24.2 & 24.12 & 19.79 & 5.00 & 21.50 & 8.31 \\

AM1401-324S & --- & --- & 22.75 & 4.30 & --- & ---- & 22.12 & 4.00 & --- & --- & 20.75 & 3.5 \\

AM1448-262NE & 22.55 & 7.34 & 22.61 & 13.56 & 21.79 & 9.93 & 21.64 & 8.10 & 20.75 & 12.63 & 19.51 & 5.73 \\

AM1448-262SW & --- & --- & 22.55 & 8.68 & --- & --- & 21.99 & 9.25 & --- & --- & 20.45 & 8.45 \\

AM2030-303SW & 21.10 & 1.10  & 21.22 & 5.30 & 22.72 & 6.20 & 20.80 & 5.00 & 20.90 & 4.50 & 19.87 & 5.99 \\

AM2030-303NE & --- & --- & 21.83 & 3.80 & --- & --- & 21.83 & 4.70 & --- & --- & 20.44 & 5.00 \\

AM2058-381N & 24.00 & 5.10  & 21.00 & 5.20 & 22.55 & 5.10 & 20.41 & 5.20 & 21.70 & 4.50 & 18.90 & 4.40 \\

AM2058-381S & --- & --- & 22.27 & 5.77 & --- & --- & 22.17 & 6.53 & --- & --- & 21.22 & 6.50 \\

AM2105-332SE & 21.00 & 5.00 & 21.95 & 16.50 & 21.00 & 8.50 & 22.00 & 16.0 & 19.25 & 8.05 & 20.45 & 20.0 \\

AM2105-332NW & 19.65 & 1.20 & 21.75 & 6.20 & 19.65 & 1.70 & 21.75 & 7.65 & 19.45 & 3.50 & 21.30 & 15.65 \\

AM2229-735E & --- & --- & 20.54 & 3.20 & --- & --- & 19.84 & 3.26 & --- & --- & 18.59 & 3.16 \\

AM2229-735W & --- & ---  & 23.30 & 6.6 & --- & --- & 22.48 & 5.60 & --- & --- & 21.38 & 5.90 \\

AM2238-575W & 24.10 & 12.5 & 21.45 & 13.37 & 22.21 & 10.00 & 20.25 & 11.50 & 20.35 & 7.26 & 19.15 & 10.52 \\

AM2238-575E & --- & --- & 20.71 & 2.67  & --- & --- & 19.66 & 2.90  & --- & --- & 18.09 & 3.15  \\

AM2306-721S & 23.50 & 6.50 & 20.30 & 7.20  & 22.80 & --- & --- & --- & --- & --- & --- & --- \\

AM2306-721N & --- & --- & 20.81 & 4.91 & --- & --- & 20.68 & 5.02 & --- & --- & 20.27 & 5.50  \\

AM2322-821SE & 22.85 & 6.51 & 22.06 & 27.02 & 21.65 & 5.50 & 21.02 & 21.30 & 20.60 & 6.00 & 20.12 & 21.60 \\

AM2322-821NW & --- & --- & 21.20 & 5.80 & --- & --- & 21.07 & 6.60 & --- & --- & 20.10 & 5.60 \\

AM2330-451NE & 22.67 & 27.52 & 21.12 & 5.83 & 19.5 & 11.00 & 20.1 & 22.7 & 18.4 & 11.4 & 18.73 & 22.88 \\

AM2330-451NE & --- & --- & 21.06 & 6.82 & --- & --- & 19.43 & 6.1 & --- & --- & 18.4 & 5.72 \\

\hline\hline

\end{tabular}
\end{center}
\end{table*}

\section{The galaxies}
\label{prop}

\begin{figure*}
\begin{center}
\includegraphics[width=7cm,height=7cm]{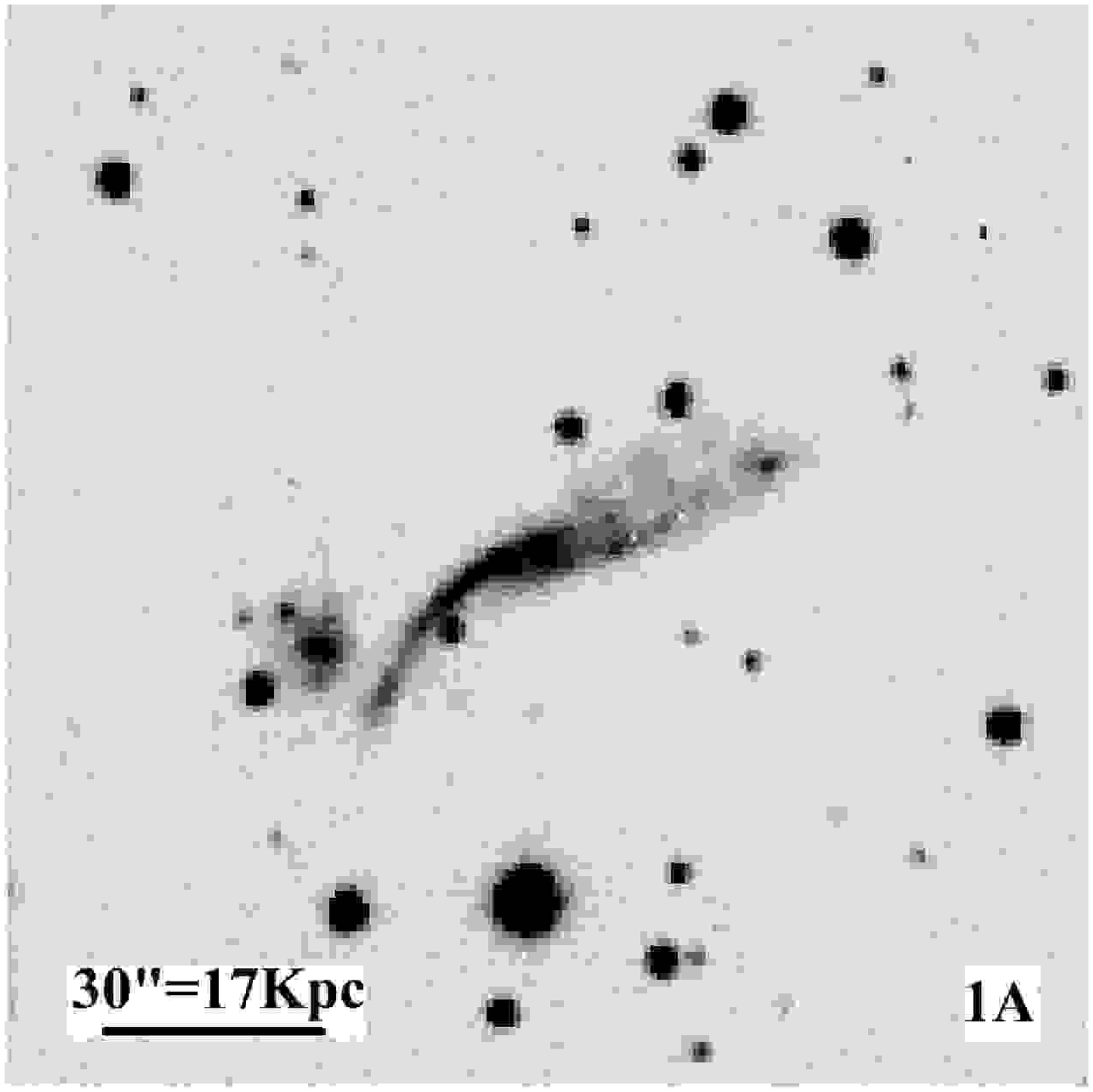}\includegraphics[width=7cm,height=7cm]{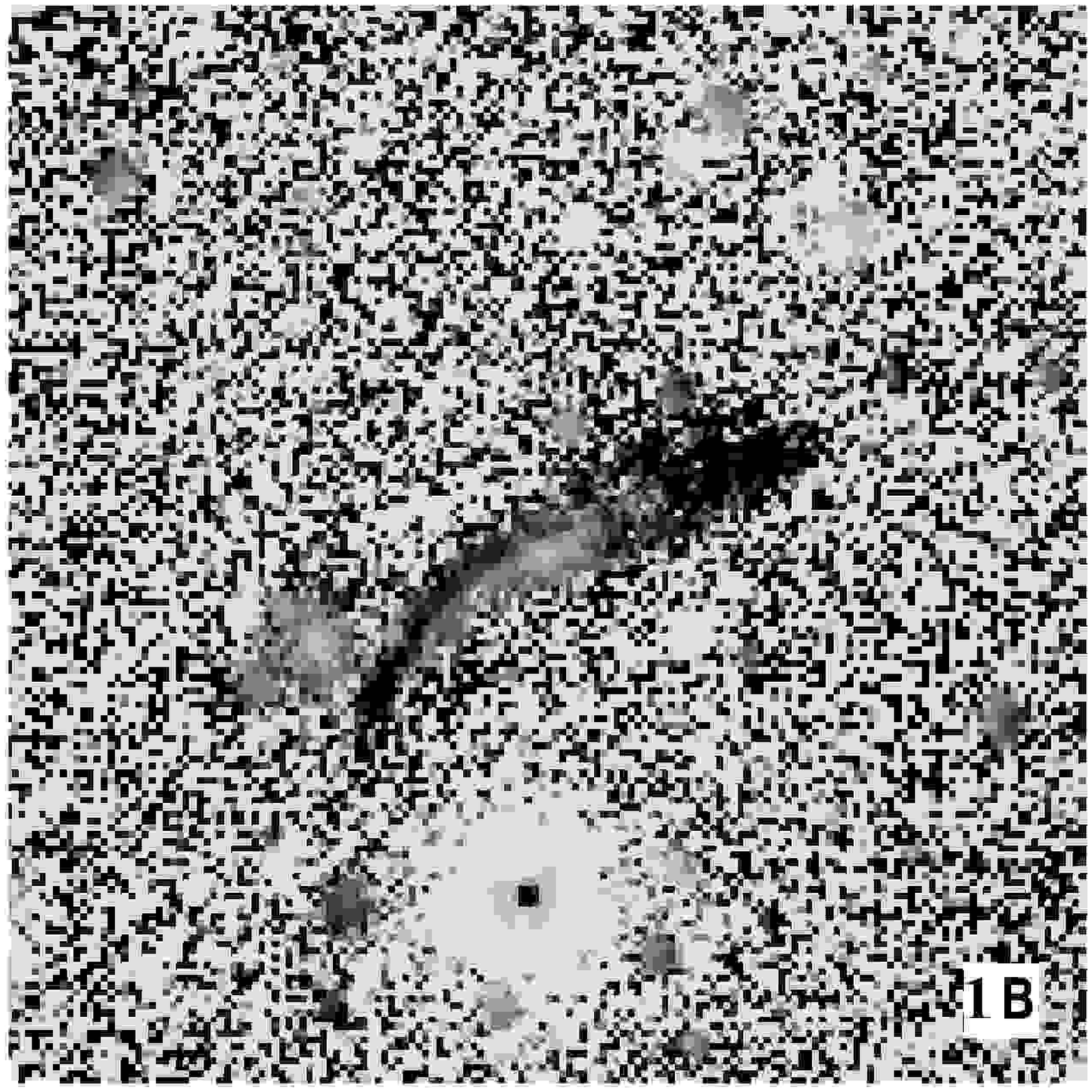}\\
\includegraphics[width=7cm,height=7cm]{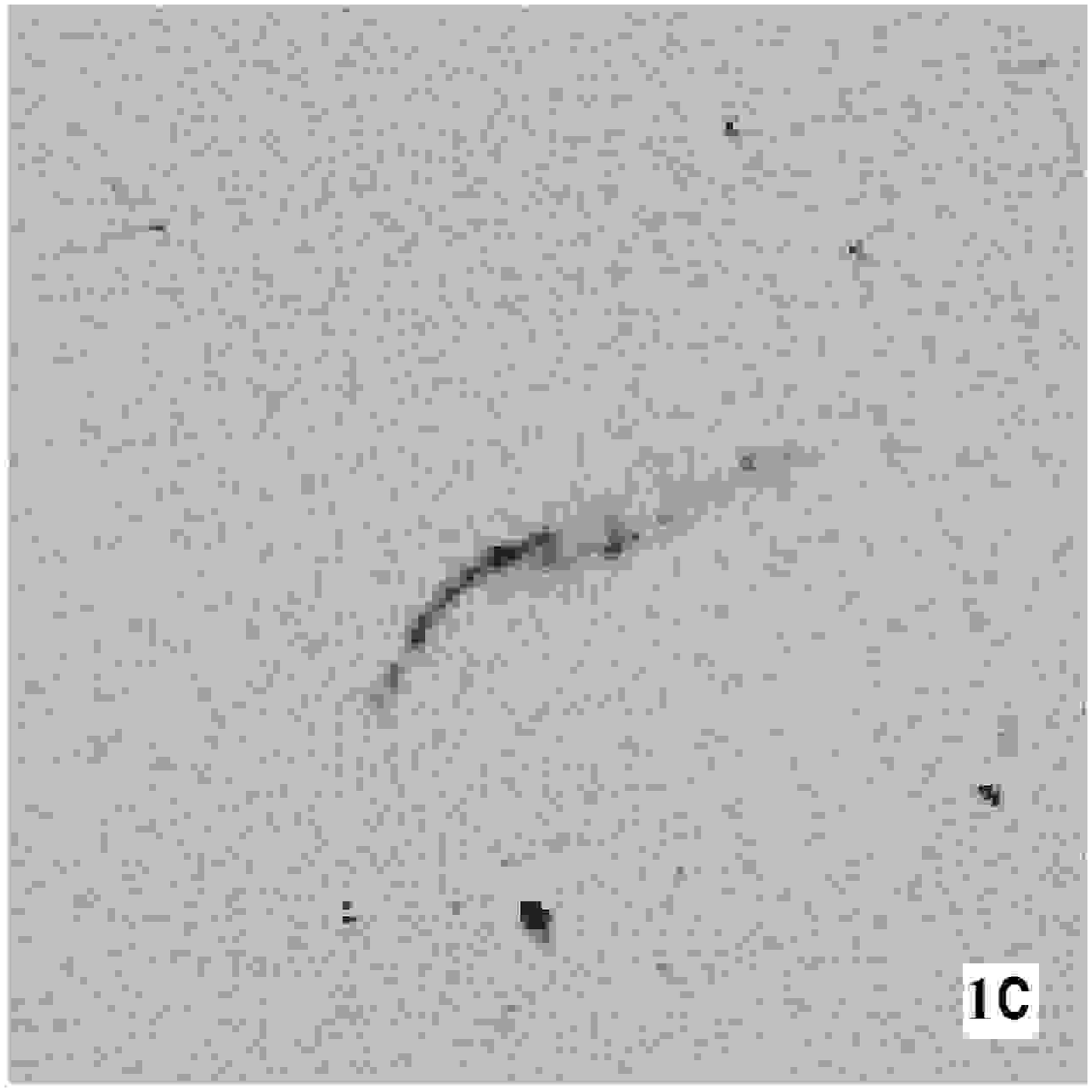}\includegraphics[width=7cm,height=7cm]{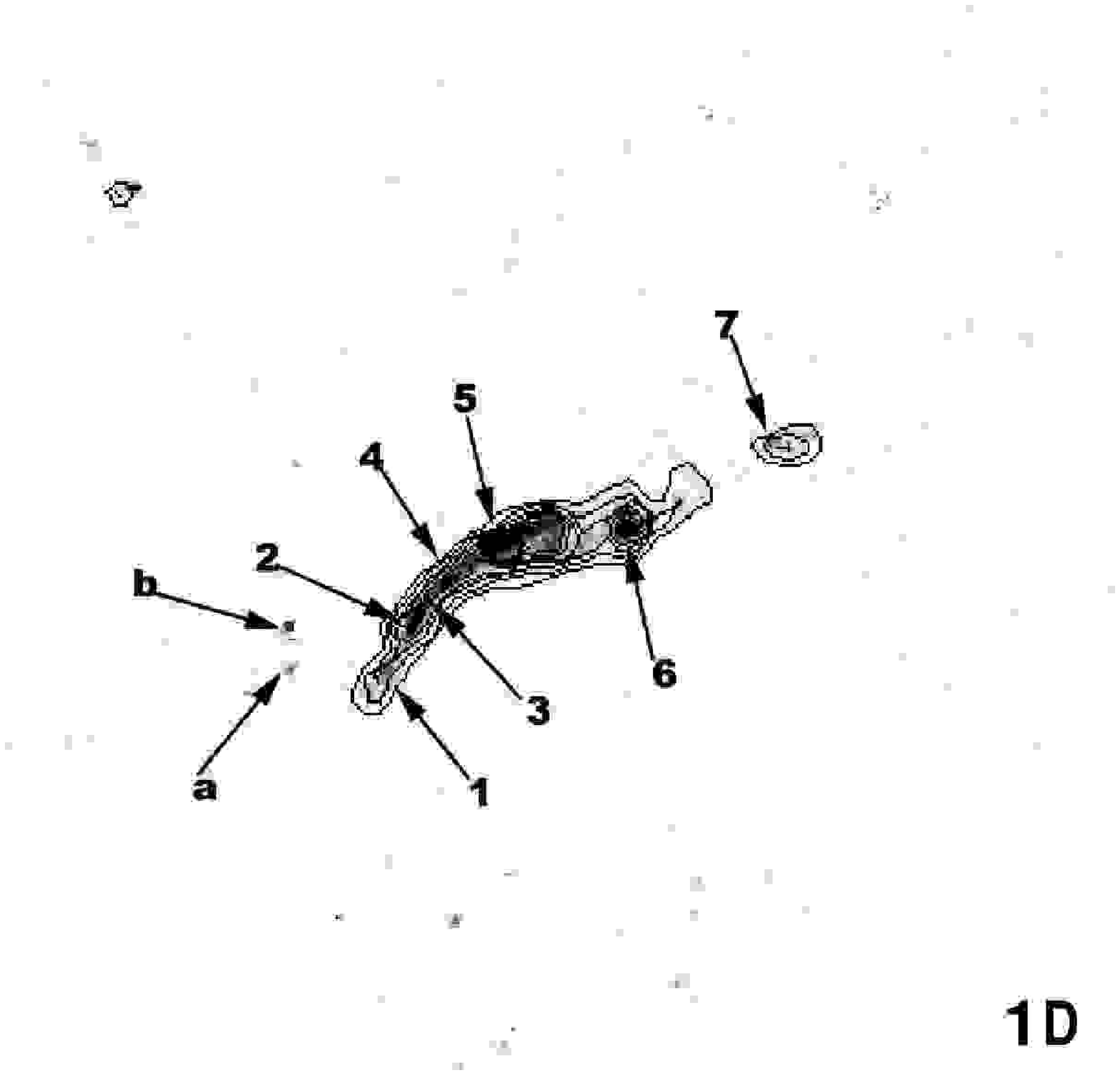}\\
\includegraphics[width=7cm,height=7cm]{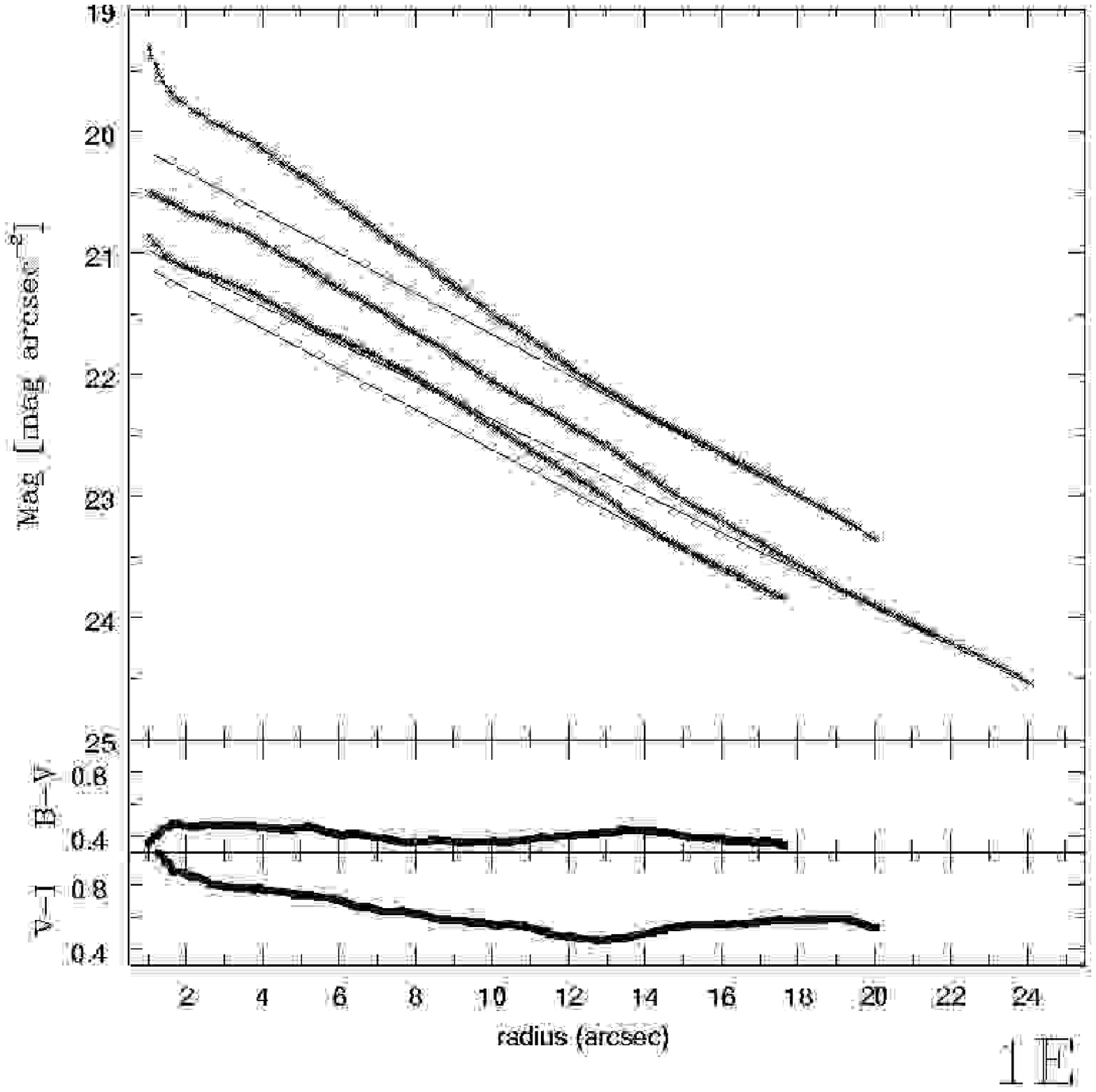}\includegraphics[width=7cm,height=7cm]{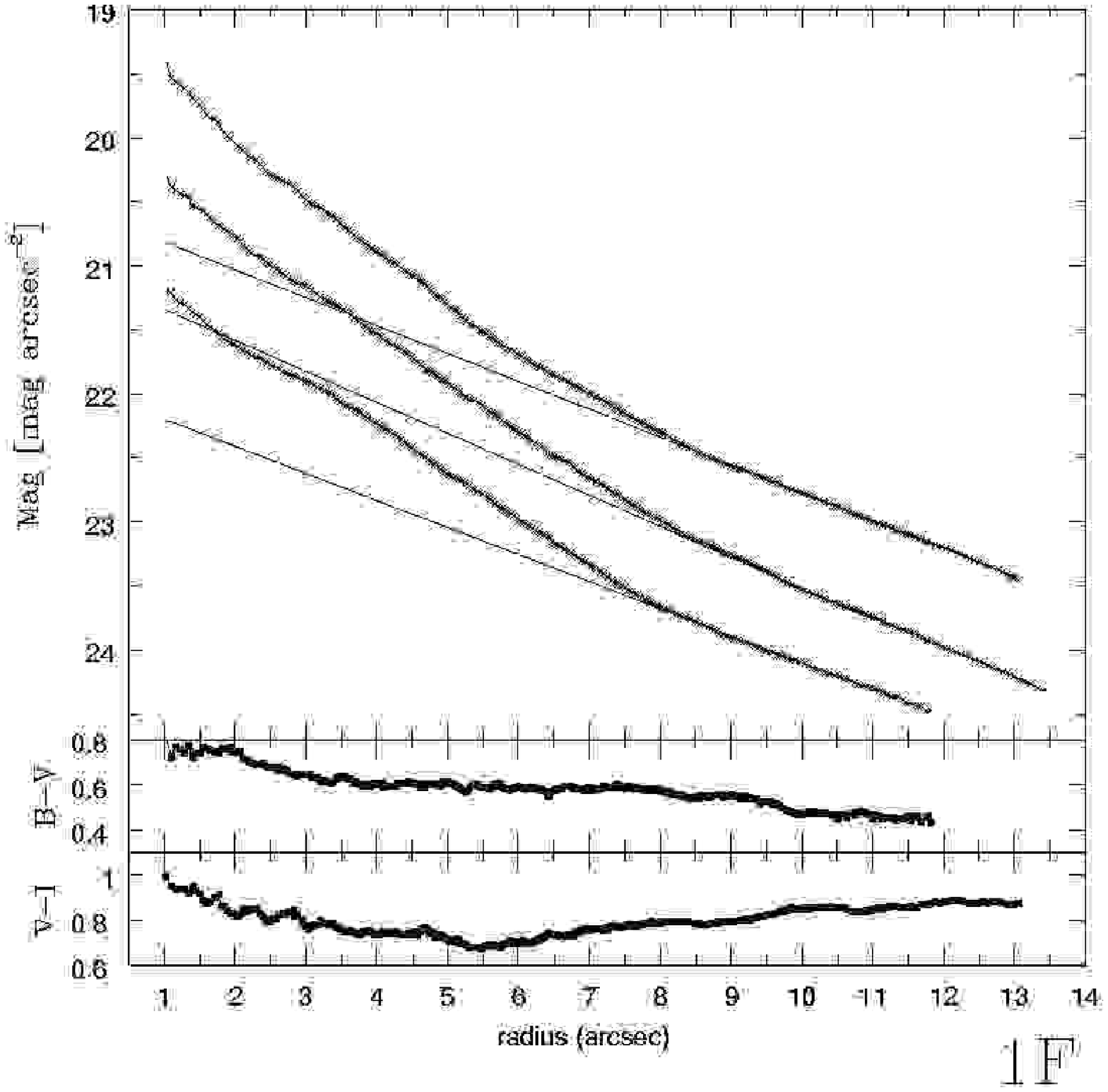}\\
\caption{ AM1256-433. Panels 1a) B image,  North is up and East left. The field is 200 x 200 arcsec. 1b) B-I image, 1c) Continuum-subtracted H$\alpha$ image and 1d) H$\alpha$ isophotes. Panels  1e) and 1f) Surface luminosity profiles of the primary and secondary components: B (down), V (middle), I (up). Solid line indicates the fitted bulge plus disk model, short dashed line is the fitted bulge and long dashed line is the exponential fitted disk. The B-V and V-I colours profiles are displaed in the lower panels.}
\label{fig1}
\end{center}
\end{figure*}

AM1256-433: This pair is composed of disk galaxies at $cz$ = 9014 $km$ $s^{-1}$ (see Fig.~\ref{fig1}). The luminosity profiles of both members are not purely exponential and show a luminosity excess in the inner 10 arcsec. The B-V colour of the main and secondary components is 0.47 and 0.61 respectively, typical values observed in Sc-Sd galaxies. The main component is very disturbed and presents two well developed arms resolved into HII regions.

The secondary galaxy shows a bar of about 2.5 arcsec long, oriented N-S. No arms are visible in this galaxy, only a tiny tail starting from the south side of the bar.\\

\begin{figure*}
\begin{center}
\includegraphics[width=7cm,height=7cm]{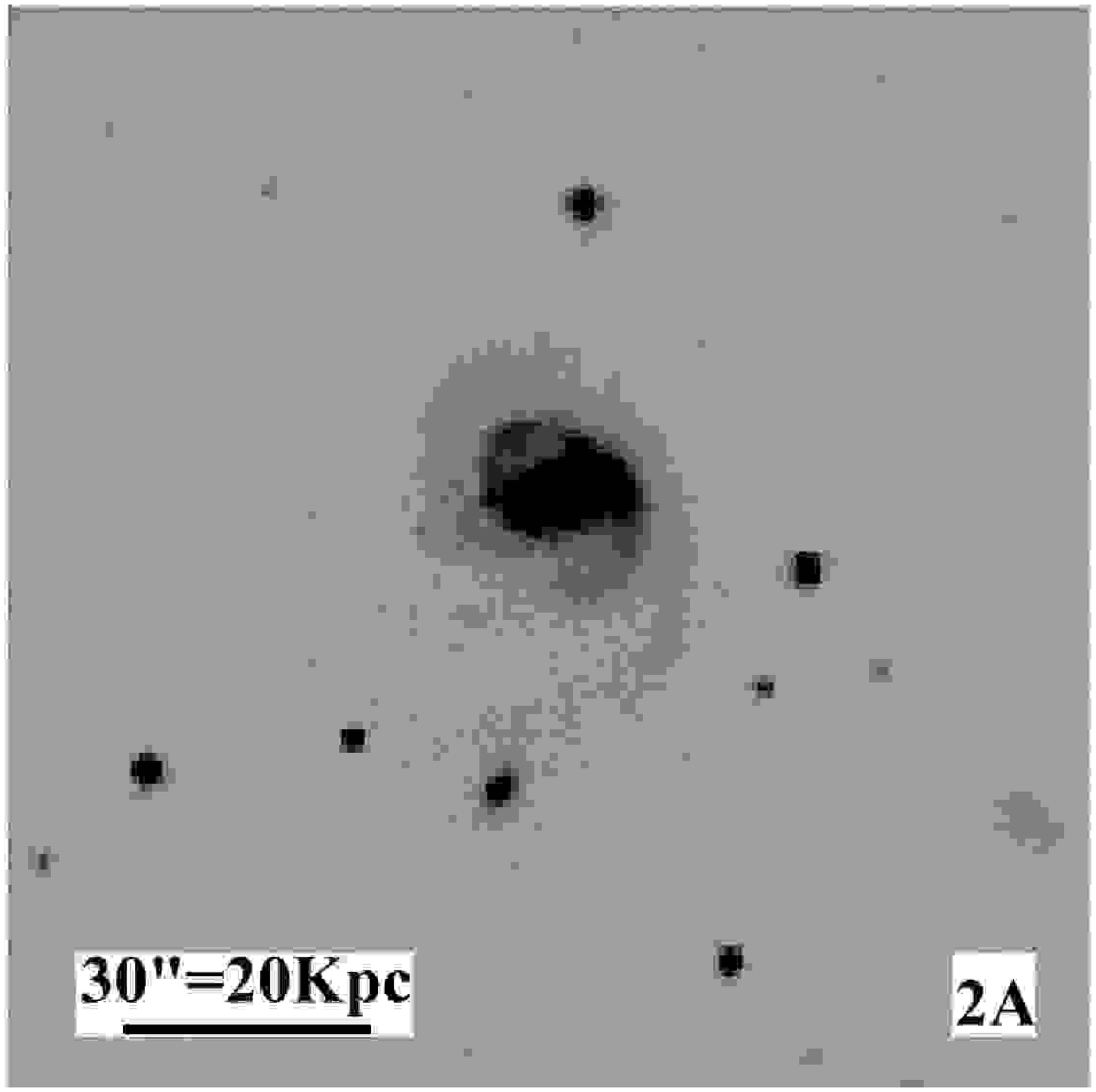}\includegraphics[width=7cm,height=7cm]{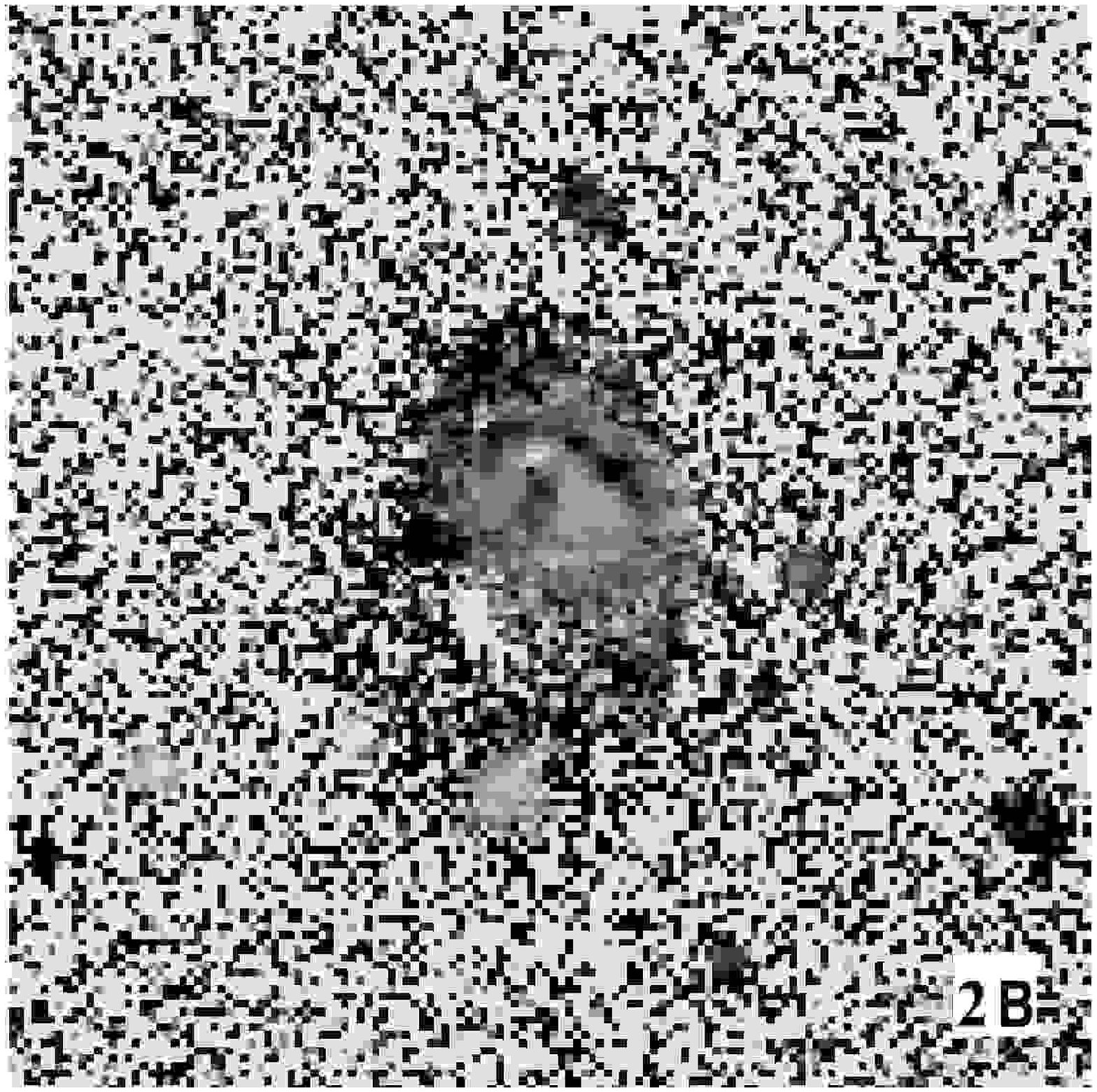}\\
\includegraphics[width=7cm,height=7cm]{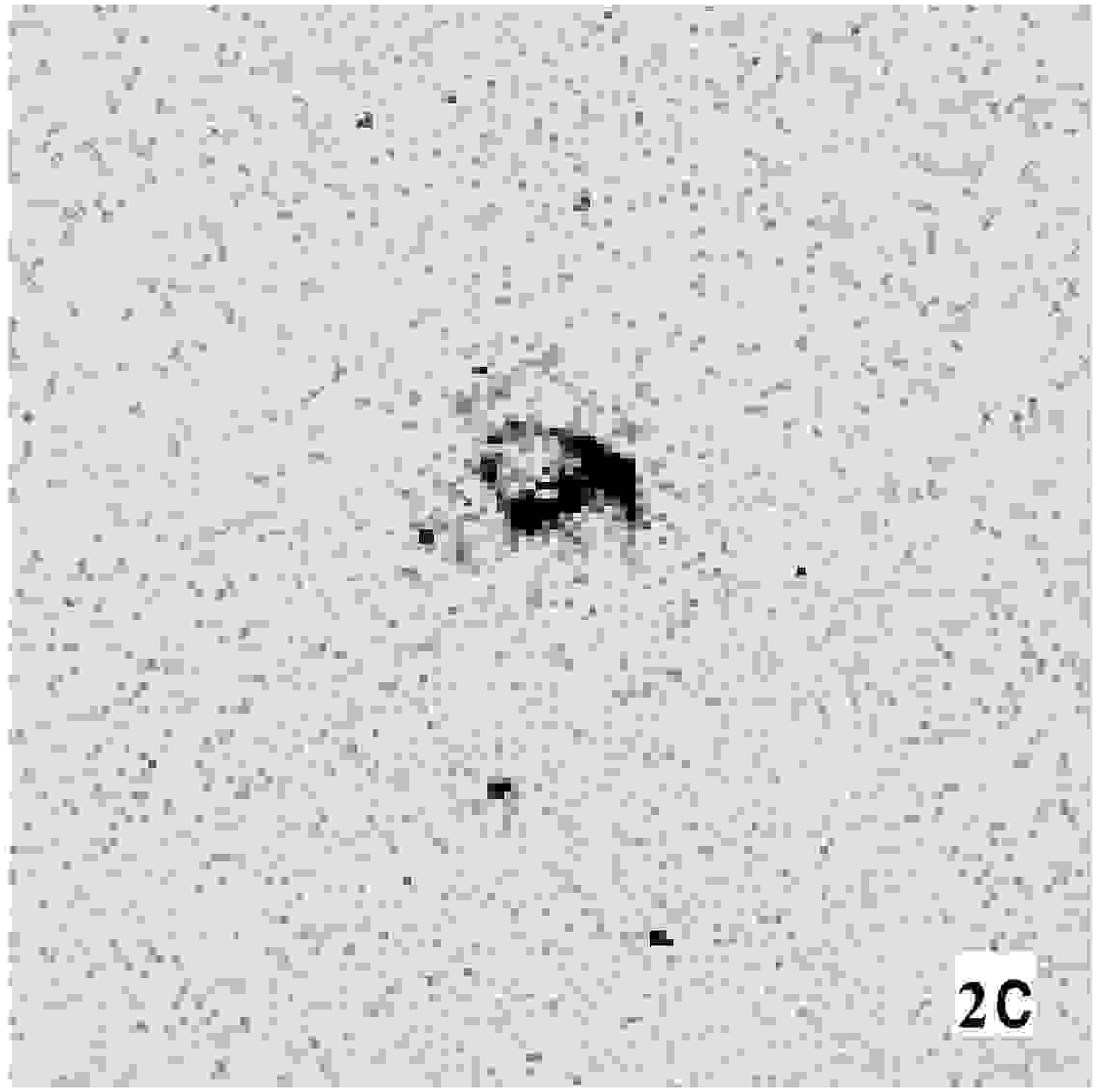}\includegraphics[width=7cm,height=7cm]{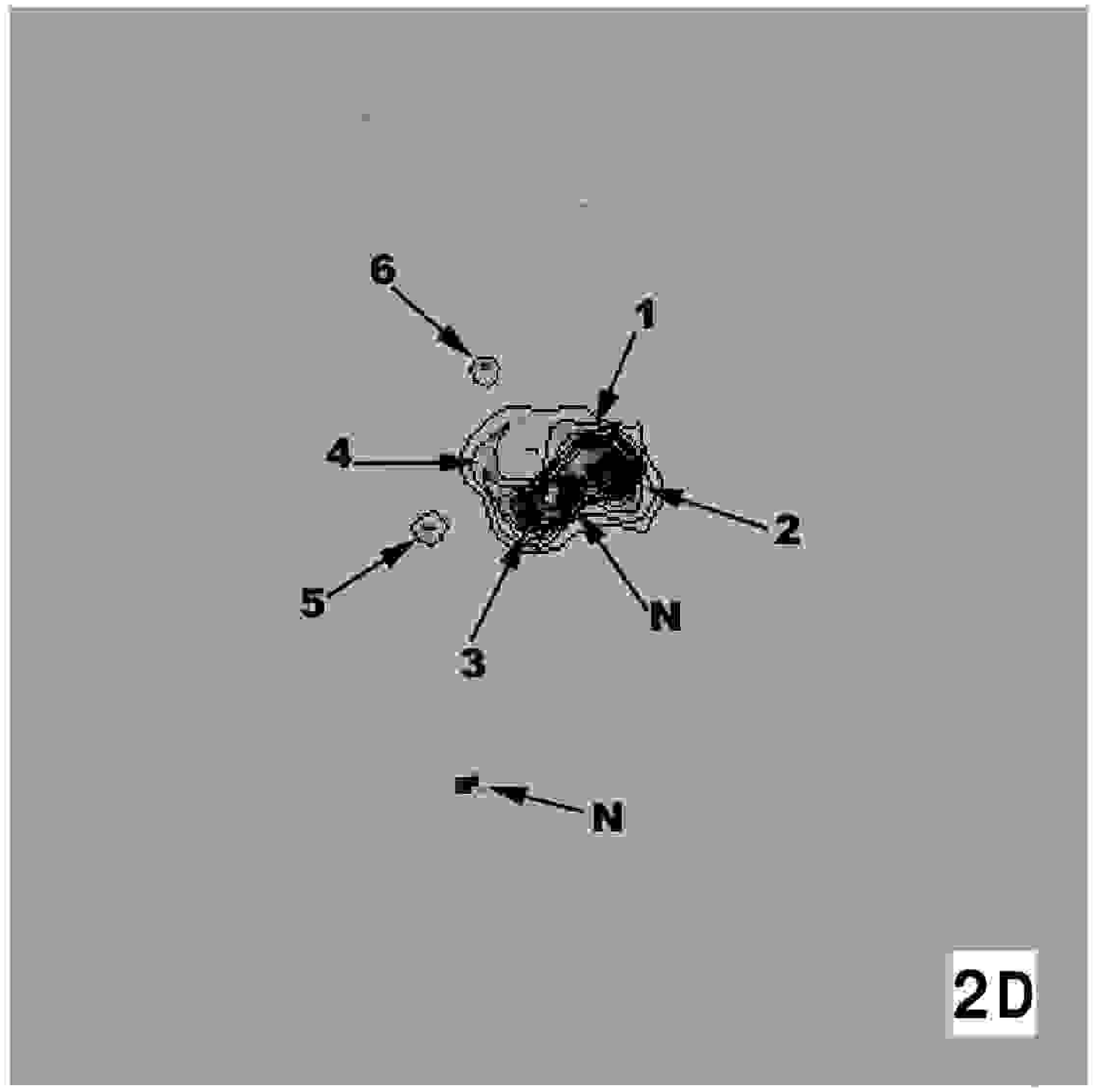}\\
\includegraphics[width=7cm,height=7cm]{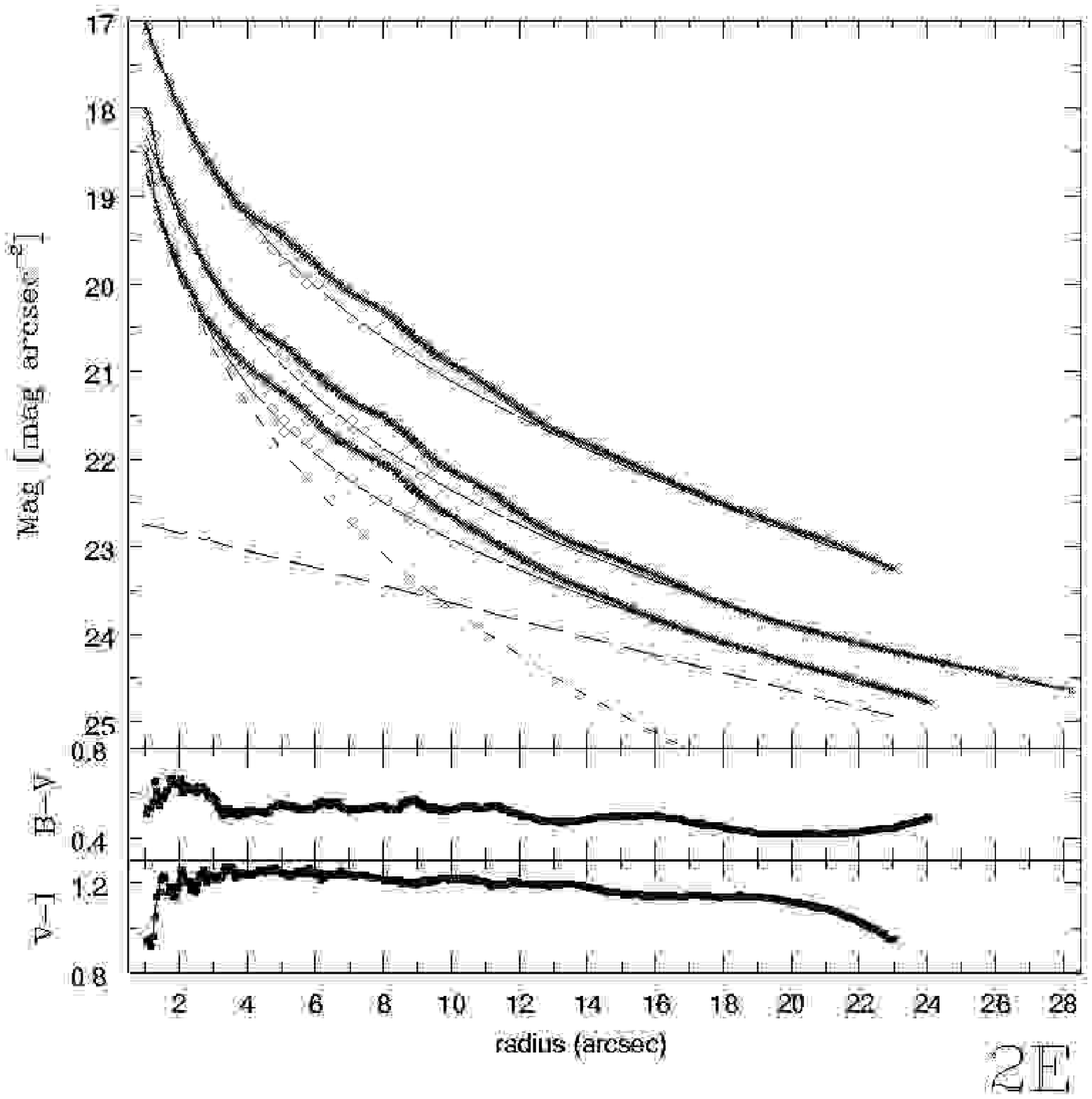}\includegraphics[width=7cm,height=7cm]{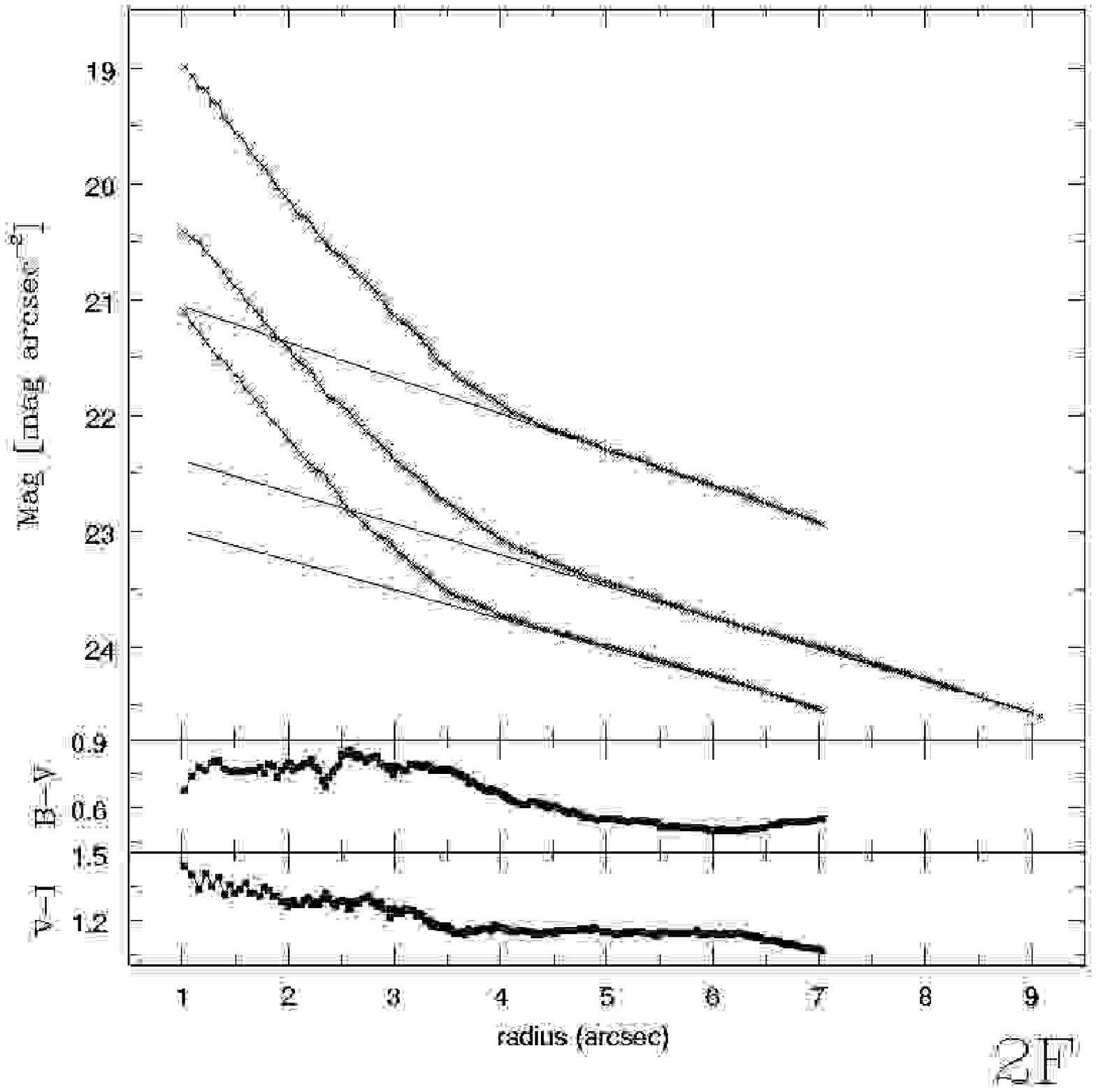}\\
\caption{AM1401-324. Same as Fig 1.}
\label{fig2}
\end{center}
\end{figure*}

AM1401-324: This is a typical M51 type pair at $cz$ = 10321 $km$  $s^{-1}$. The main galaxy has a very bright nucleus and one arm that wraps the whole galaxy, giving the appearance of a ring (see Fig.~\ref{fig2}). An incipient bar that has   several strong HII regions is also observed. This galaxy is the most luminous and bluest object in the whole sample.

The secondary galaxy is located at the very end of the main galaxy arm and looks like a dwarf elliptical. However its luminosity profile does not follow the $R^{1/4}$ law.\\

\begin{figure*}
\begin{center}
\includegraphics[width=7cm,height=7cm]{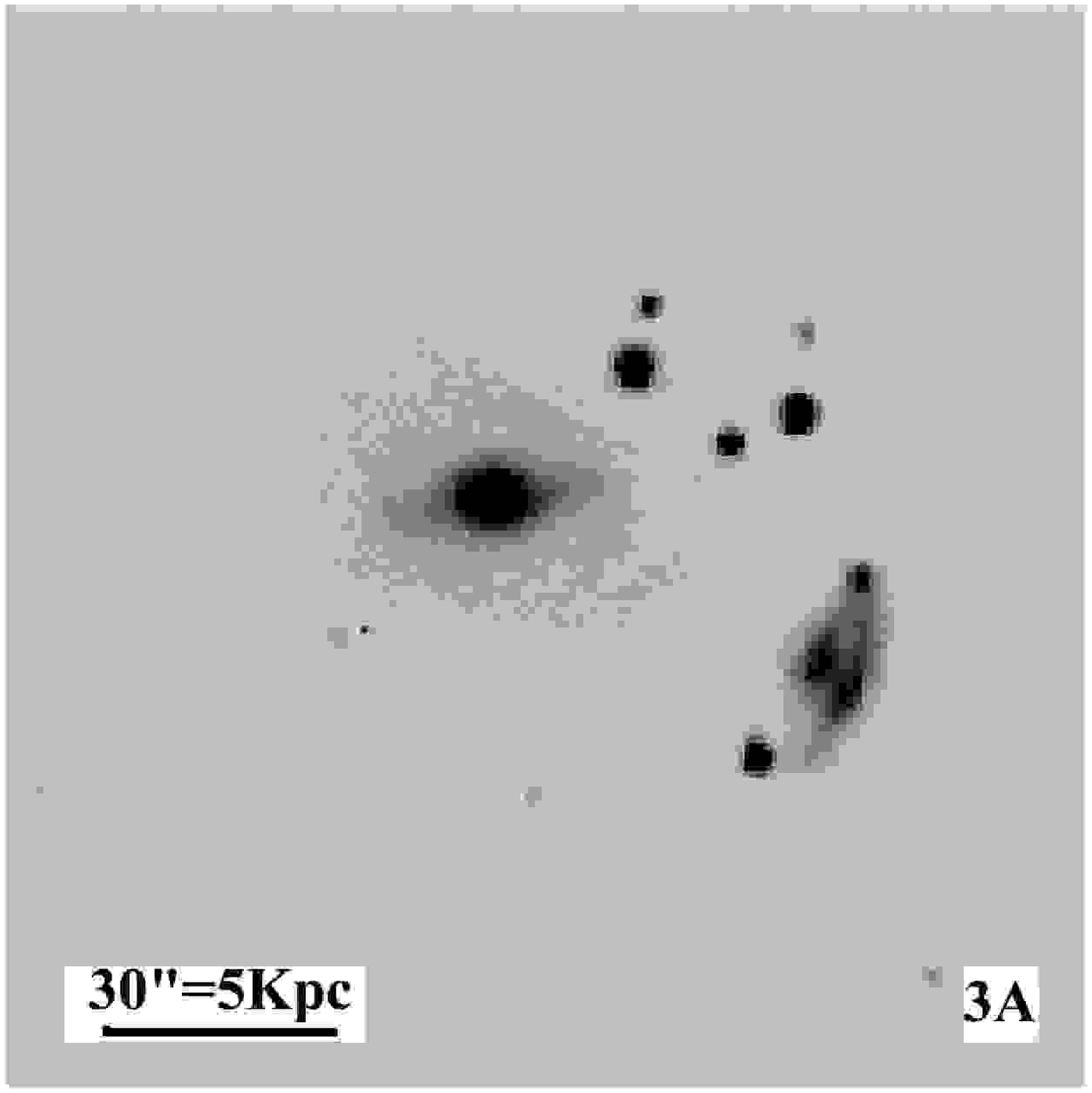}\includegraphics[width=7cm,height=7cm]{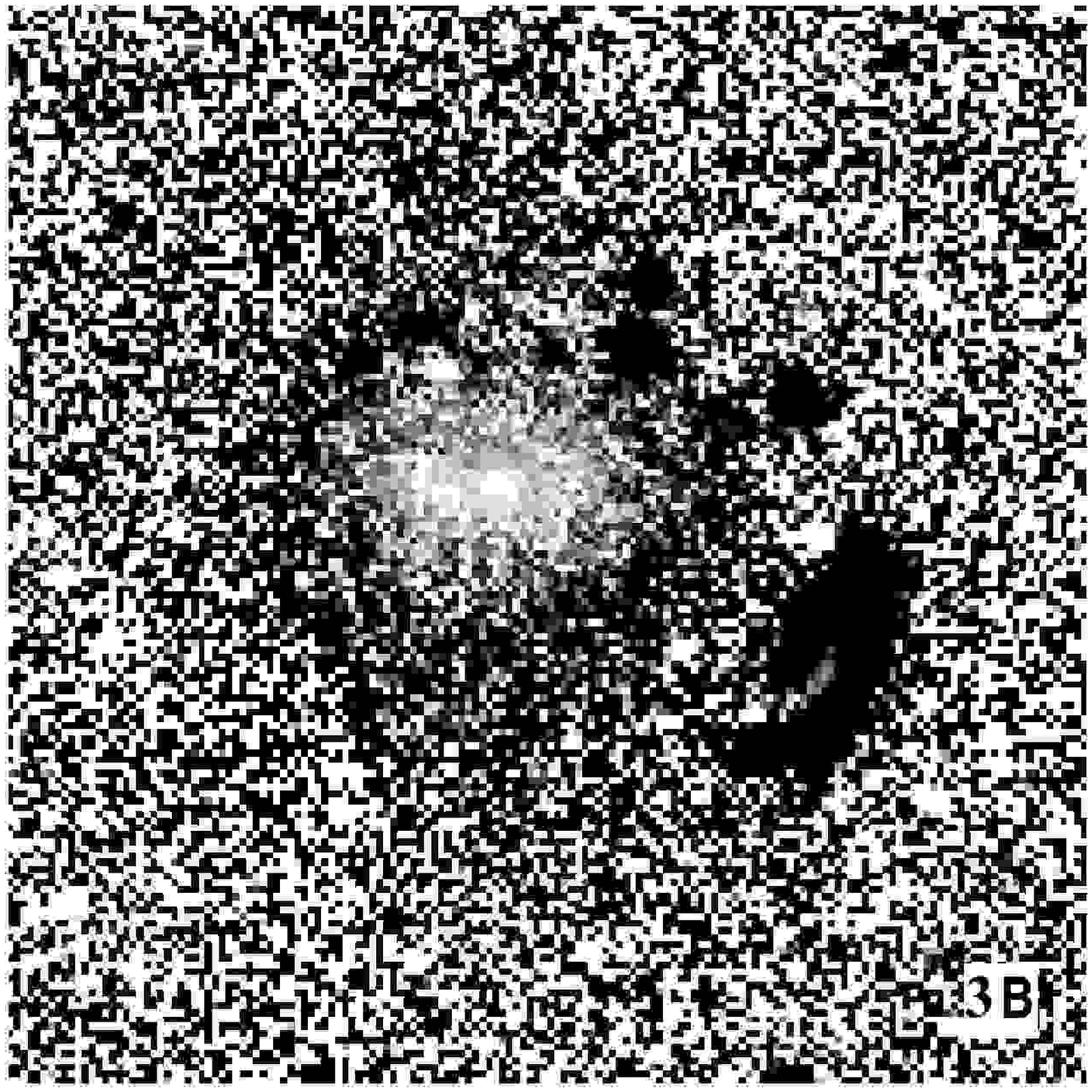}\\
\includegraphics[width=7cm,height=7cm]{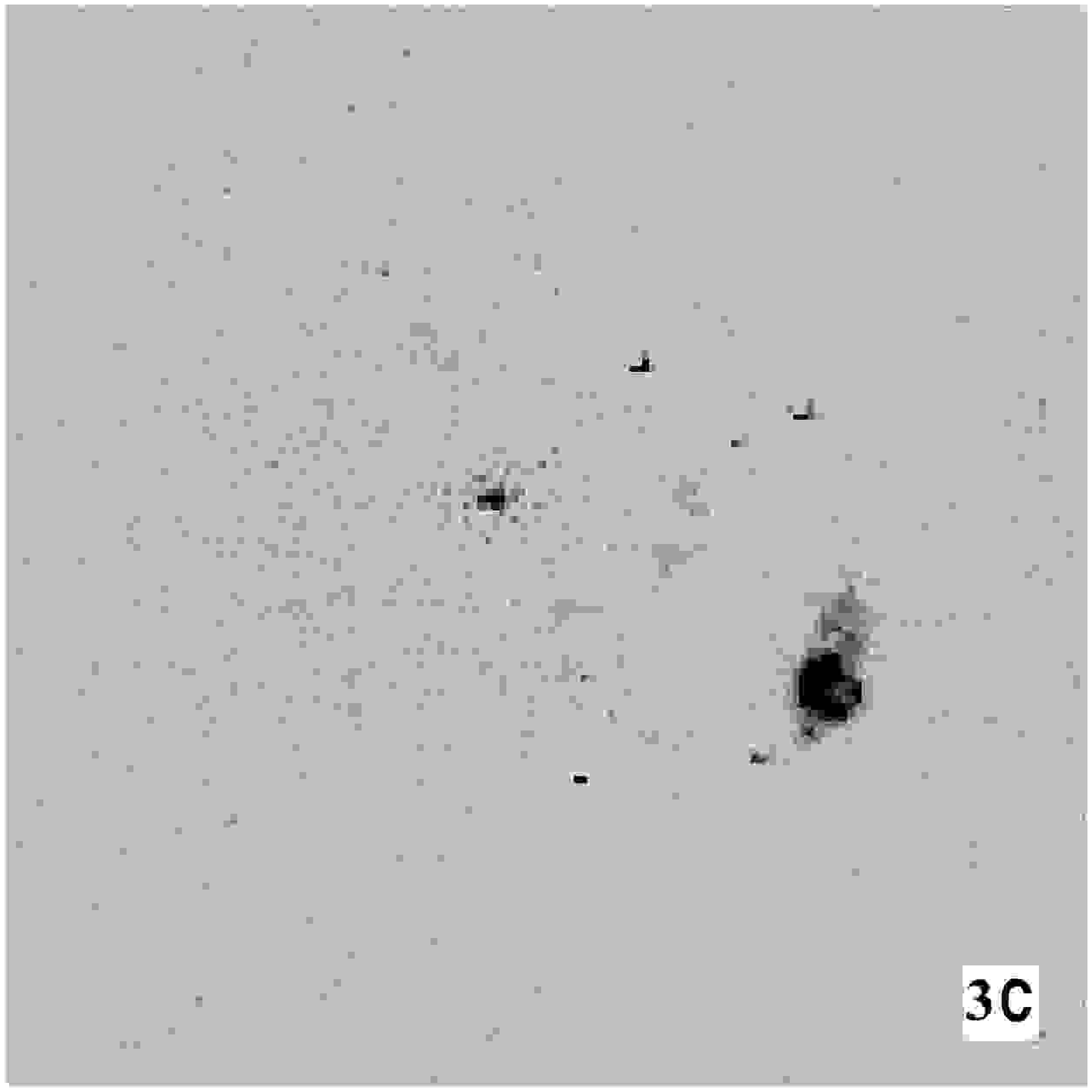}\includegraphics[width=7cm,height=7cm]{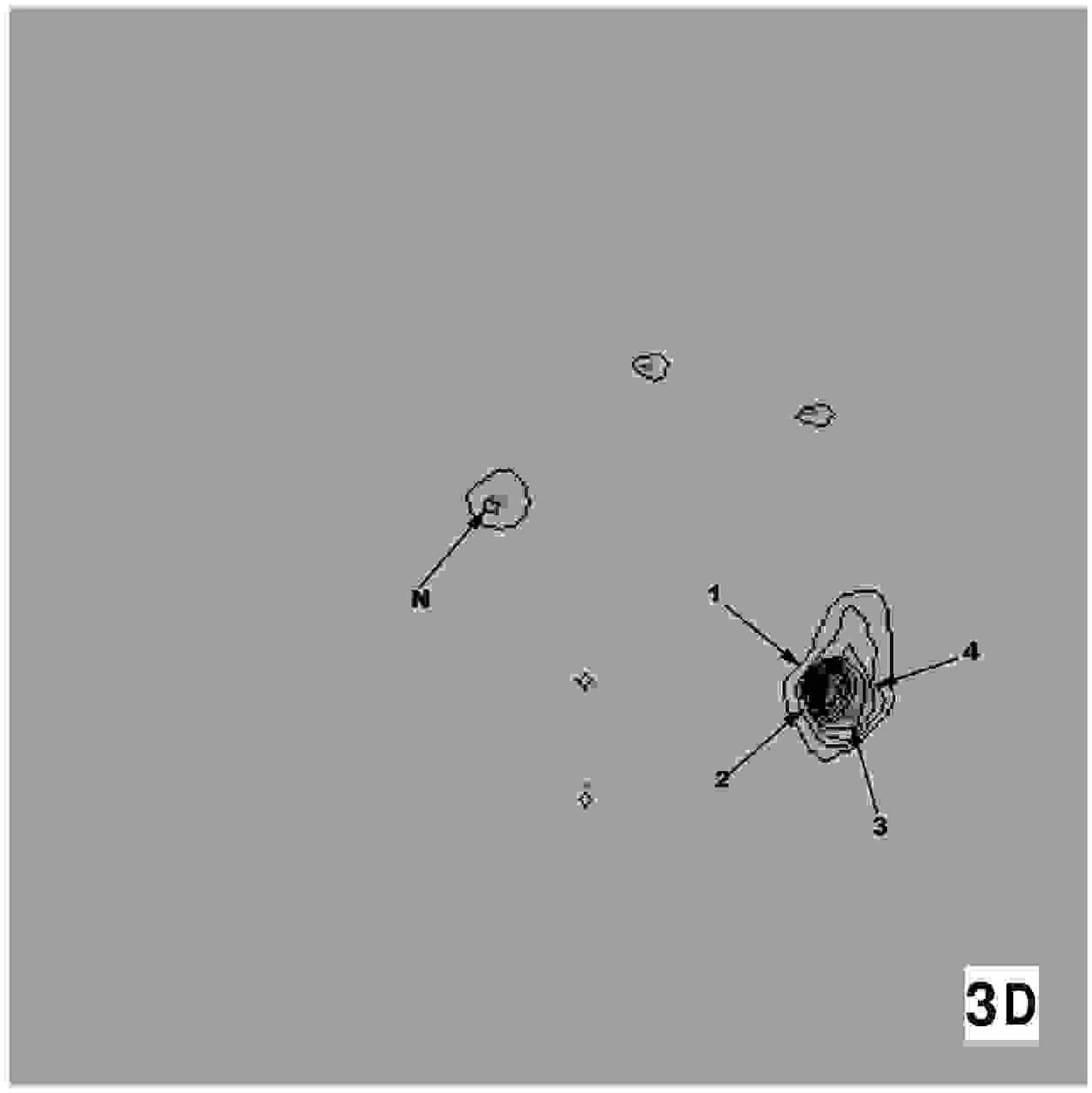}\\
\includegraphics[width=7cm,height=7cm]{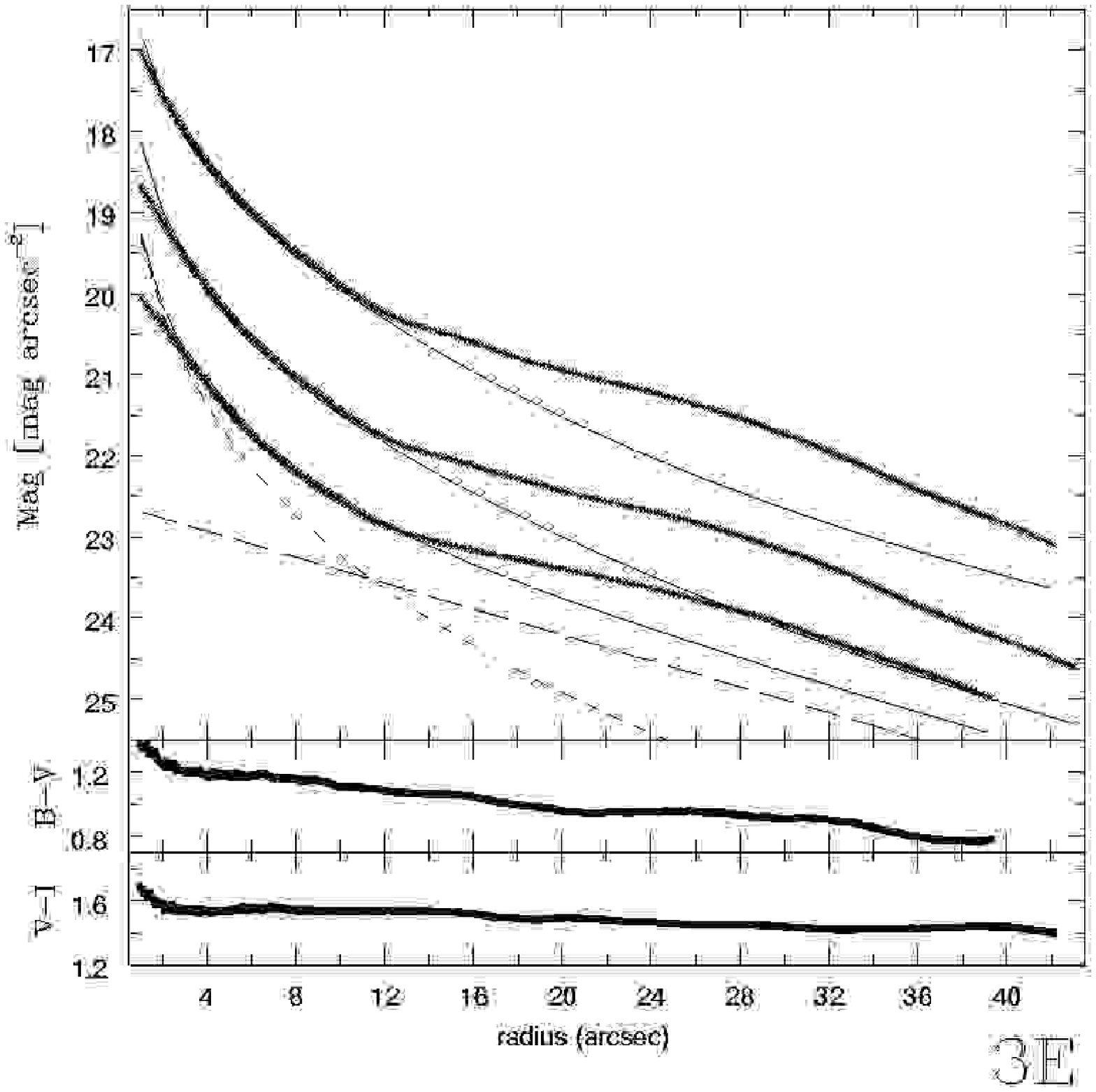}\includegraphics[width=7cm,height=7cm]{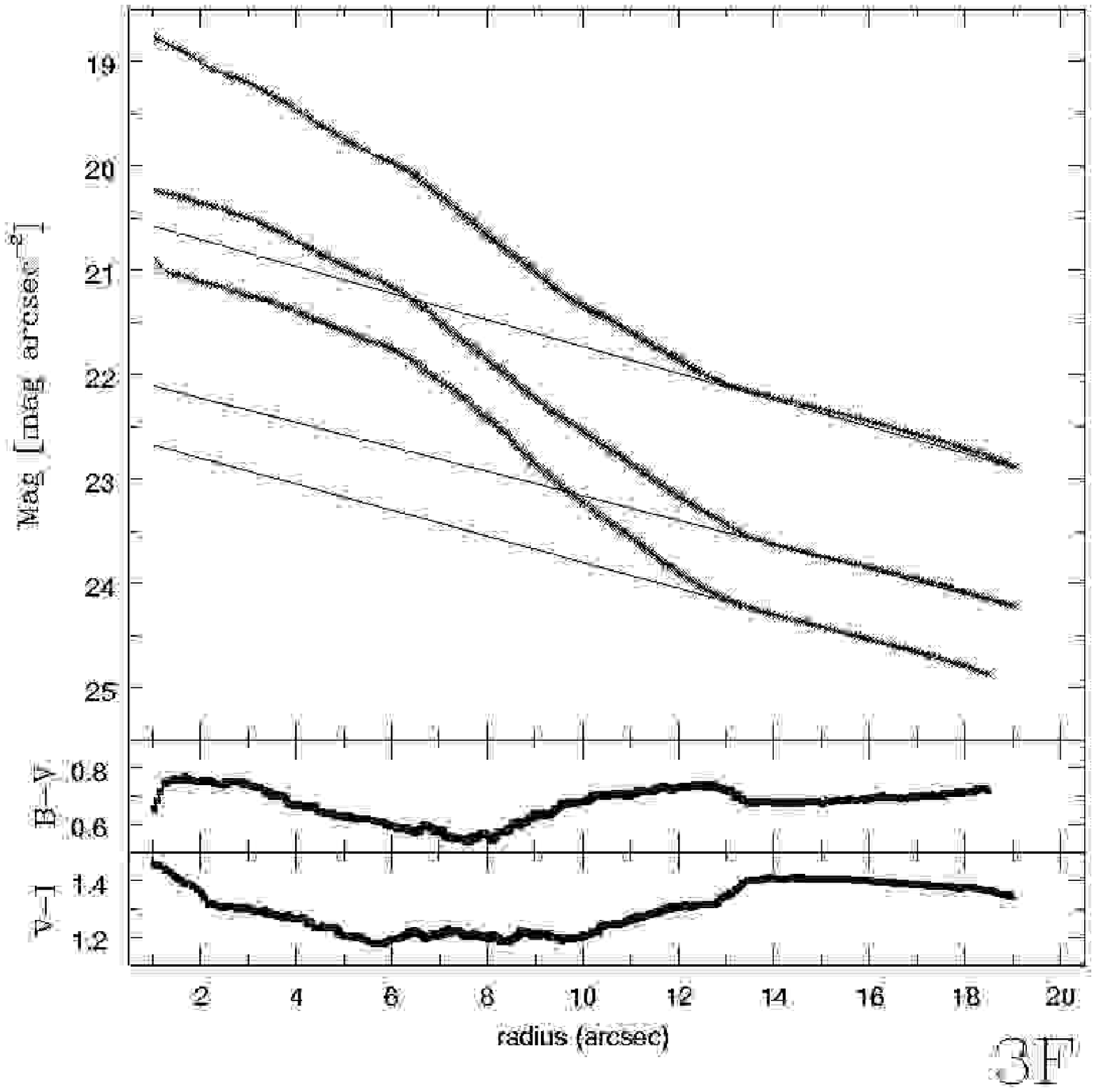}\\
\caption{AM1448-262. Same as Fig 1.}
\label{fig3}
\end{center}
\end{figure*}

AM1448-262: This pair is located at $cz$ = 2576 $km$  $s^{-1}$ and the main galaxy shows a conspicuous bar and two well-developed arms. There is no sign of a ring as is suggested in the RC3 (see Fig.~\ref{fig3}).

The secondary galaxy is clearly a peculiar disk galaxy; the outermost region of the luminosity profile fits the exponential law. This galaxy shows an elongated nucleus and a very bright region located towards the south. This region appears very bright in the H$\alpha$ image.\\

\begin{figure*}
\begin{center}
\includegraphics[width=7cm,height=7cm]{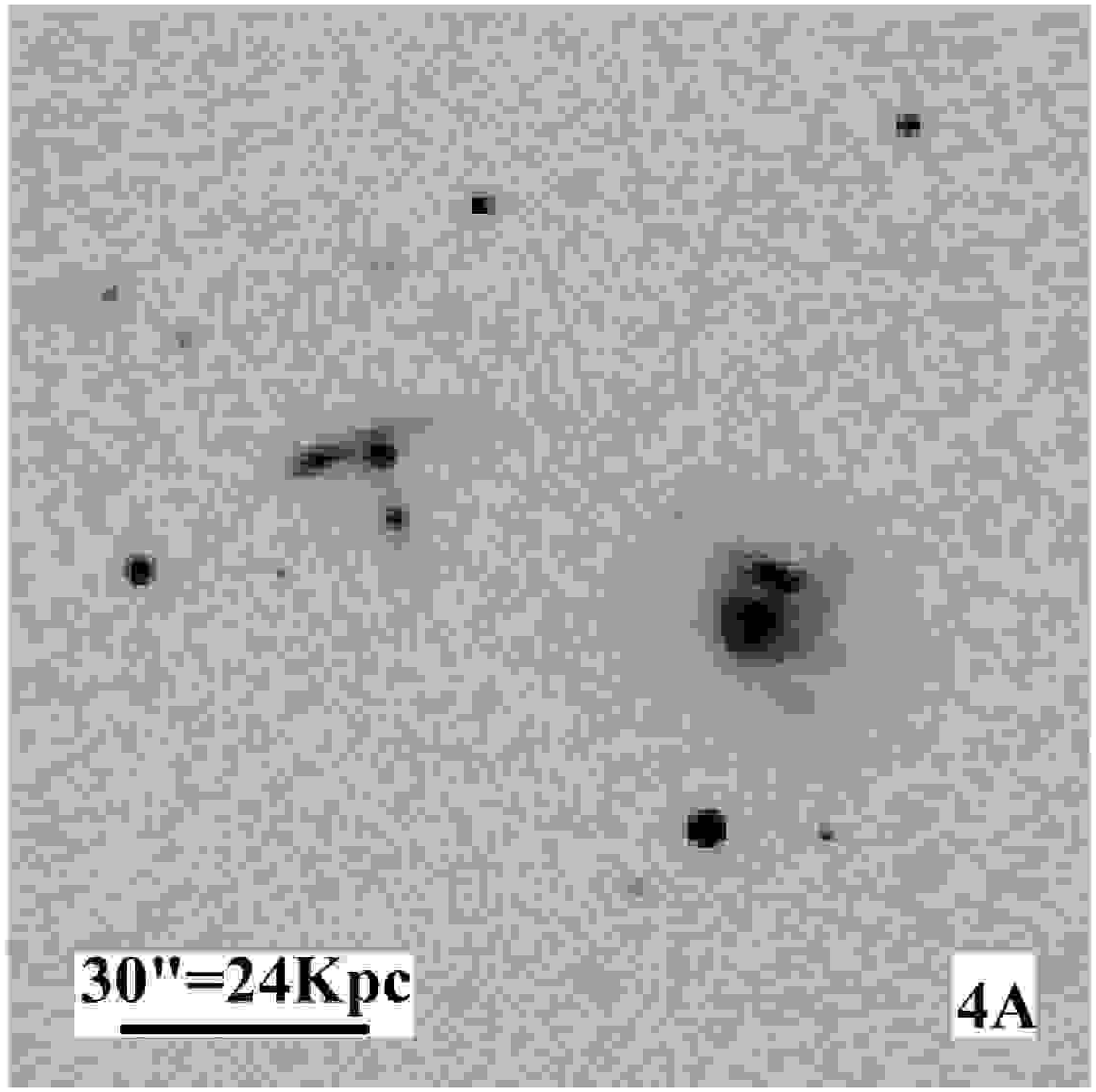}\includegraphics[width=7cm,height=7cm]{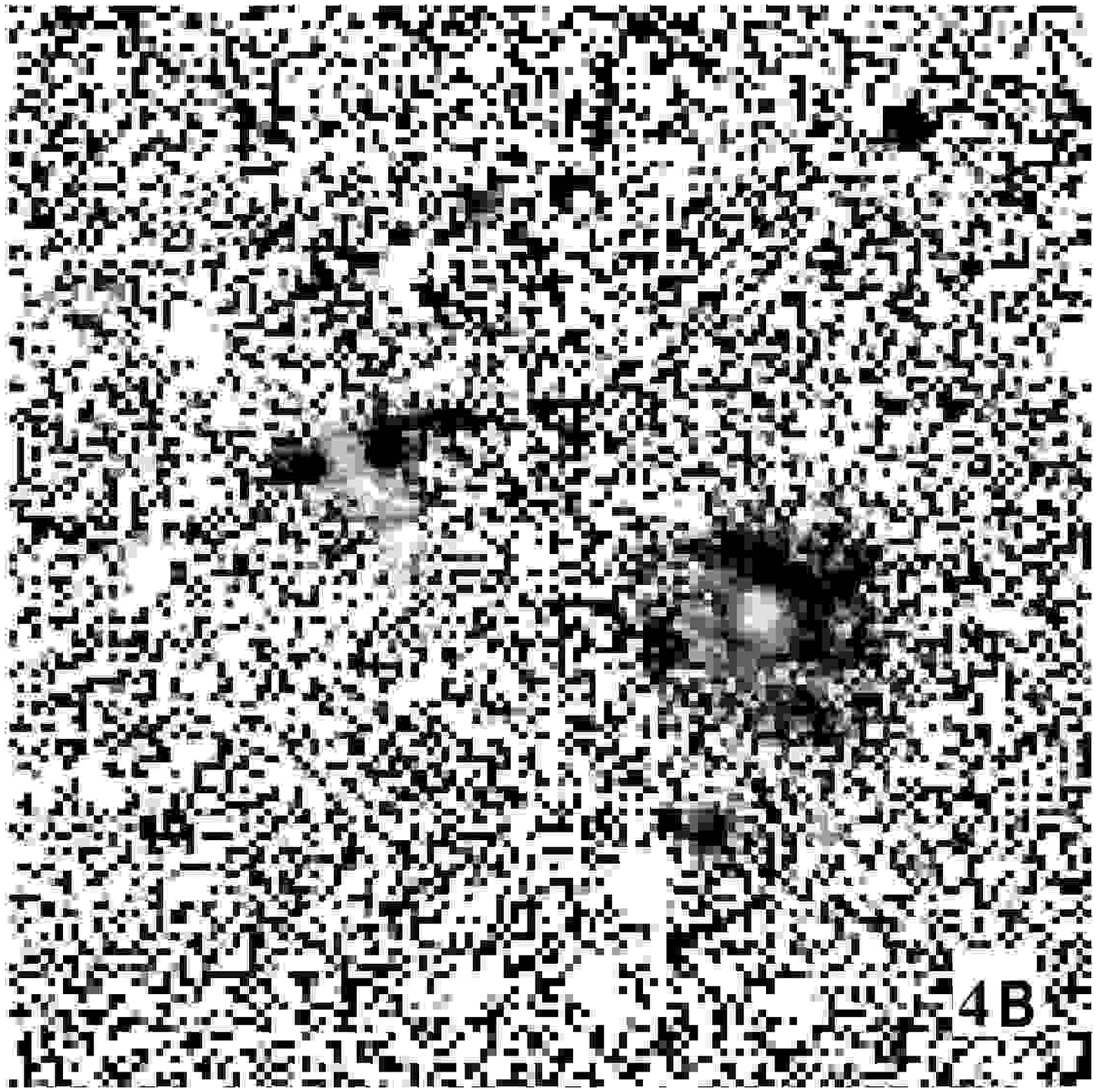}\\
\includegraphics[width=7cm,height=7cm]{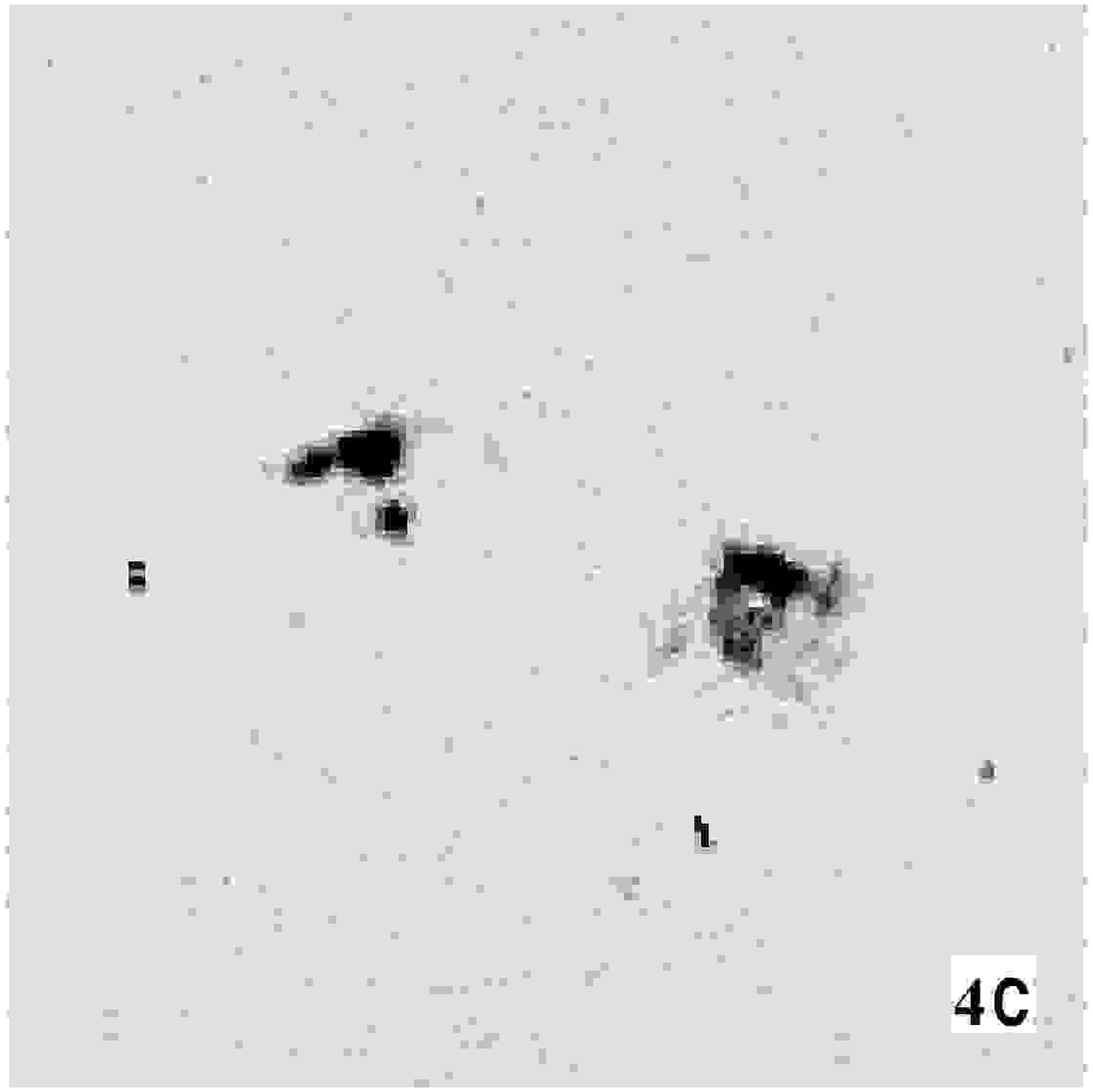}\includegraphics[width=7cm,height=7cm]{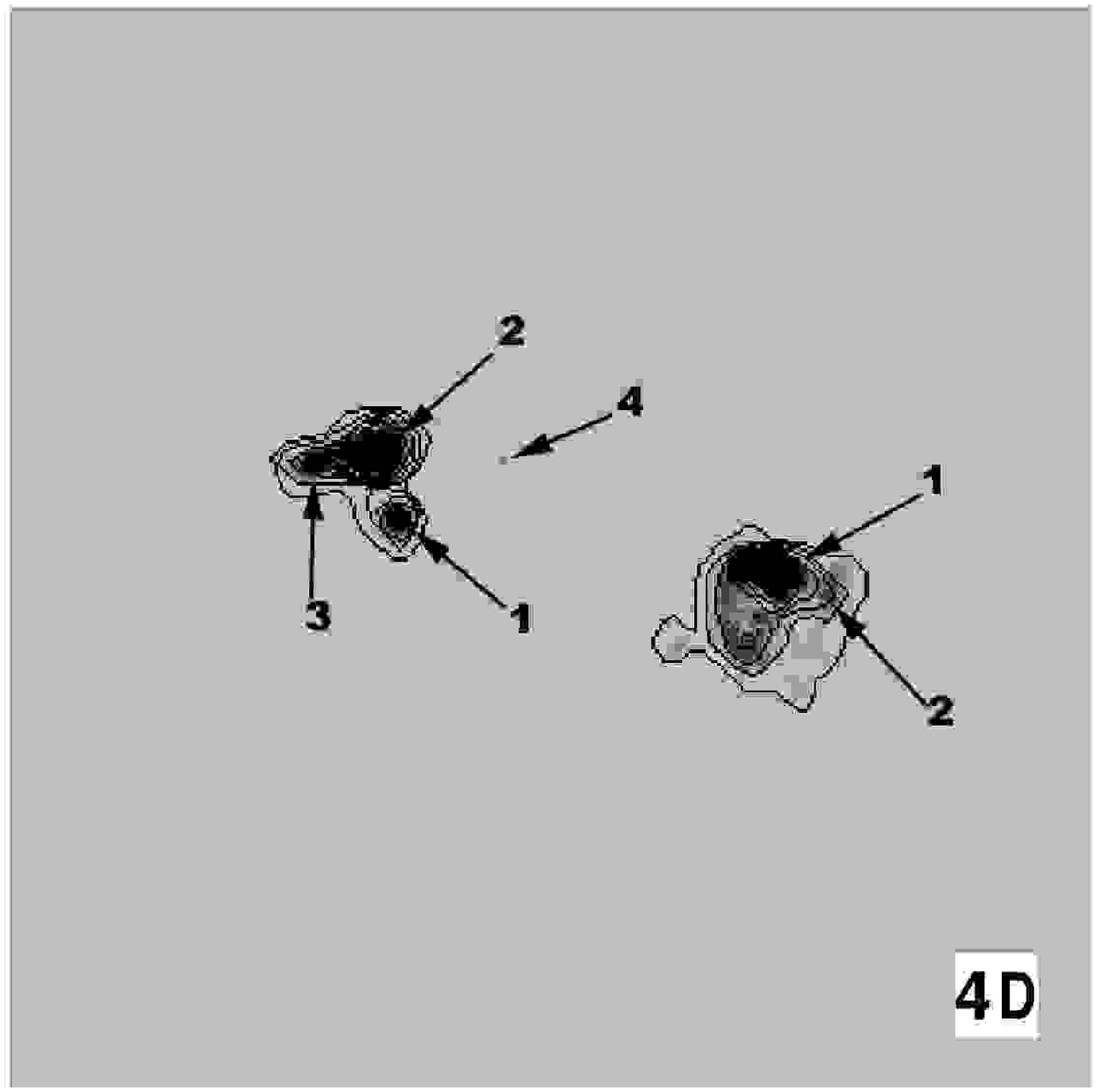}\\
\includegraphics[width=7cm,height=7cm]{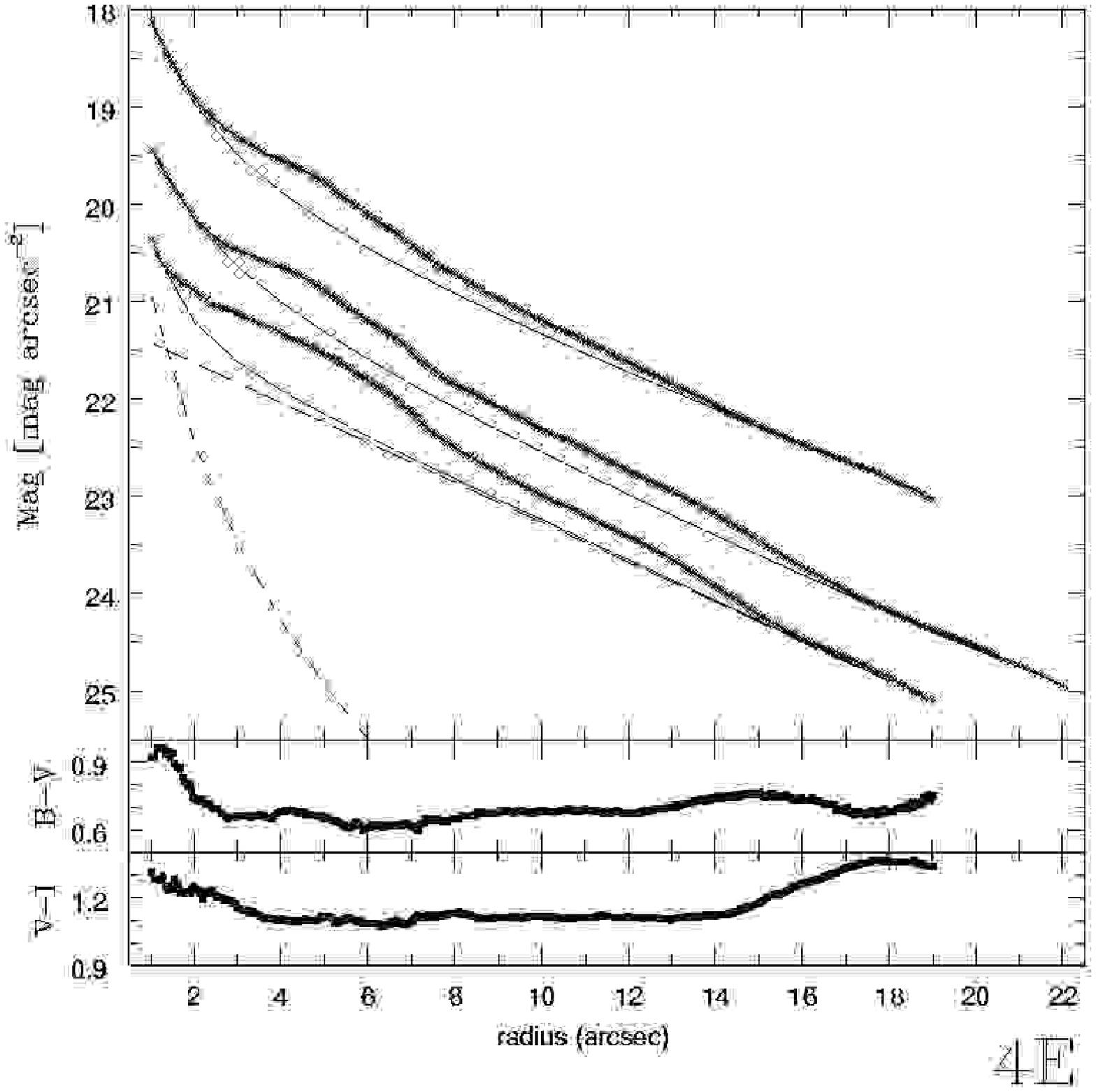}\includegraphics[width=7cm,height=7cm]{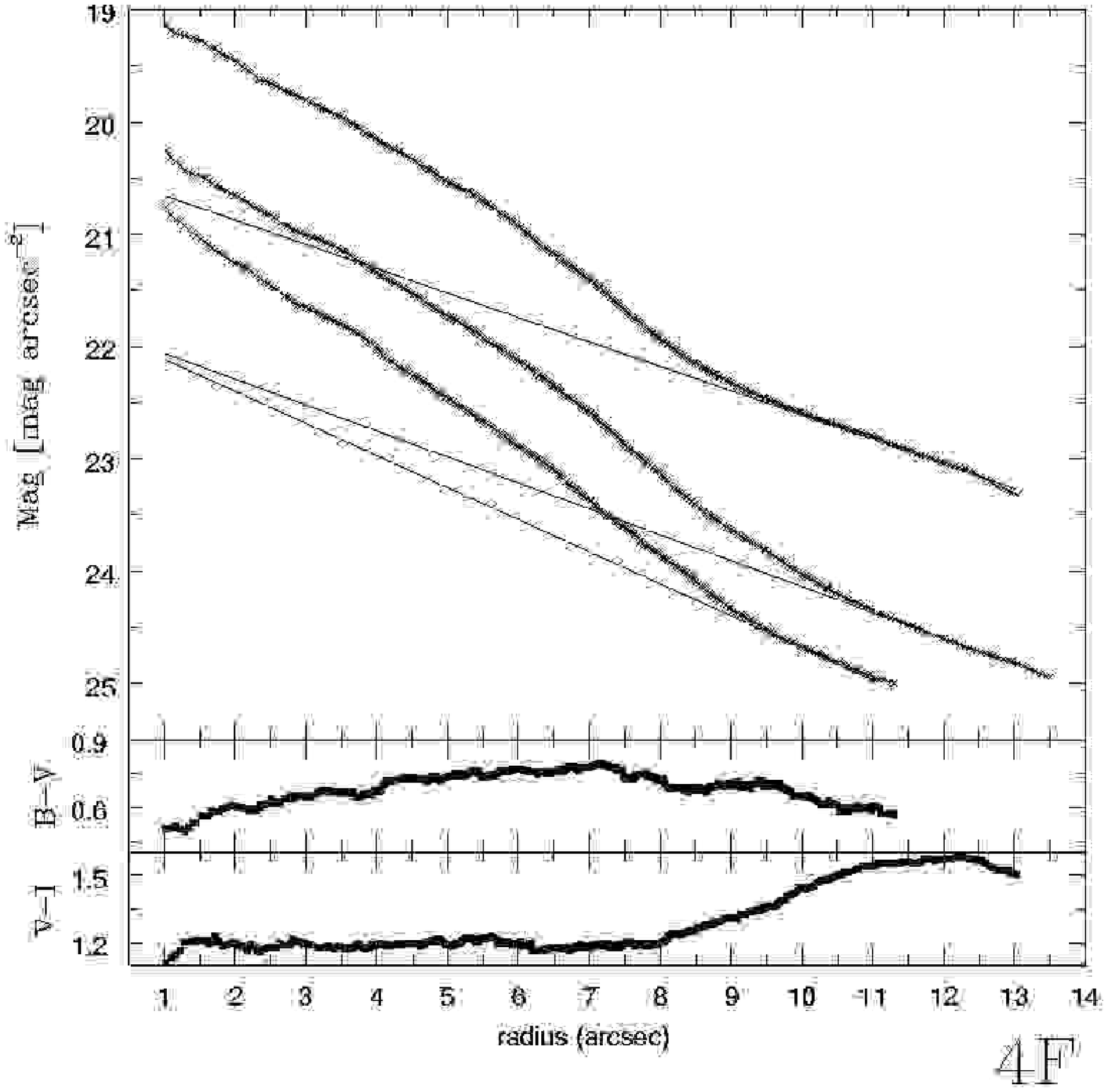}\\
\caption{AM2030-303. Same as Fig 1.}
\label{fig4}
\end{center}
\end{figure*}

AM2030-303: This system located at $cz$ = 12465 $km$  $s^{-1}$ is probably formed by more than two galaxies. The object identified as a secondary component located to the NE appears to be a triplet. The main galaxy is a bright ($M_B$ =  --21.3) late spiral with very disturbed arms. Towards the north, the galaxy exhibits a huge star-forming region, which is clearly seen in Fig.~\ref{fig4} and it is responsible for the bump in the luminosity profiles.\\

\begin{figure*}
\begin{center}
\includegraphics[width=7cm,height=7cm]{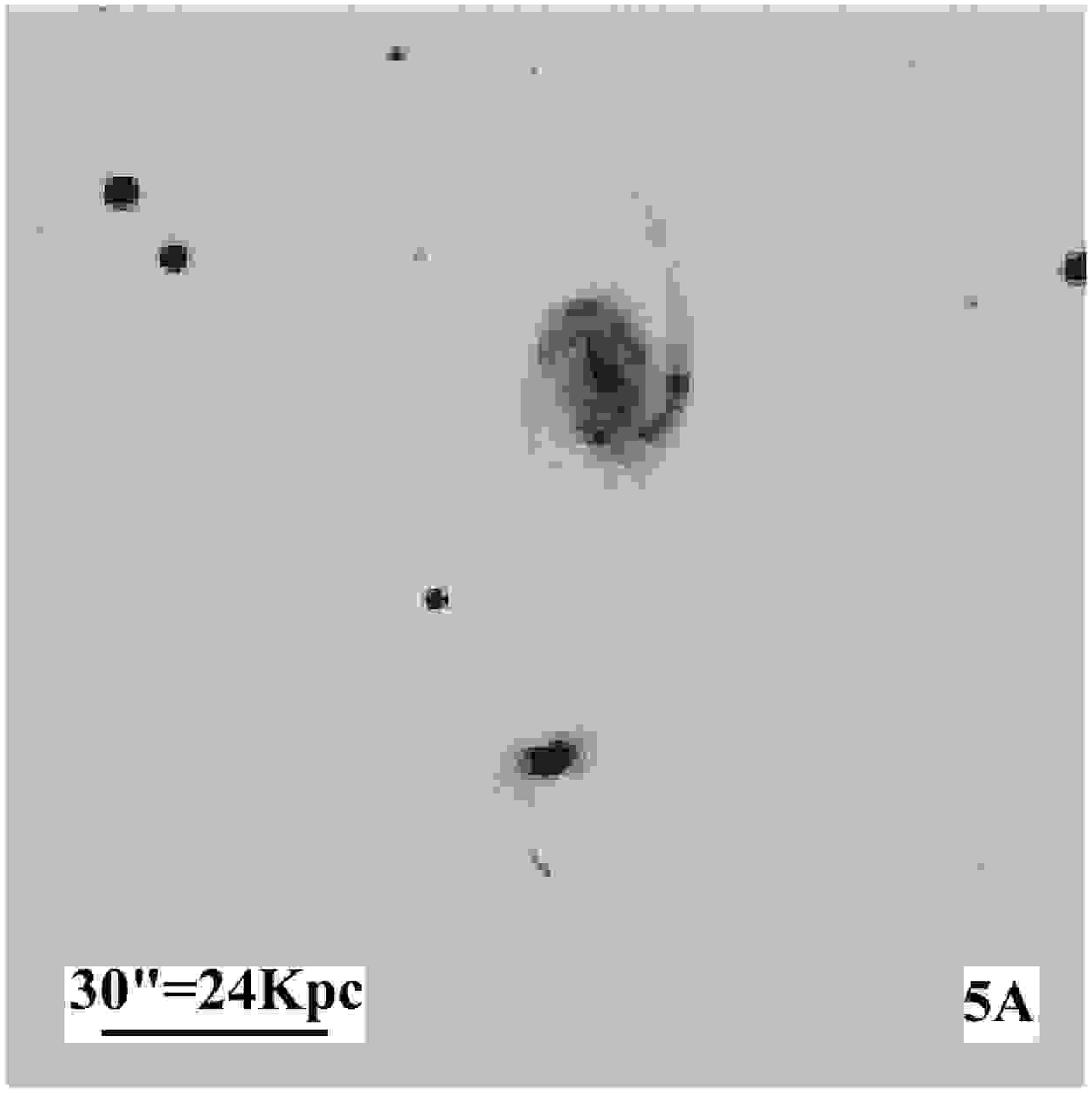}\includegraphics[width=7cm,height=7cm]{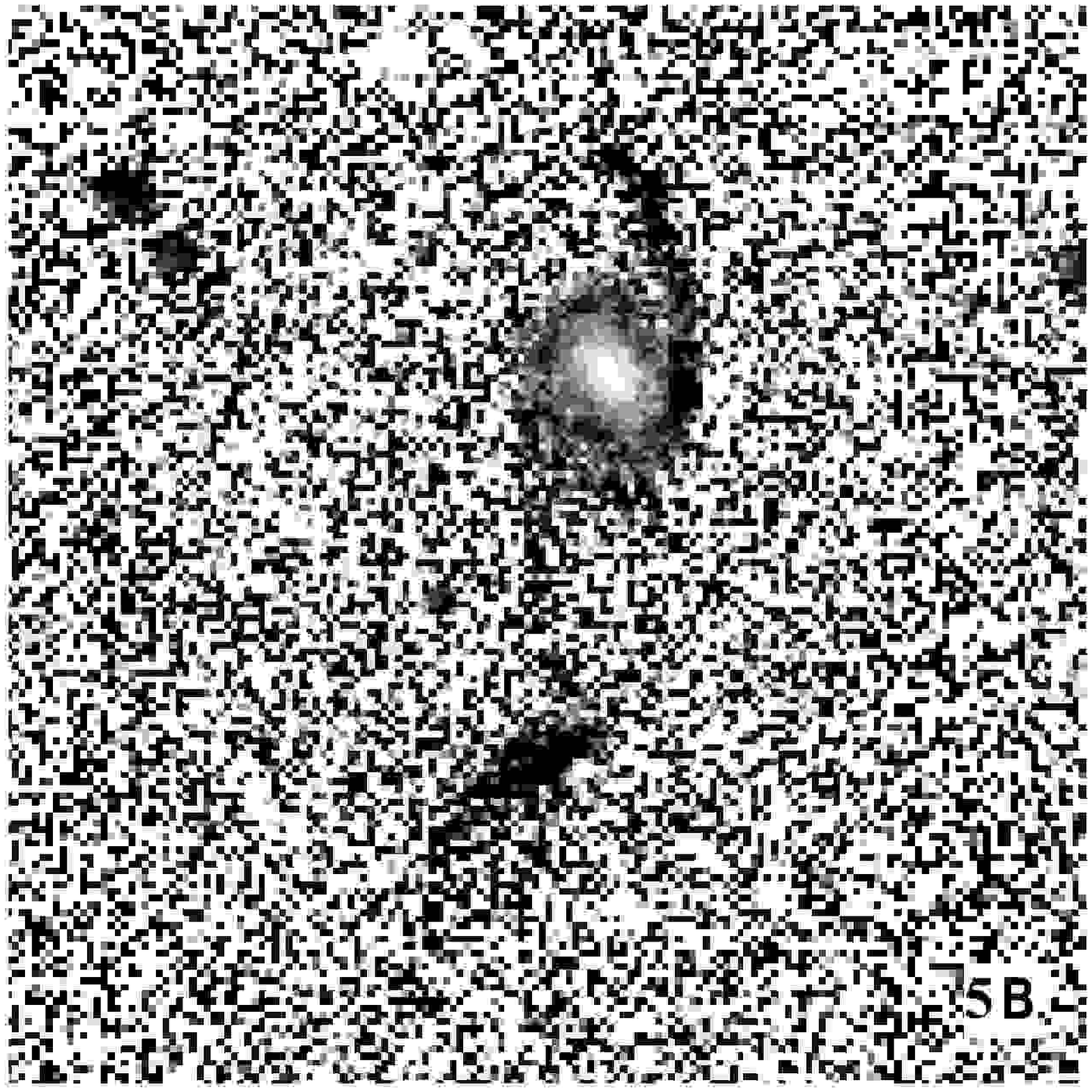}\\
\includegraphics[width=7cm,height=7cm]{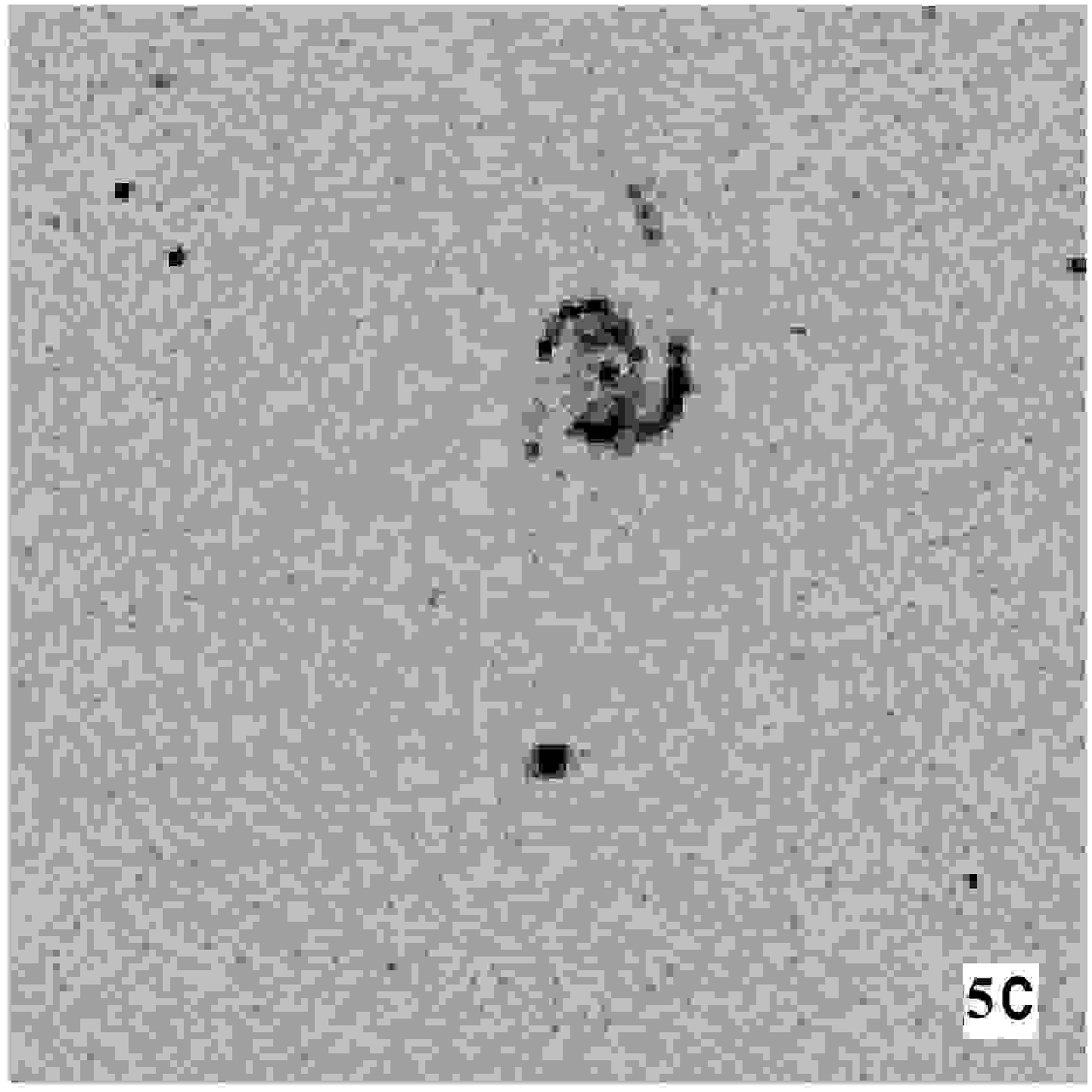}\includegraphics[width=7cm,height=7cm]{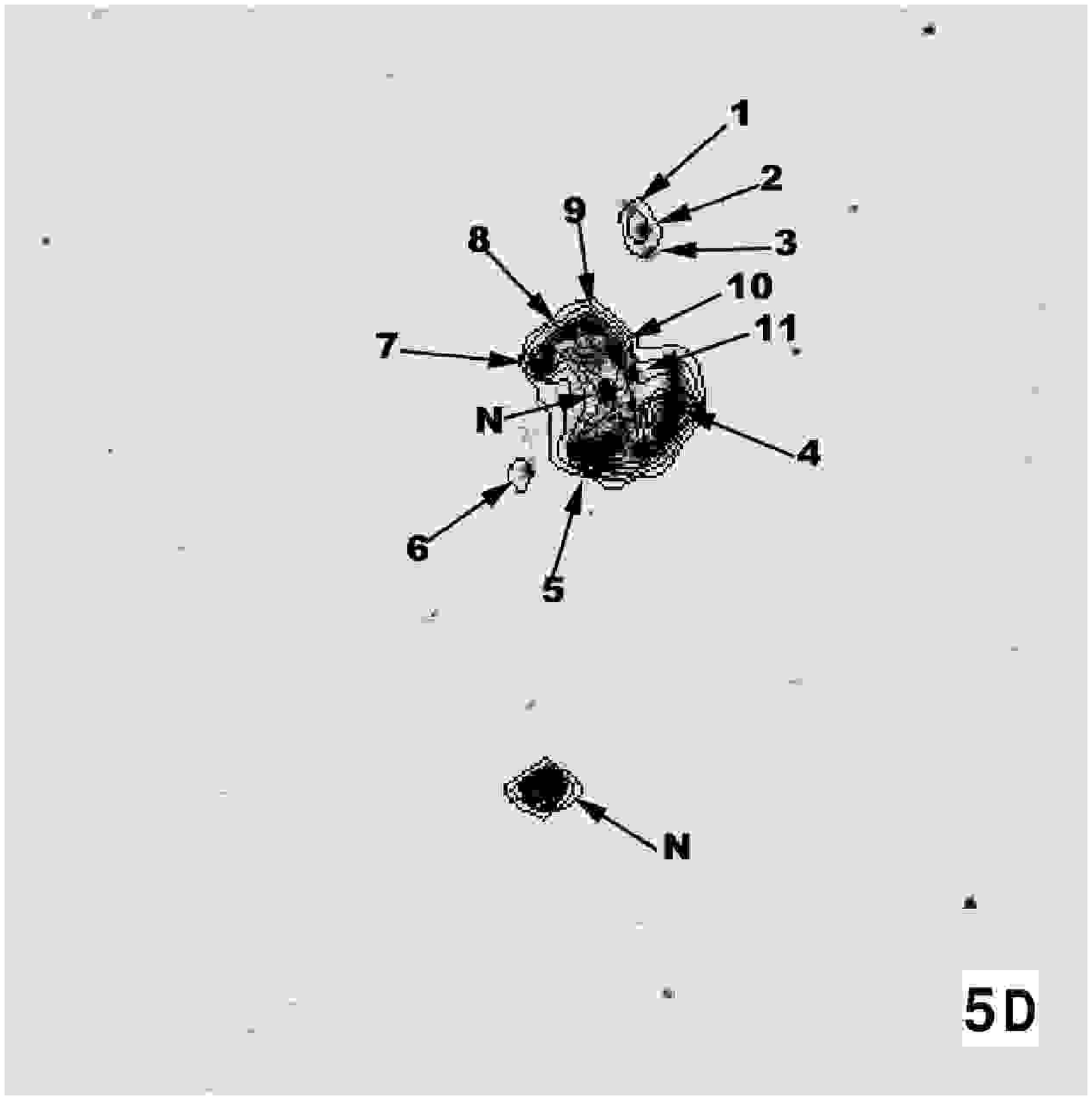}\\
\includegraphics[width=7cm,height=7cm]{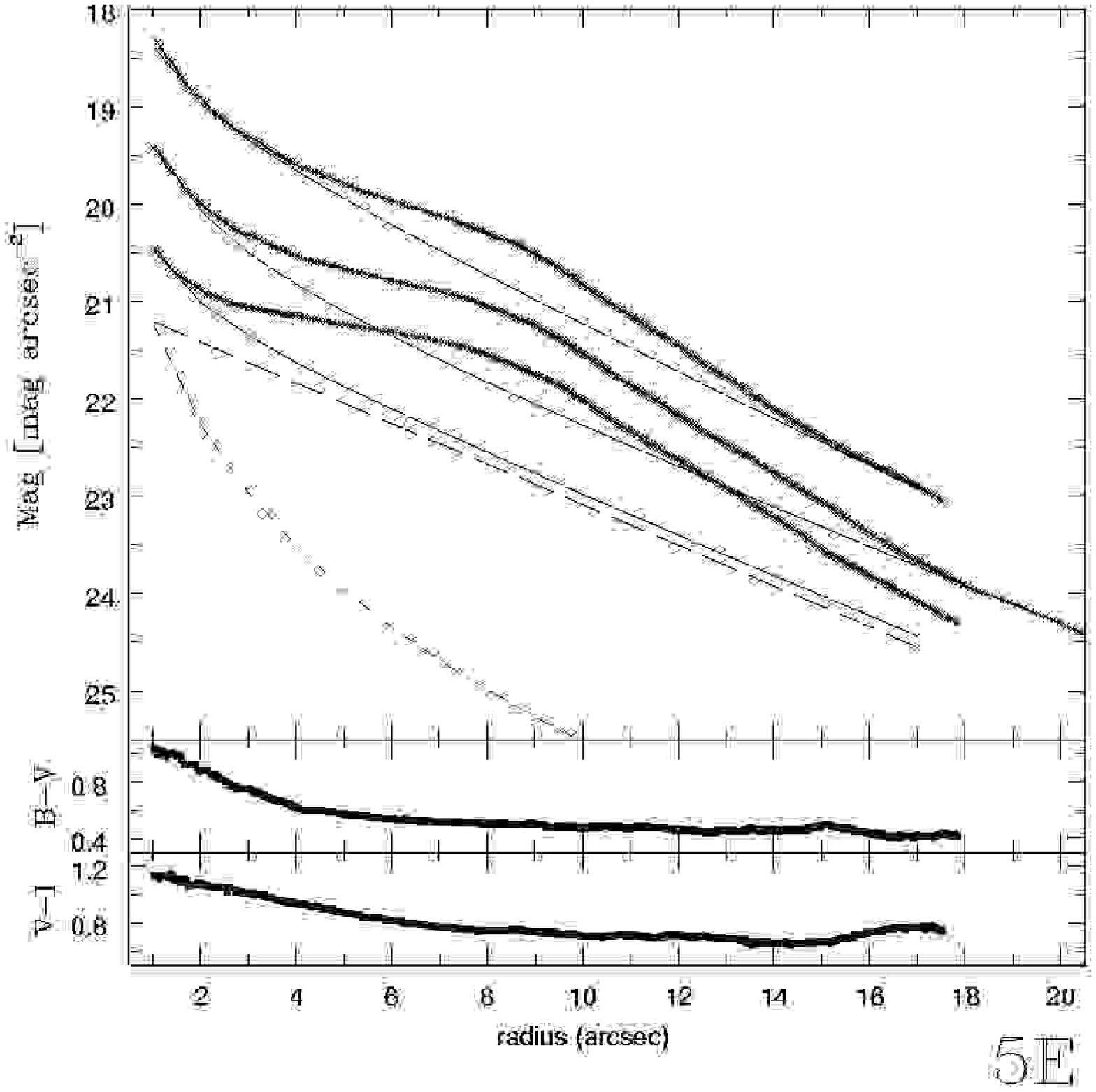}\includegraphics[width=7cm,height=7cm]{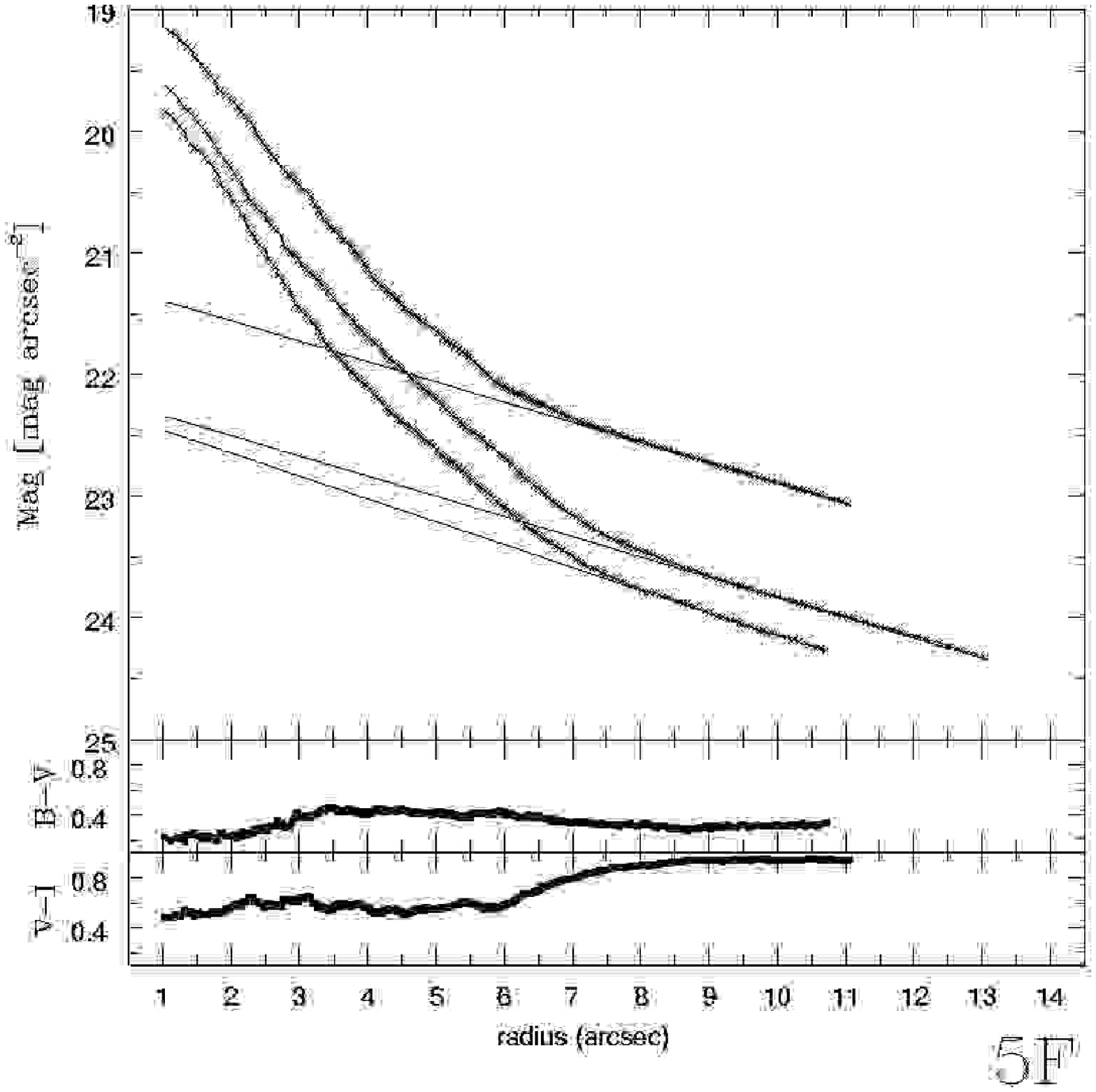}\\
\caption{AM2058-381. Same as Fig 1.}
\label{fig5}
\end{center}
\end{figure*}

AM2058-381: This is another M51 type pair located at $cz$ = 12383 $km$  $s^{-1}$. The main component presents many bright HII regions distributed along the arms (Fig.~\ref{fig5}). The integrated colour of this galaxy is rather blue (B-V) = 0.6.

The secondary is a compact galaxy peanut shape and shows enhanced star formation. The integrated colour is rather blue (B-V) = 0.4.\\

AM2105-332: These two galaxies located at $cz$ = 5449 $km$ $s^{-1}$ are peculiar. The luminosity profile parameters and colours correspond to those found in SO galaxies. However both galaxies show long tidal tails apparently in contact.

\begin{figure*}
\begin{center}
\includegraphics[width=7cm,height=7cm]{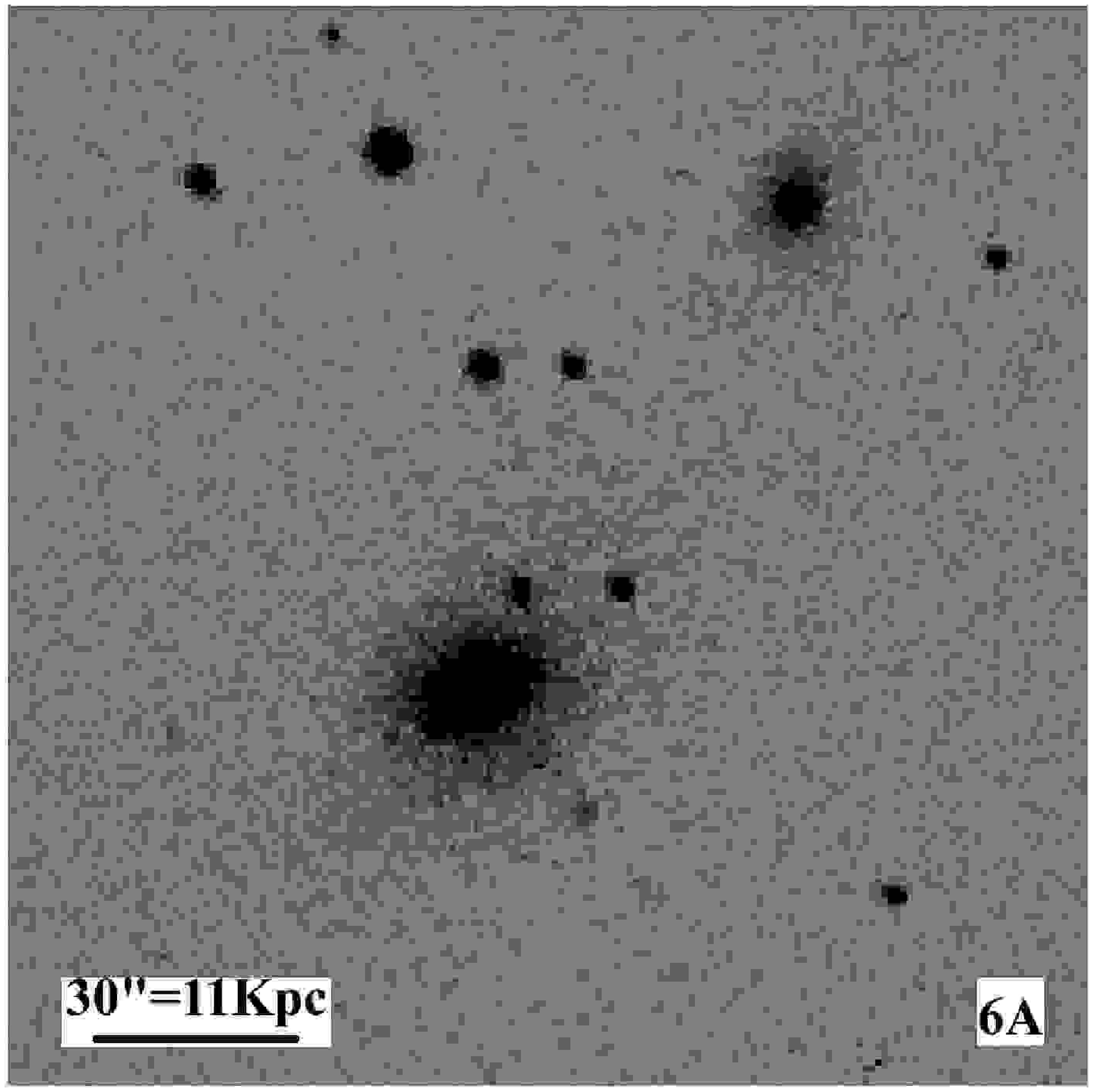}\includegraphics[width=7cm,height=7cm]{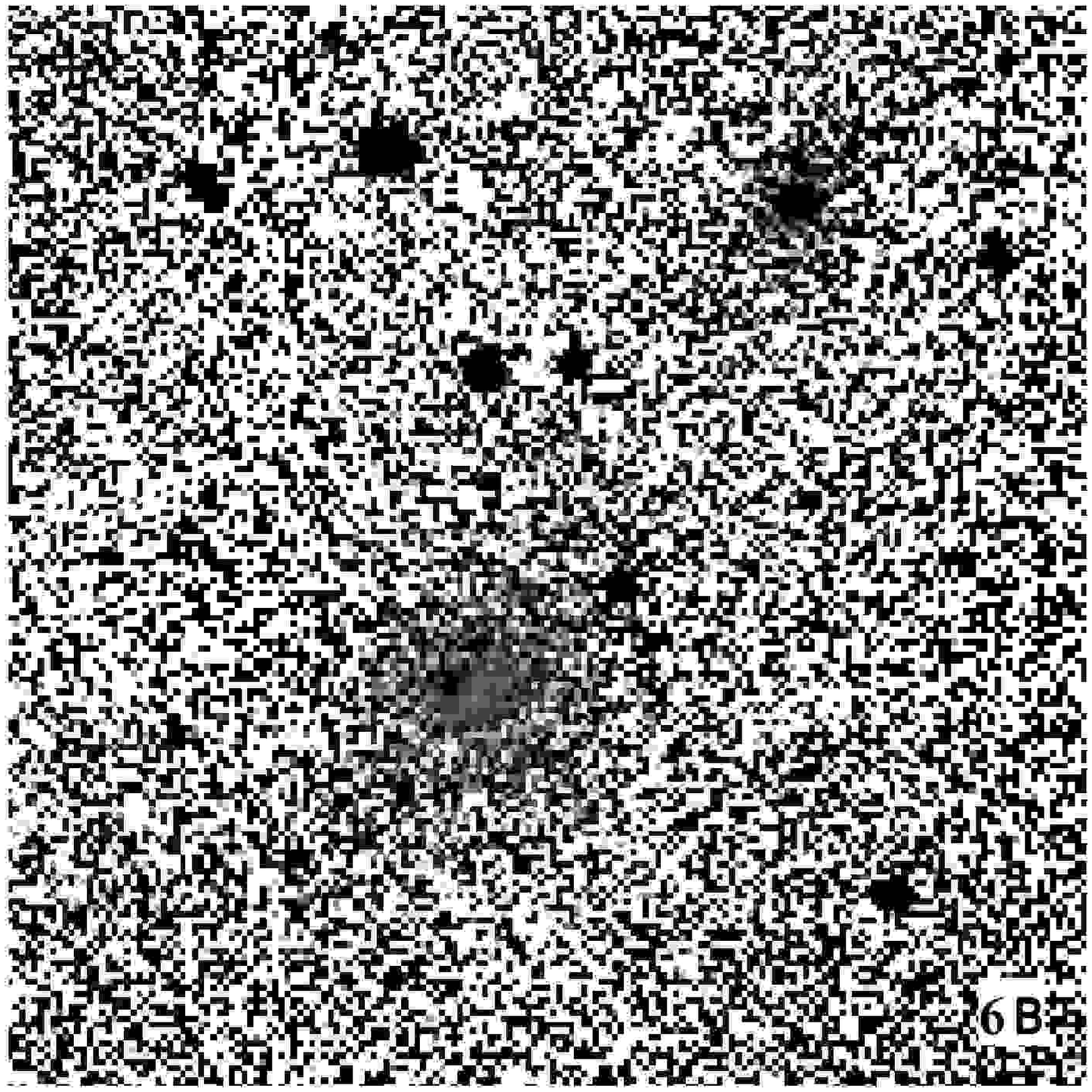}\\
\includegraphics[width=7cm,height=7cm]{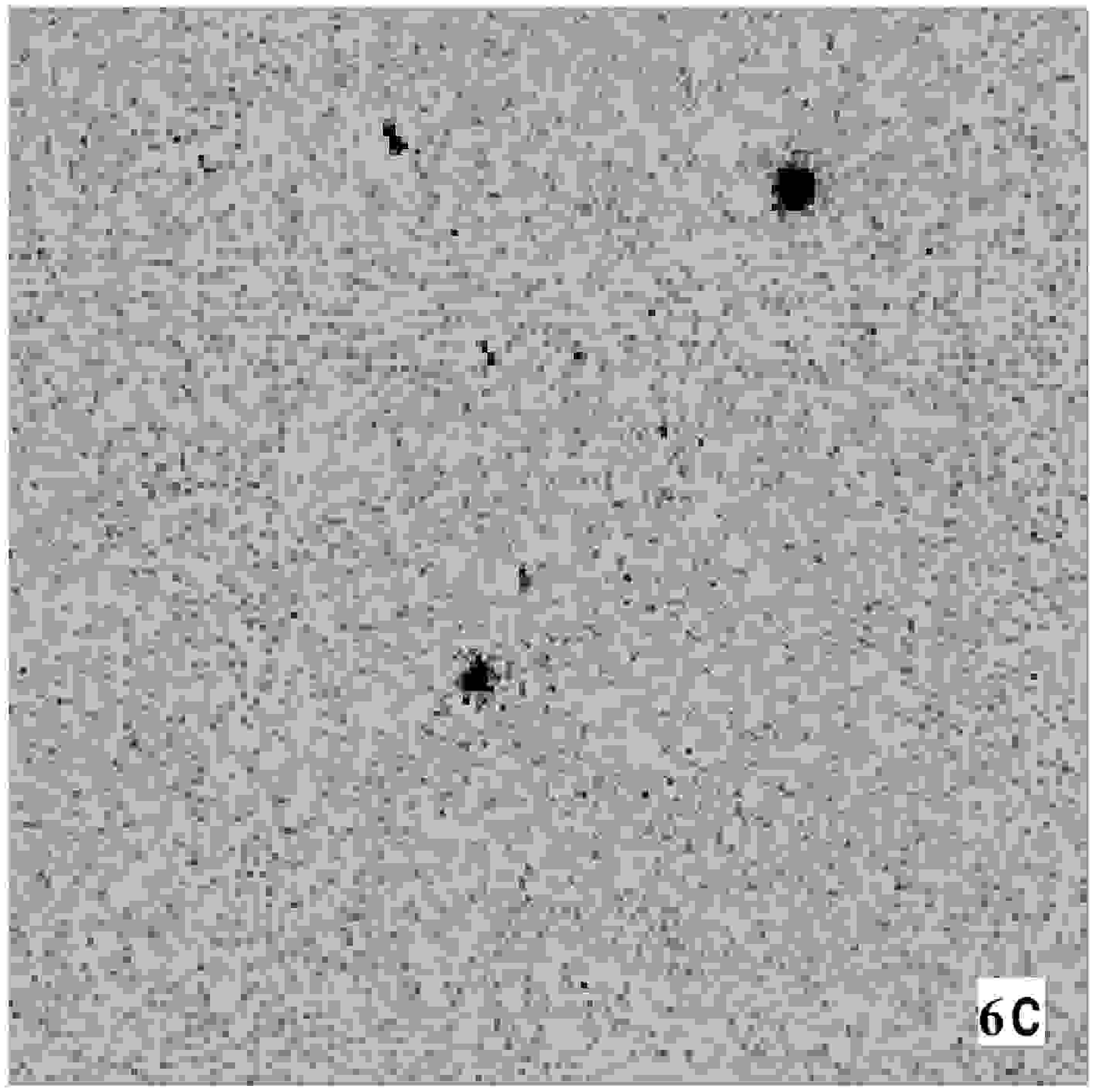}\includegraphics[width=7cm,height=7cm]{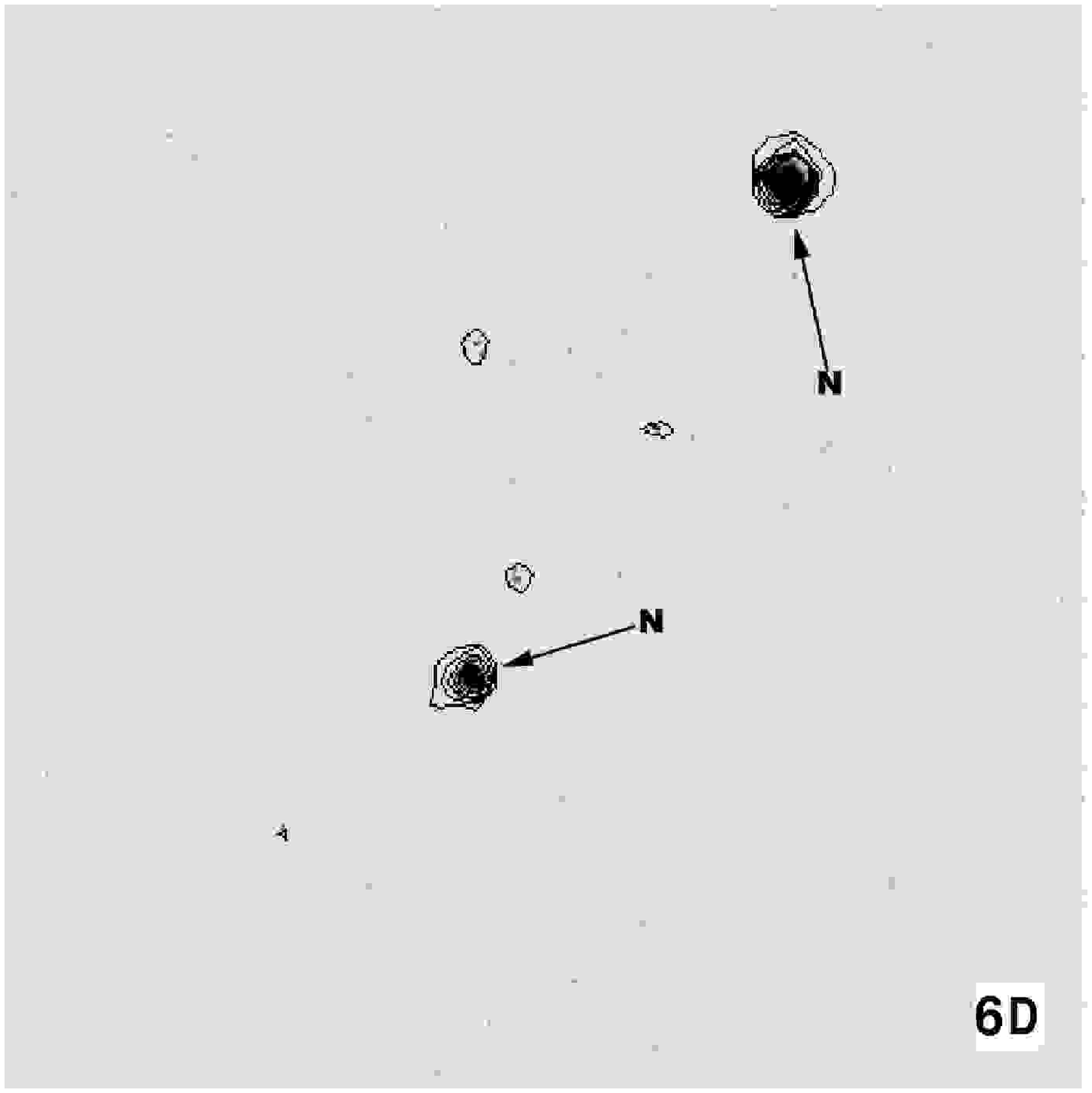}\\
\includegraphics[width=7cm,height=7cm]{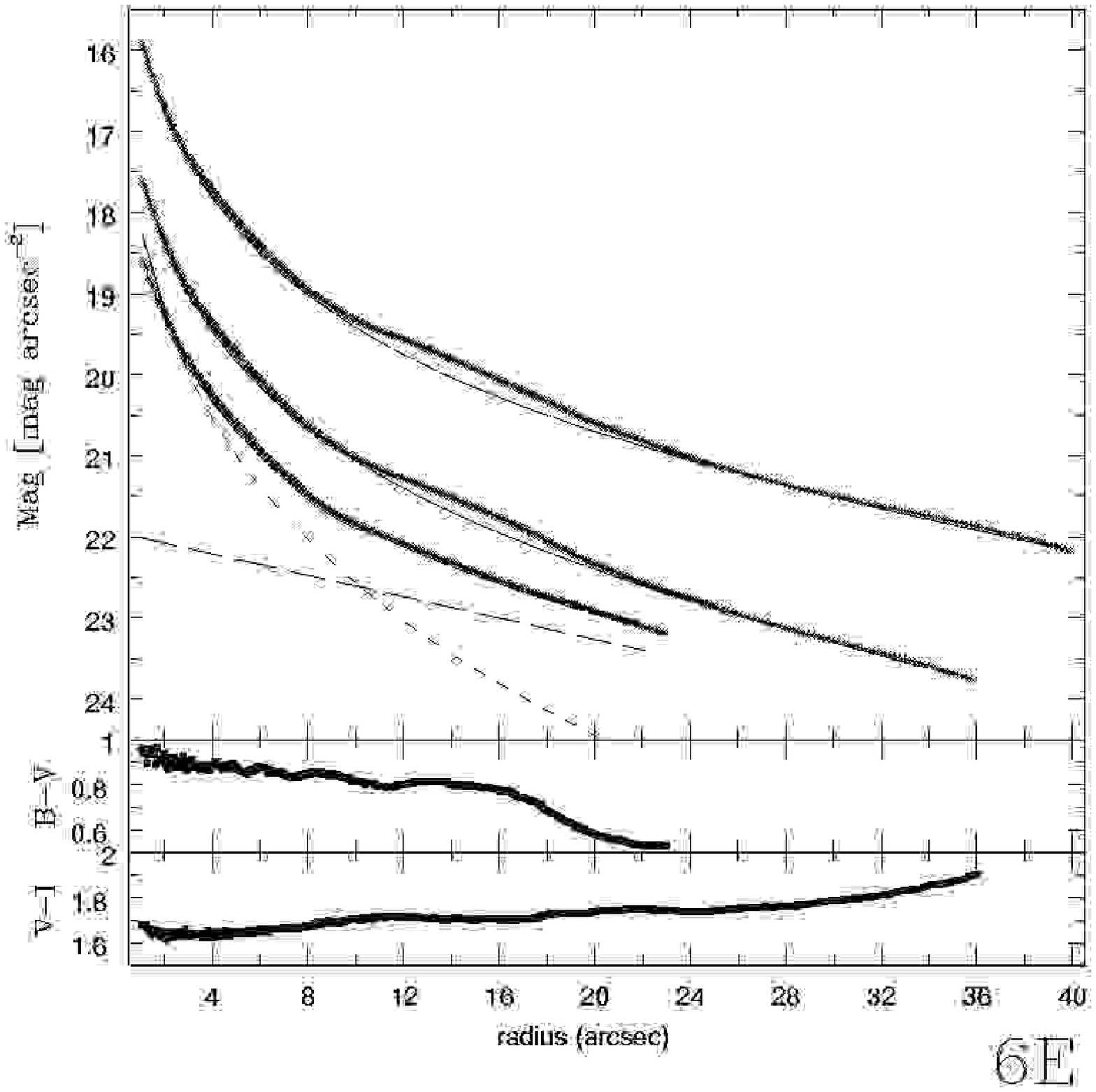}\includegraphics[width=7cm,height=7cm]{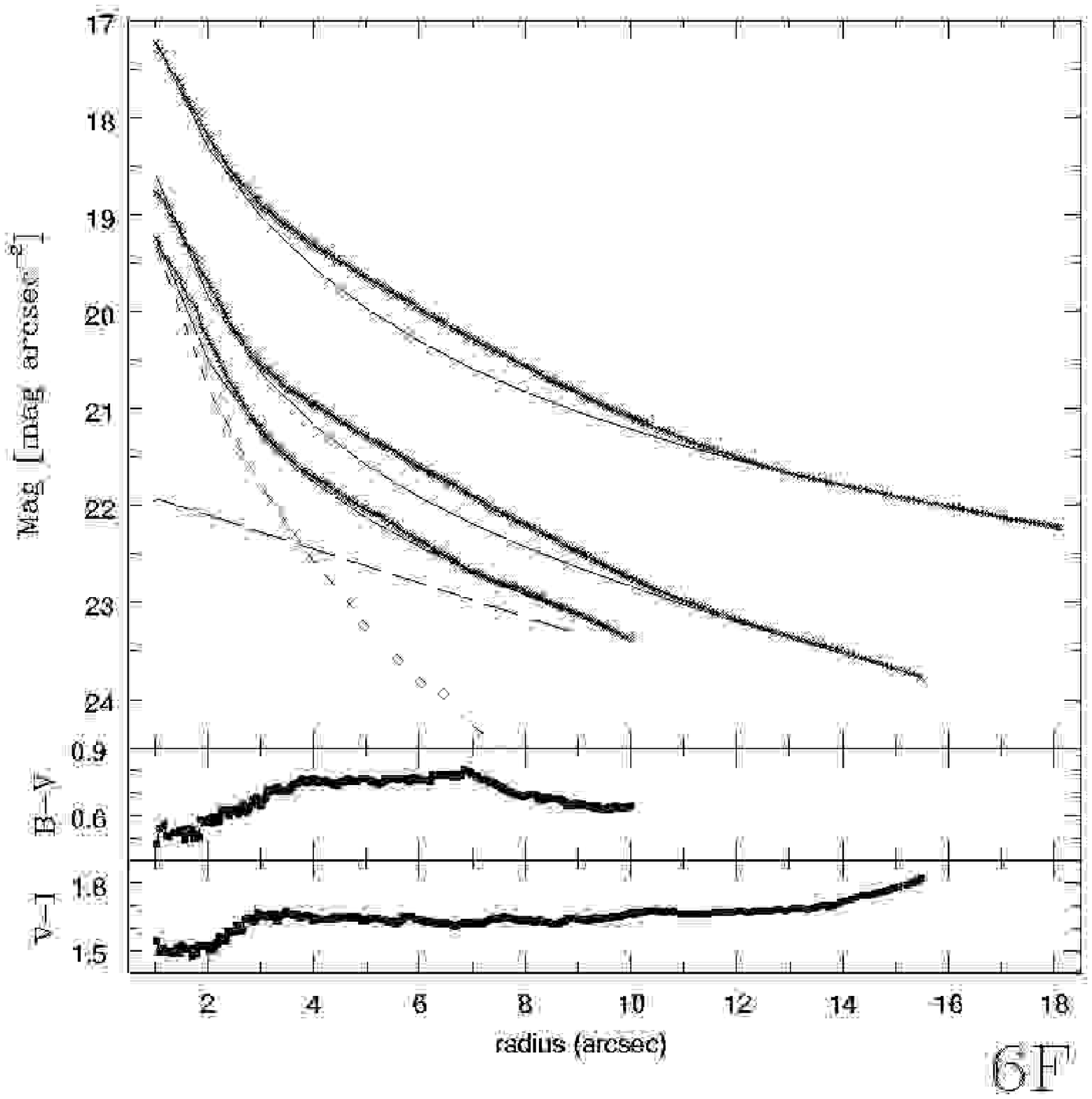}\\
\caption{AM2105-332. Same as Fig 1.}
\label{fig6}
\end{center}
\end{figure*}

The main galaxy also shows dust lanes in the southern region and four bright knots, which are very well aligned with the disk axis  (Fig.~\ref{fig6}).\\

\begin{figure*}
\begin{center}
\includegraphics[width=7cm,height=7cm]{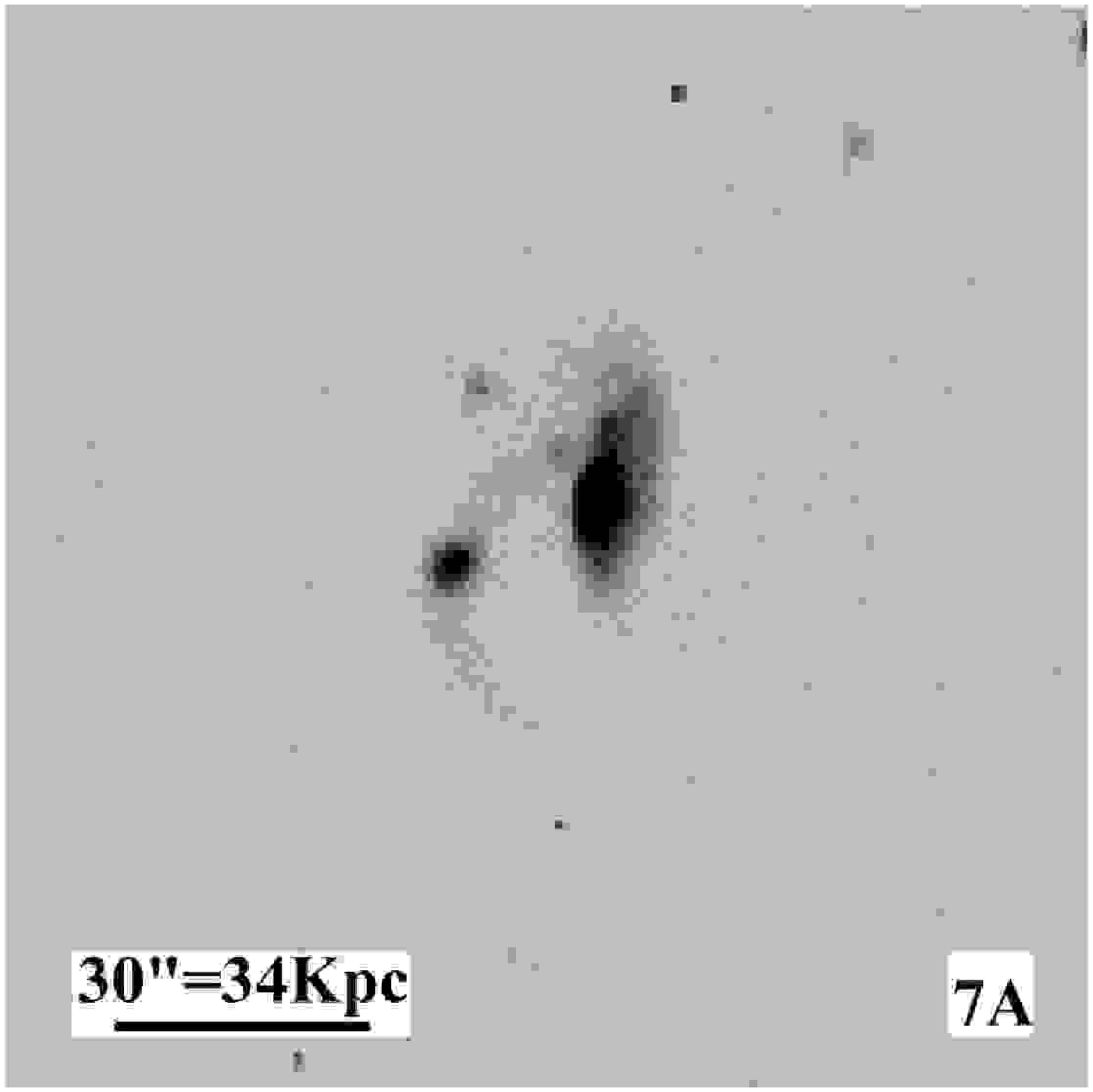}\includegraphics[width=7cm,height=7cm]{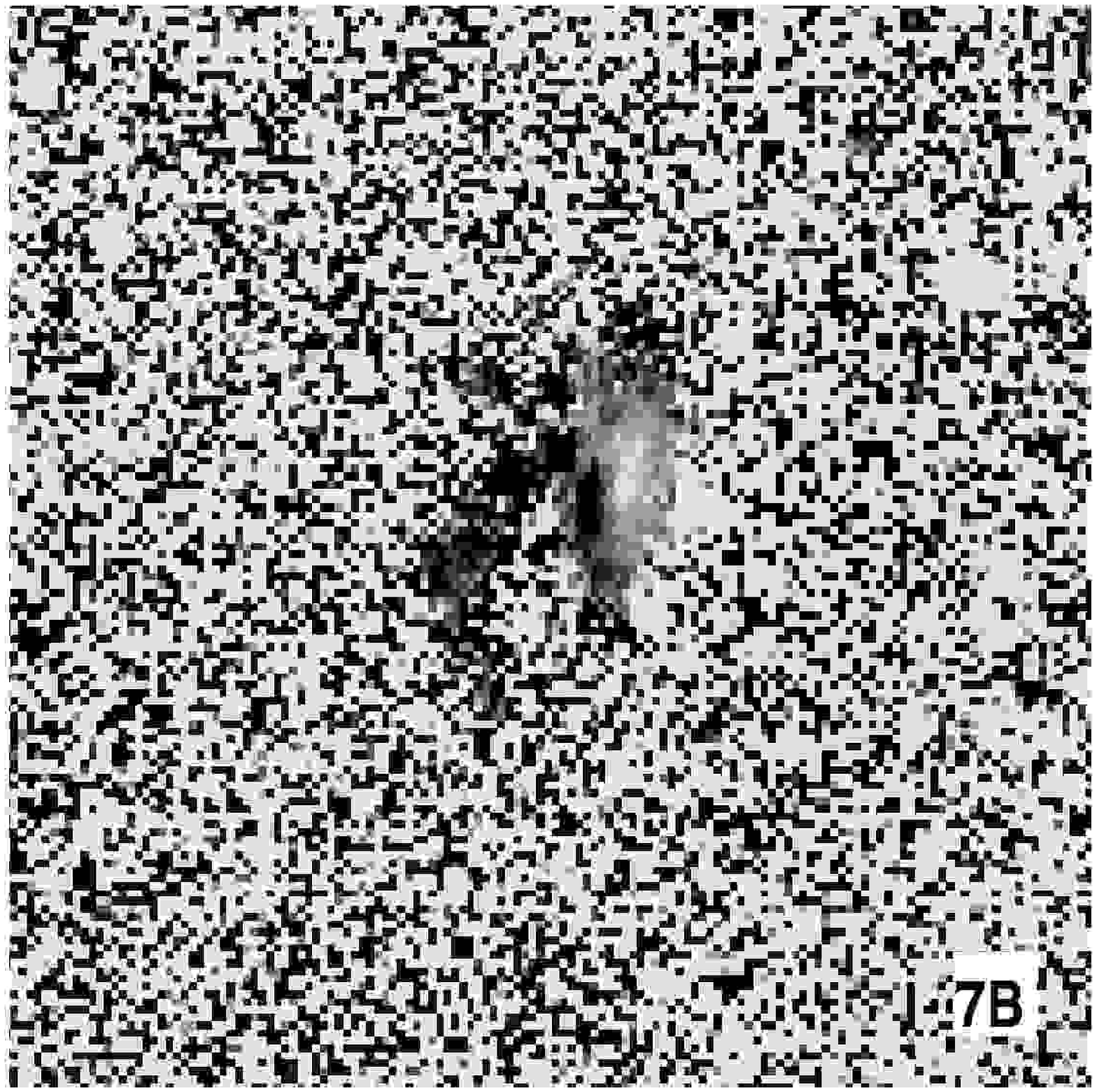}\\
\includegraphics[width=7cm,height=7cm]{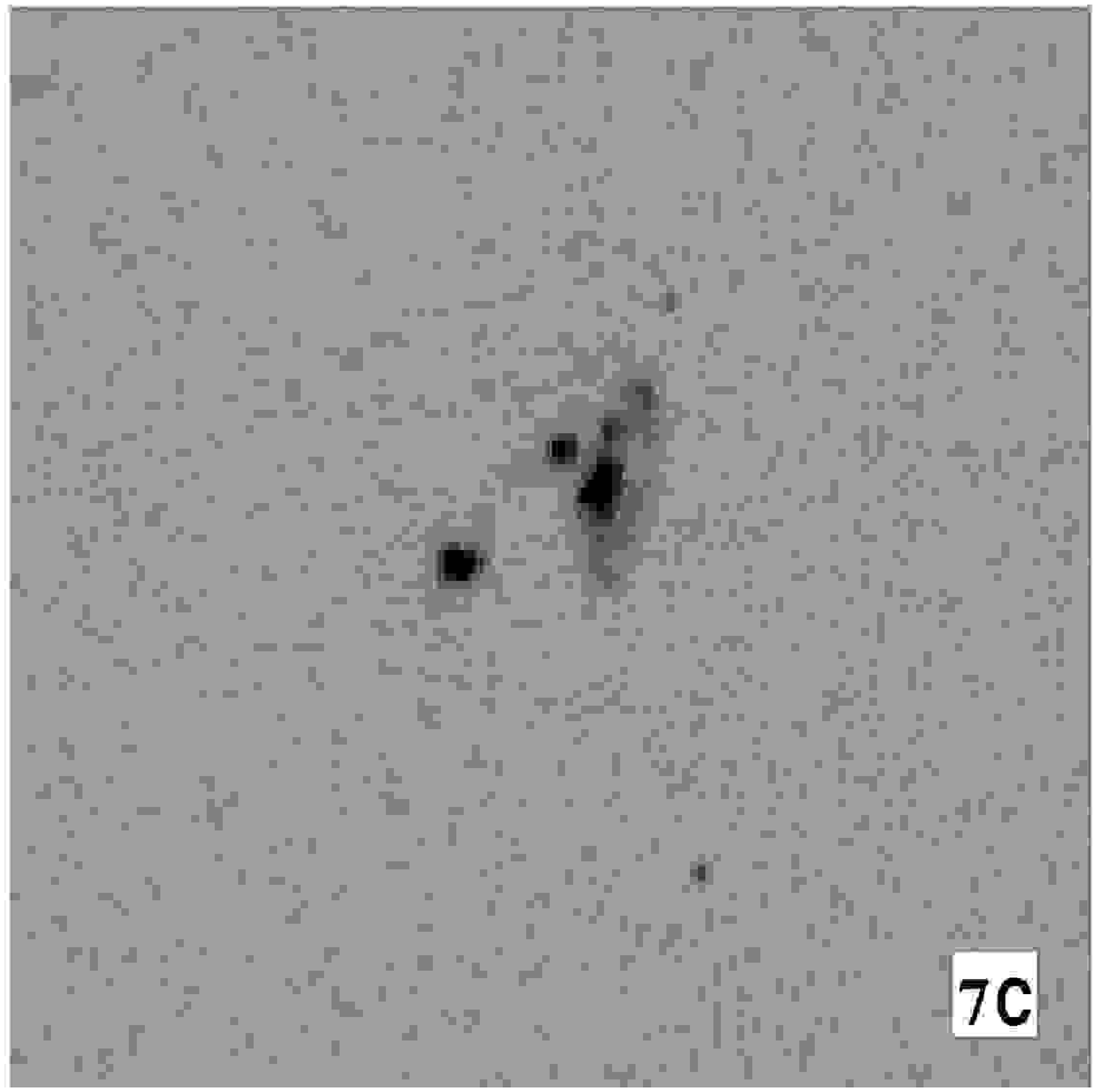}\includegraphics[width=7cm,height=7cm]{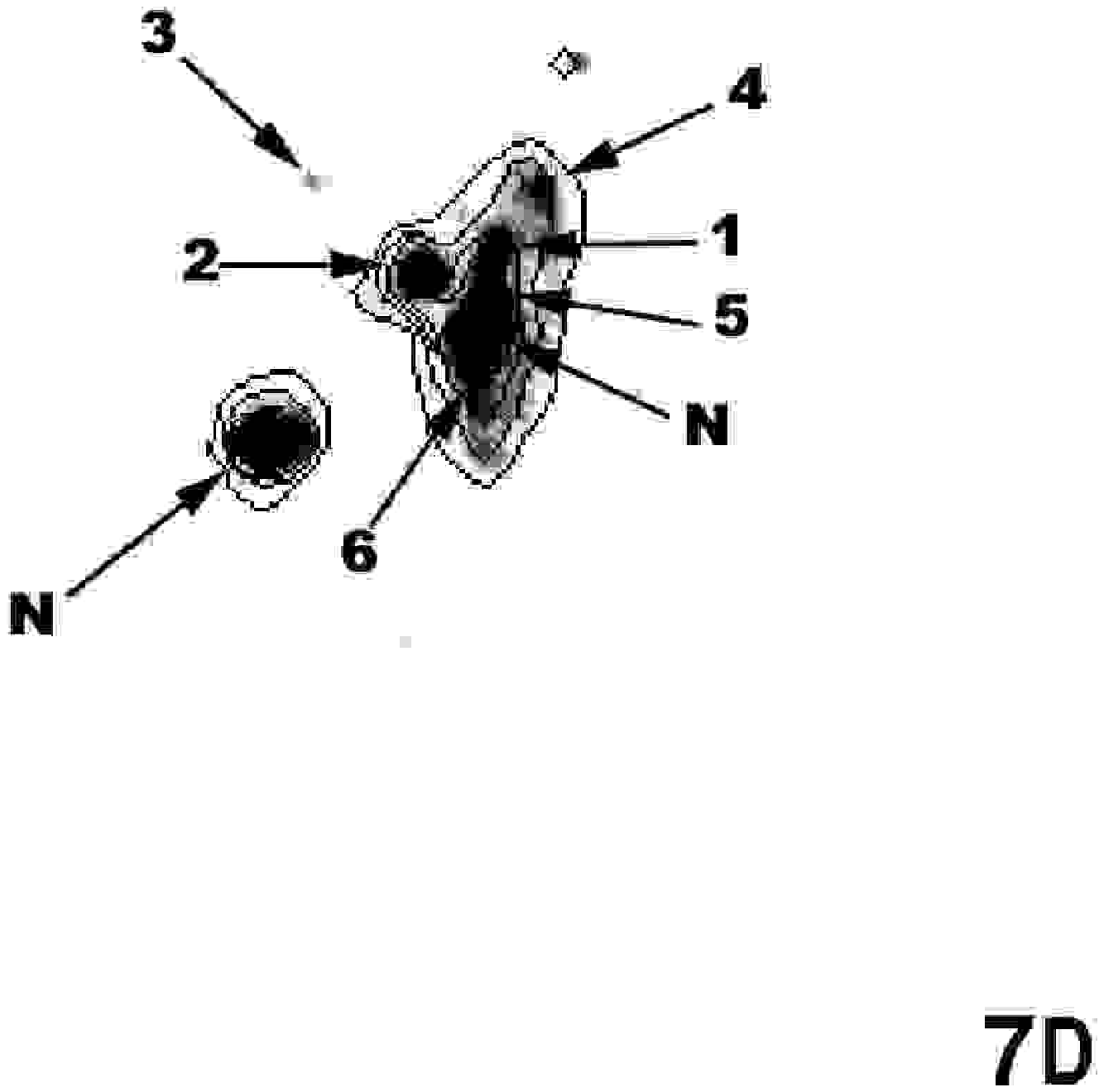}\\
\includegraphics[width=7cm,height=7cm]{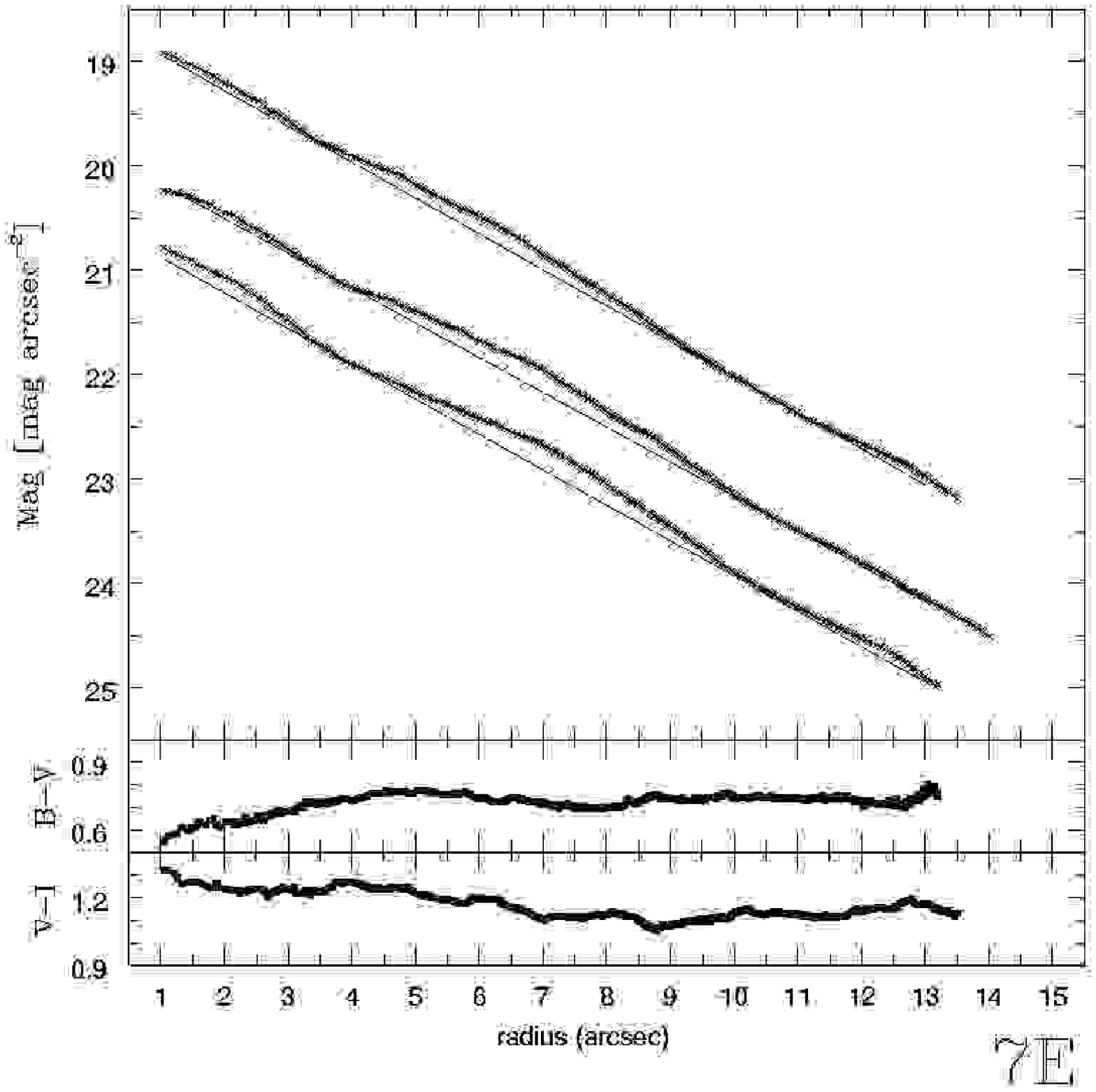}\includegraphics[width=7cm,height=7cm]{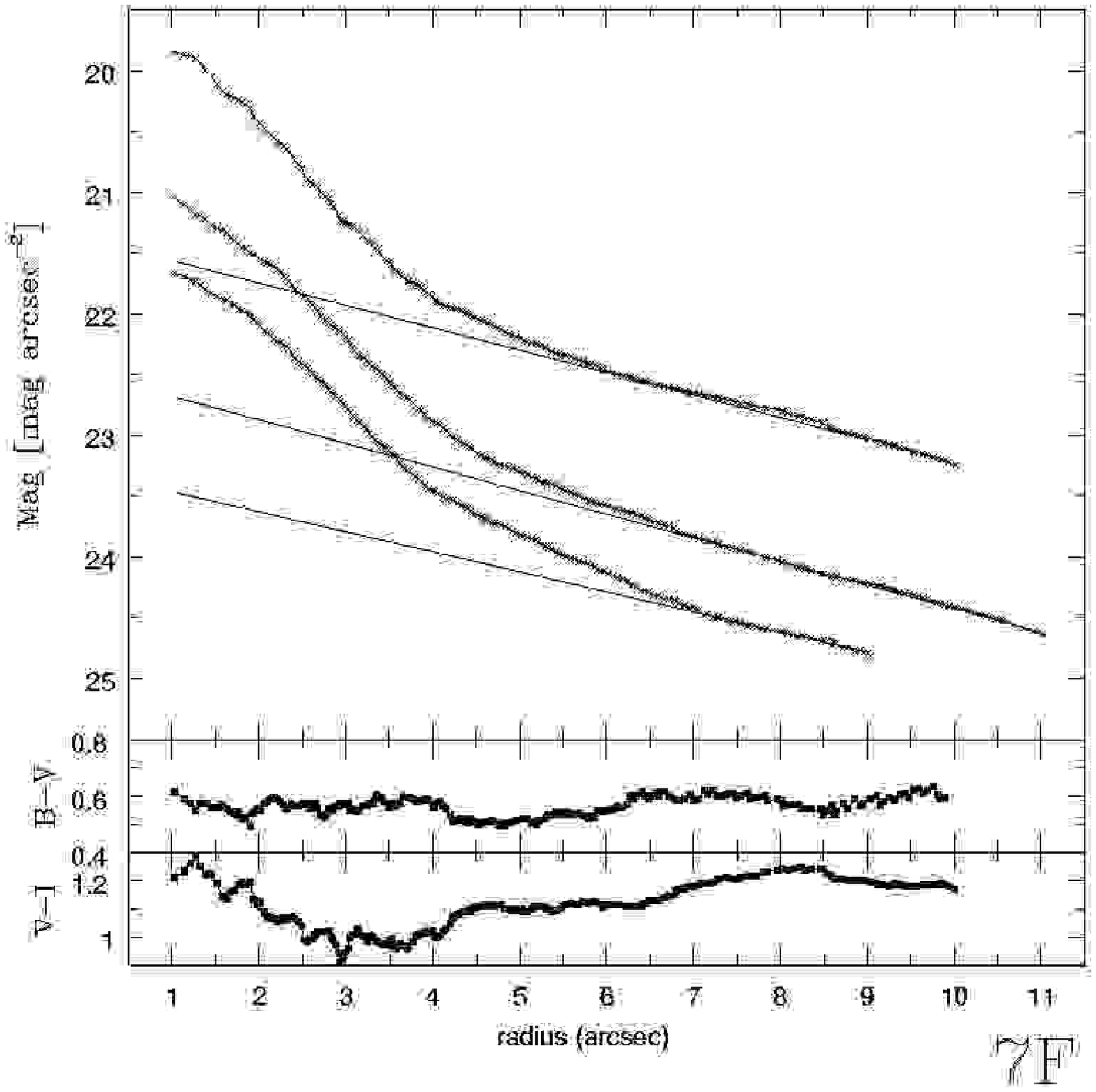}\\
\caption{AM2229-735. Same as Fig 1.}
\label{fig7}
\end{center}
\end{figure*}

AM2229-735: This pair located at $cz$ = 17535 $km$ $s^{-1}$ is formed by a disk galaxy and a compact elliptical. The galaxies are connected by a very luminous bridge that contains 17\% of the total luminosity of the system. The main galaxy is very disturbed and shows a typical exponential luminosity profile. It exhibits several star forming regions throughout the galaxy (Fig.~\ref{fig7}).\\

\begin{figure*}
\begin{center}
\includegraphics[width=7cm,height=7cm]{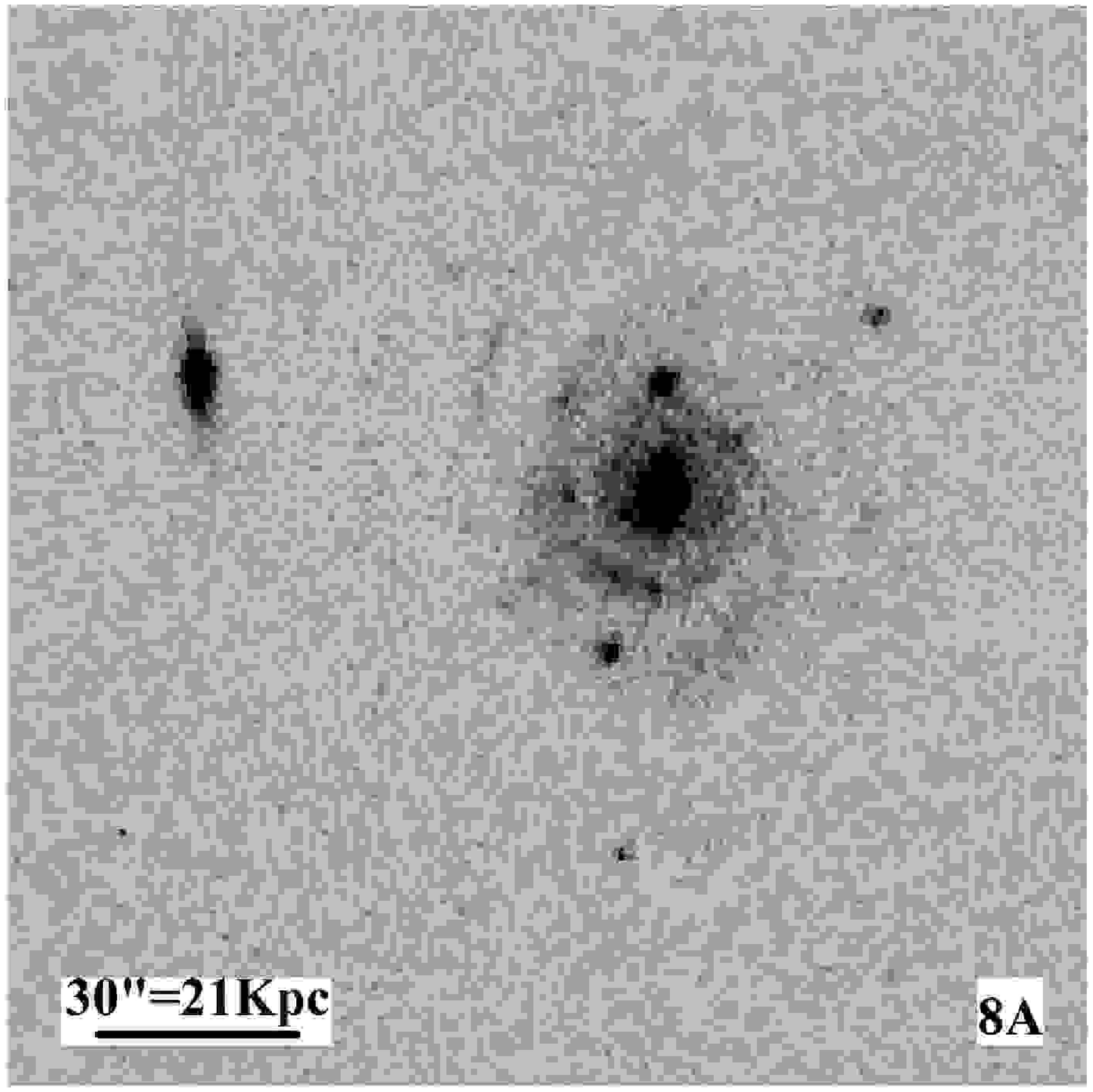}\includegraphics[width=7cm,height=7cm]{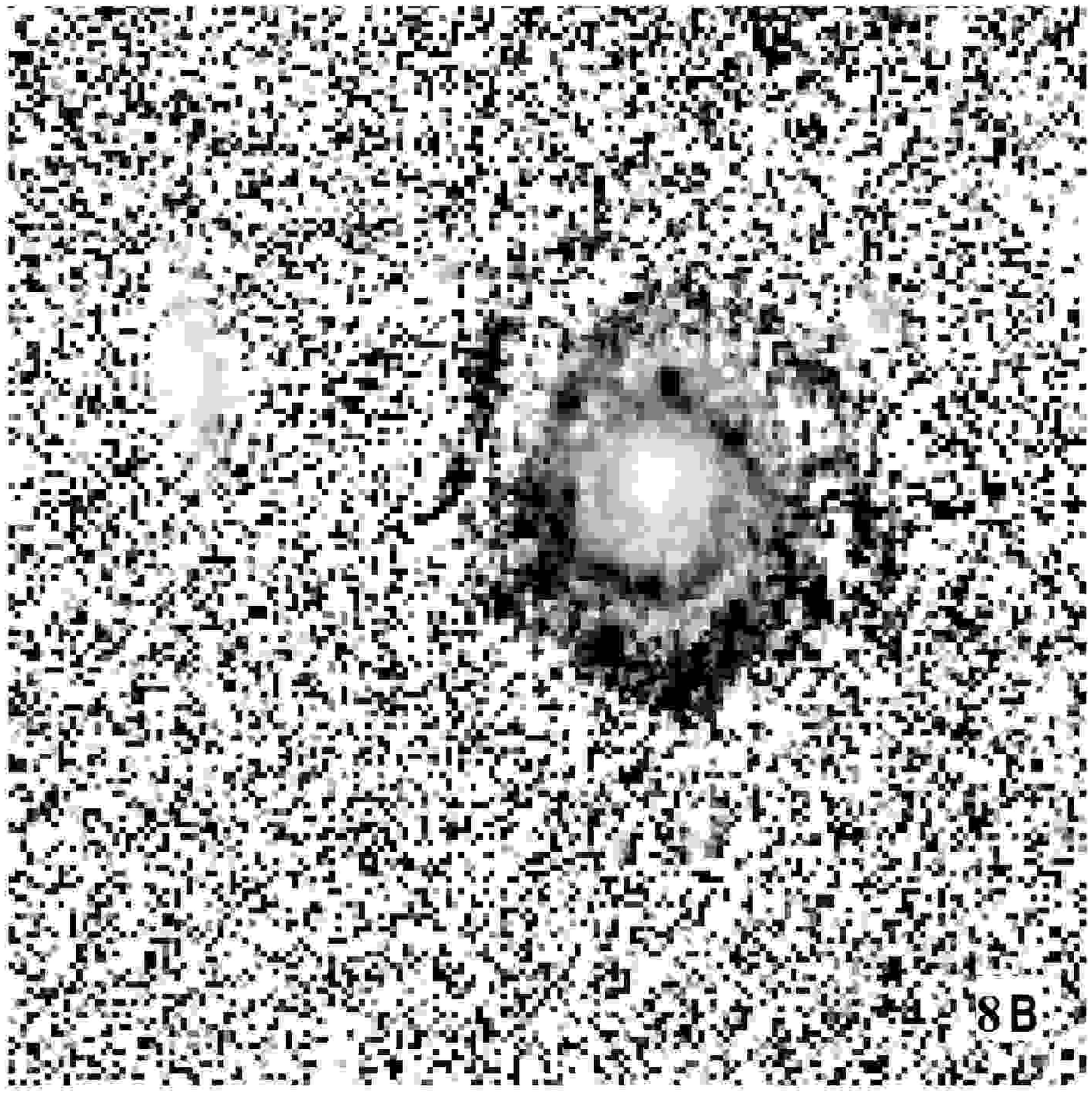}\\
\includegraphics[width=7cm,height=7cm]{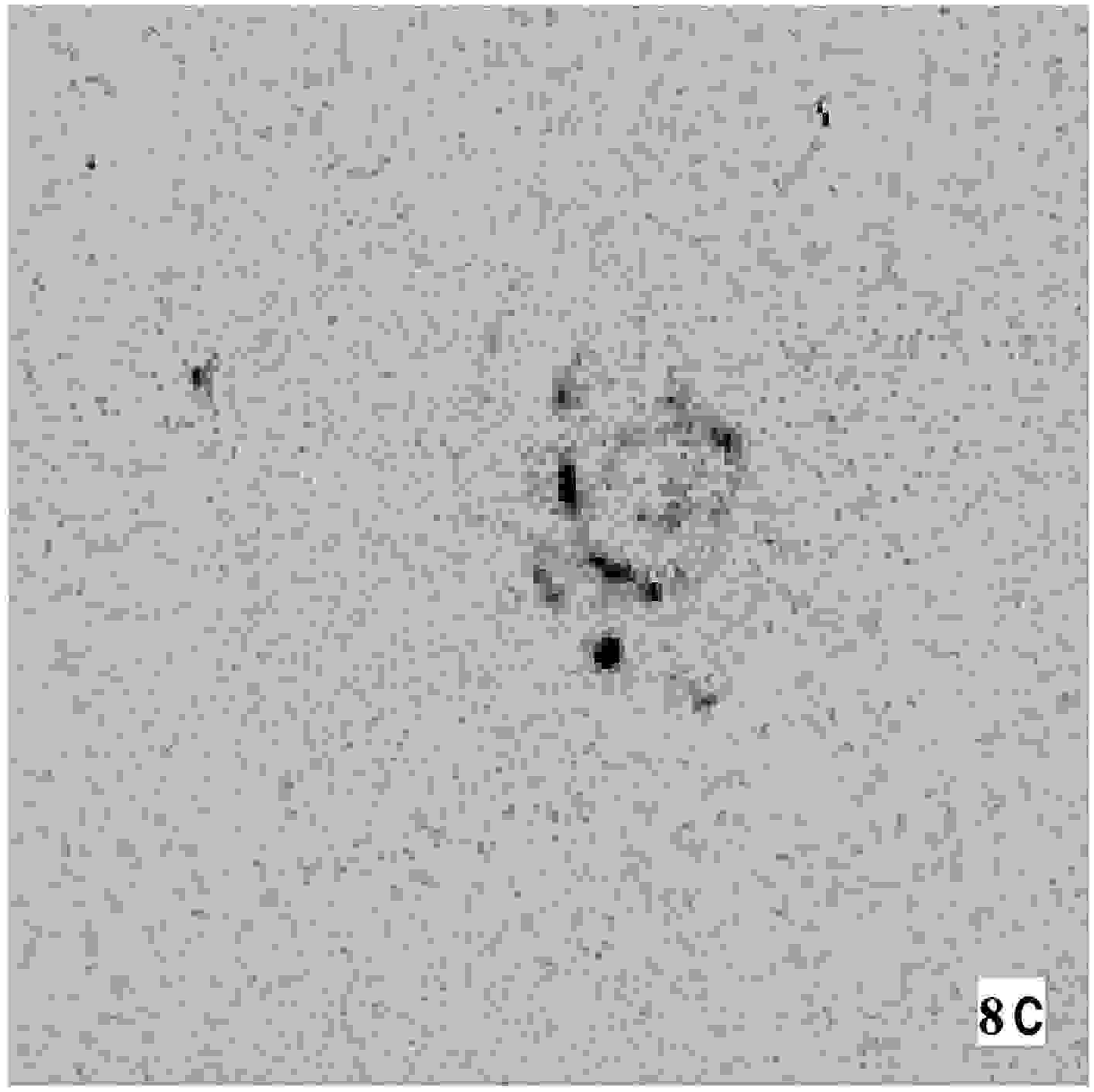}\includegraphics[width=7cm,height=7cm]{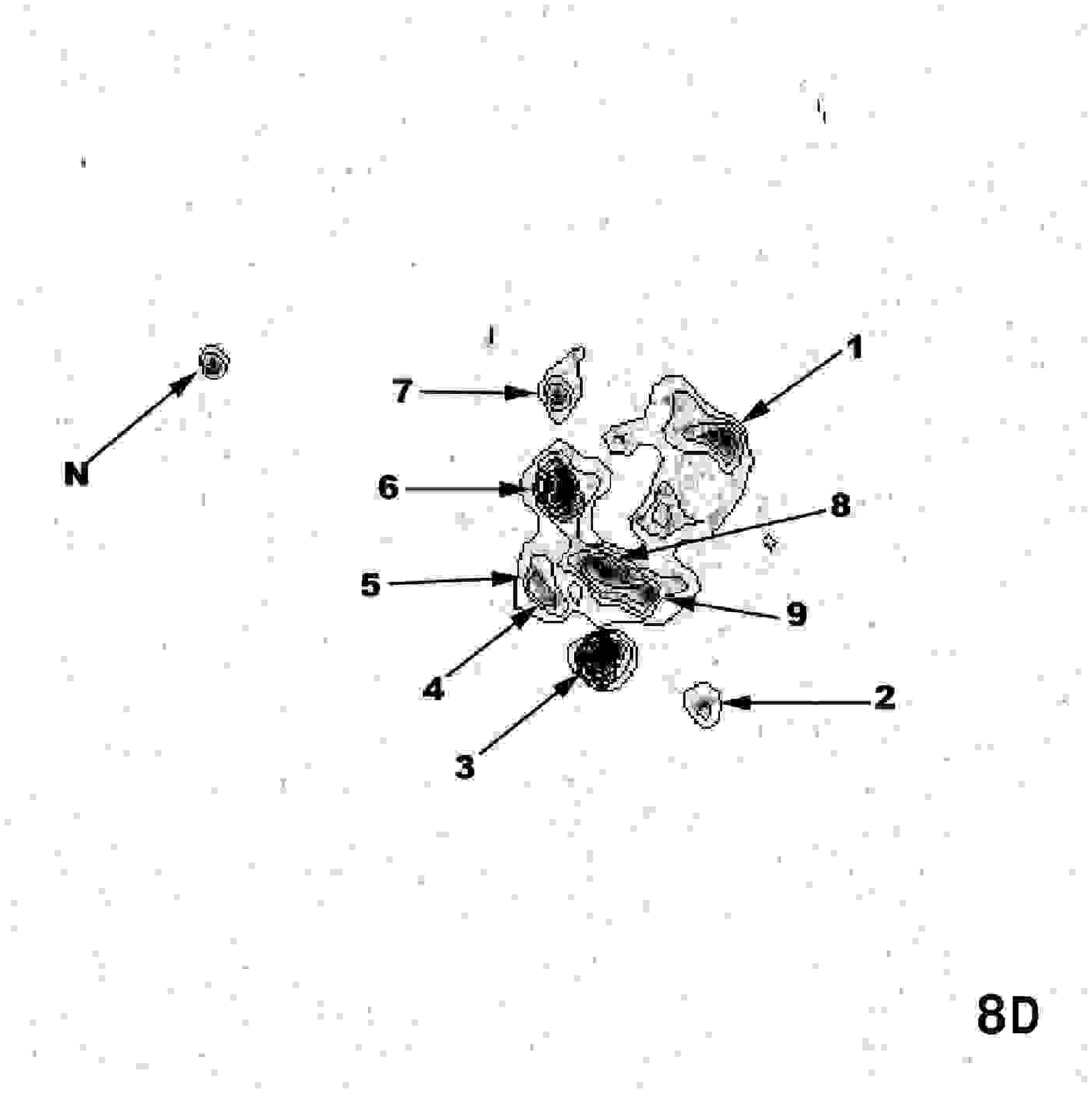}\\
\includegraphics[width=7cm,height=7cm]{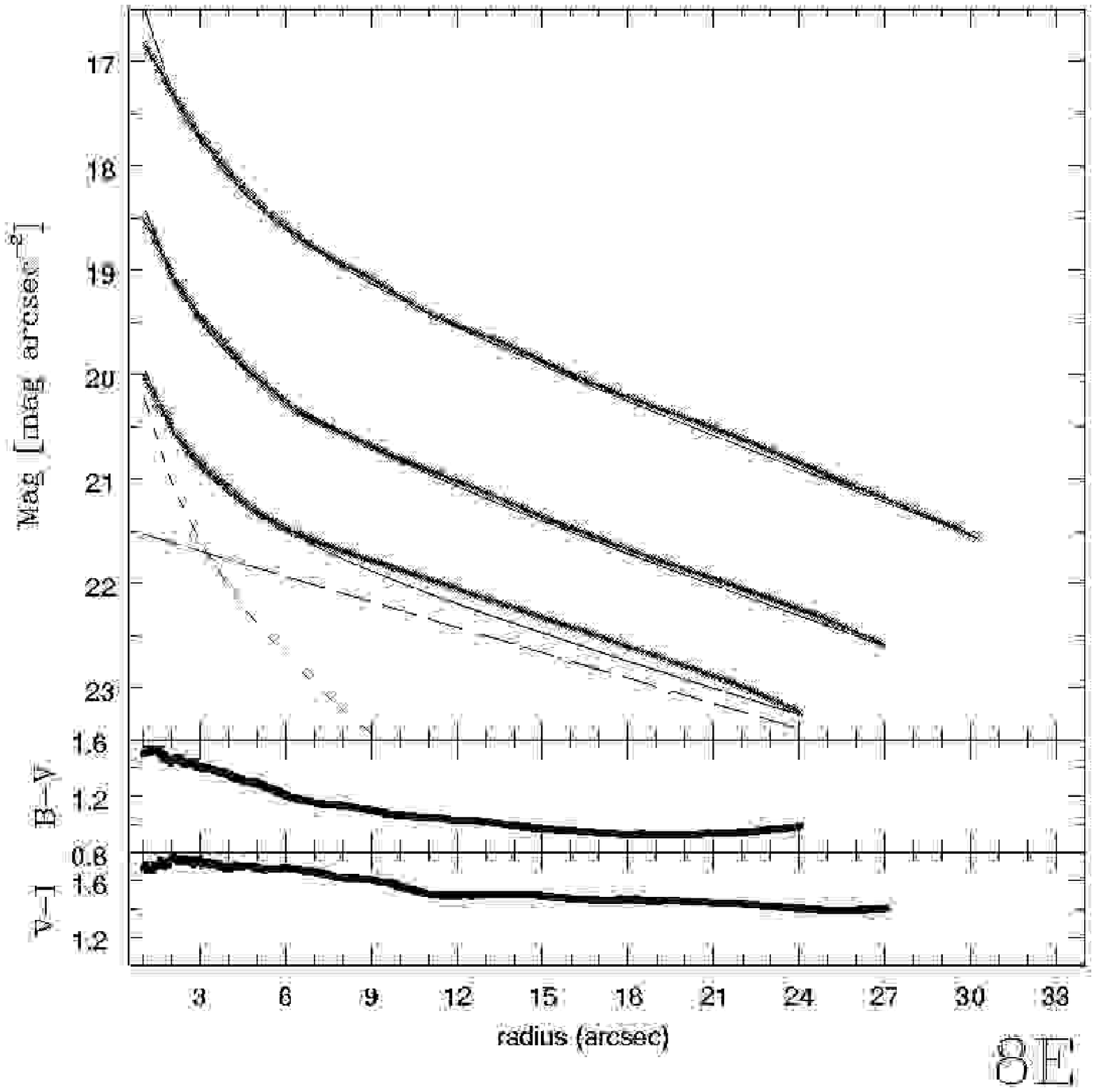}\includegraphics[width=7cm,height=7cm]{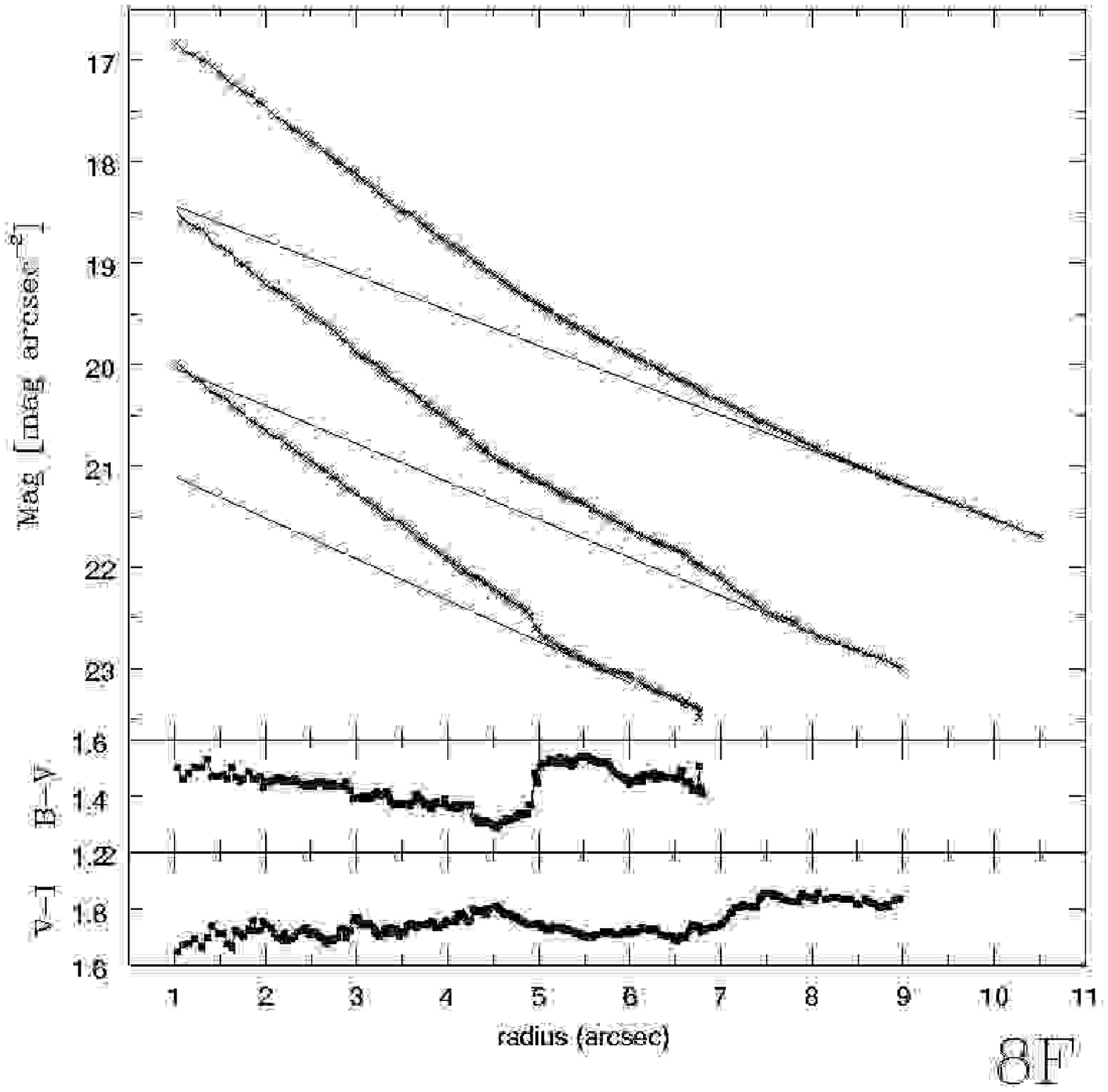}\\
\caption{AM2238-575. Same as Fig 1.}
\label{fig8}
\end{center}
\end{figure*}

AM2238-575: This system is listed as an apparent pair in the Arp-Madore Catalogue (1987). Spectroscopic data from DP97 shows that it is a physical pair located at $cz$ = 10838 $km$ $s^{-1}$. The main galaxy is a giant Sb/c, almost face-on. The luminosity profile is fitted by a bulge plus a disk component. Several HII regions are observed along the east arm (Fig.~\ref{fig8}).

The secondary is an edge-on disk galaxy, probably SO type.\\

\begin{figure*}
\begin{center}
\includegraphics[width=7cm,height=7cm]{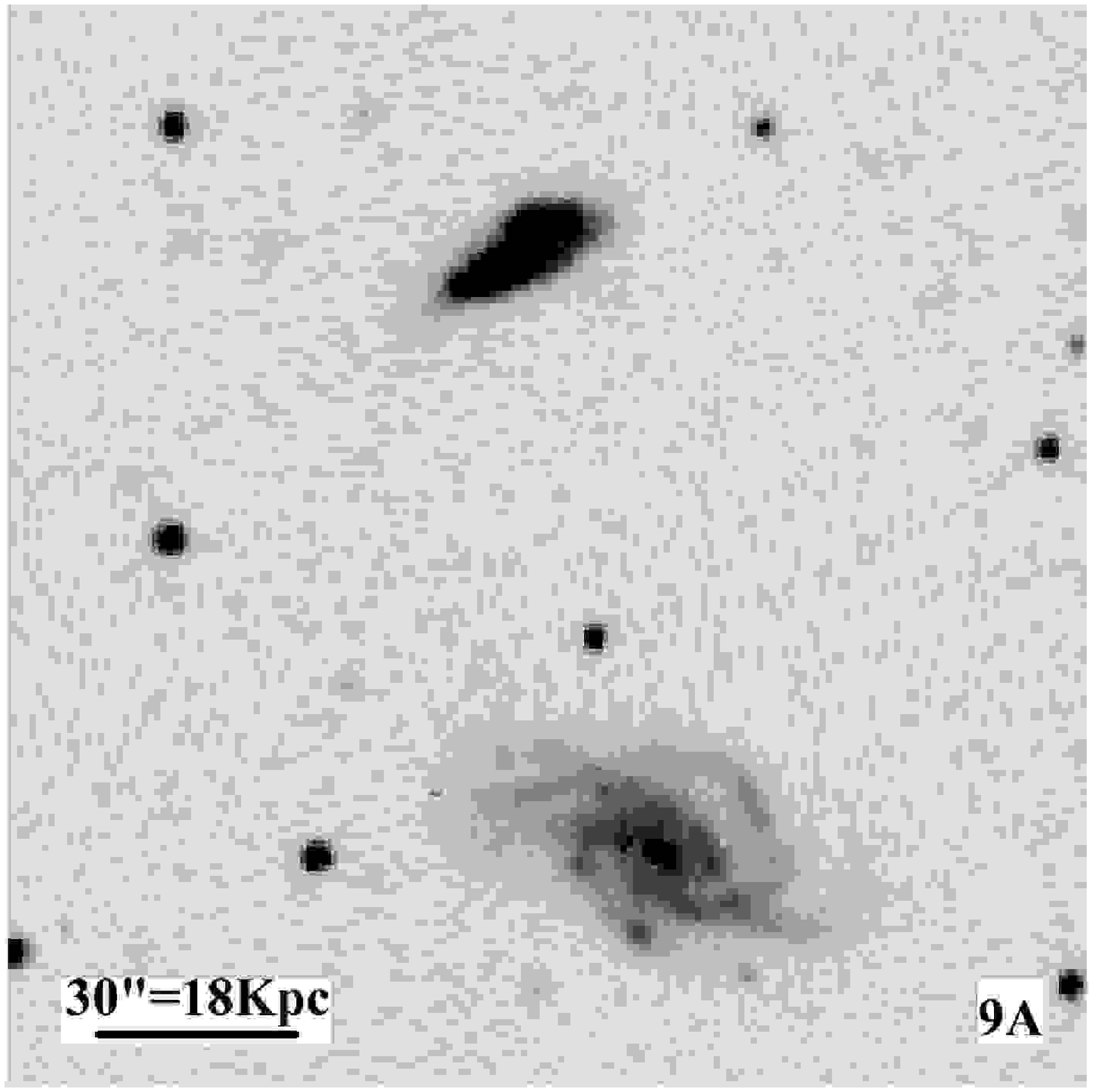}\includegraphics[width=7cm,height=7cm]{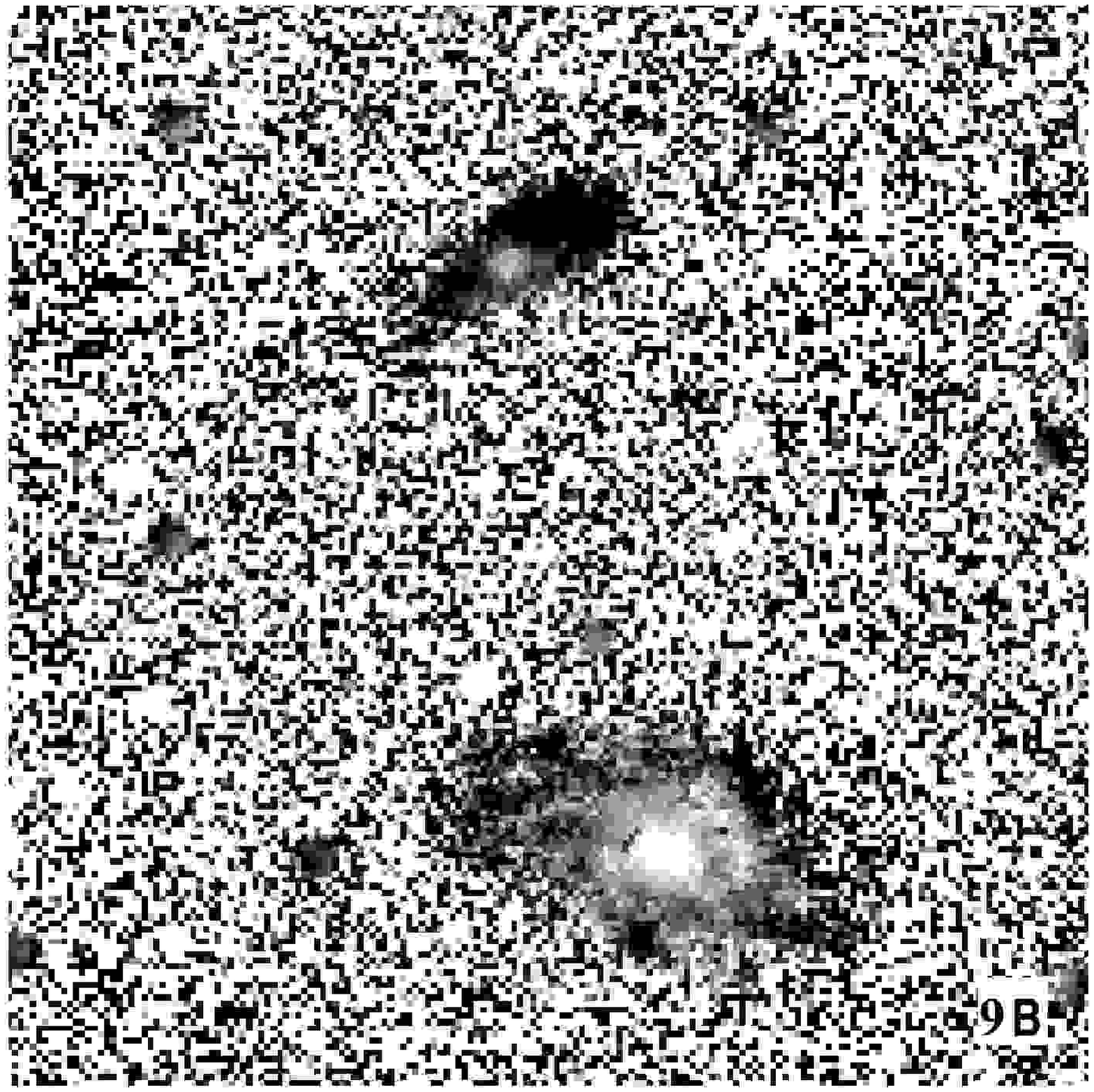}\\
\includegraphics[width=7cm,height=7cm]{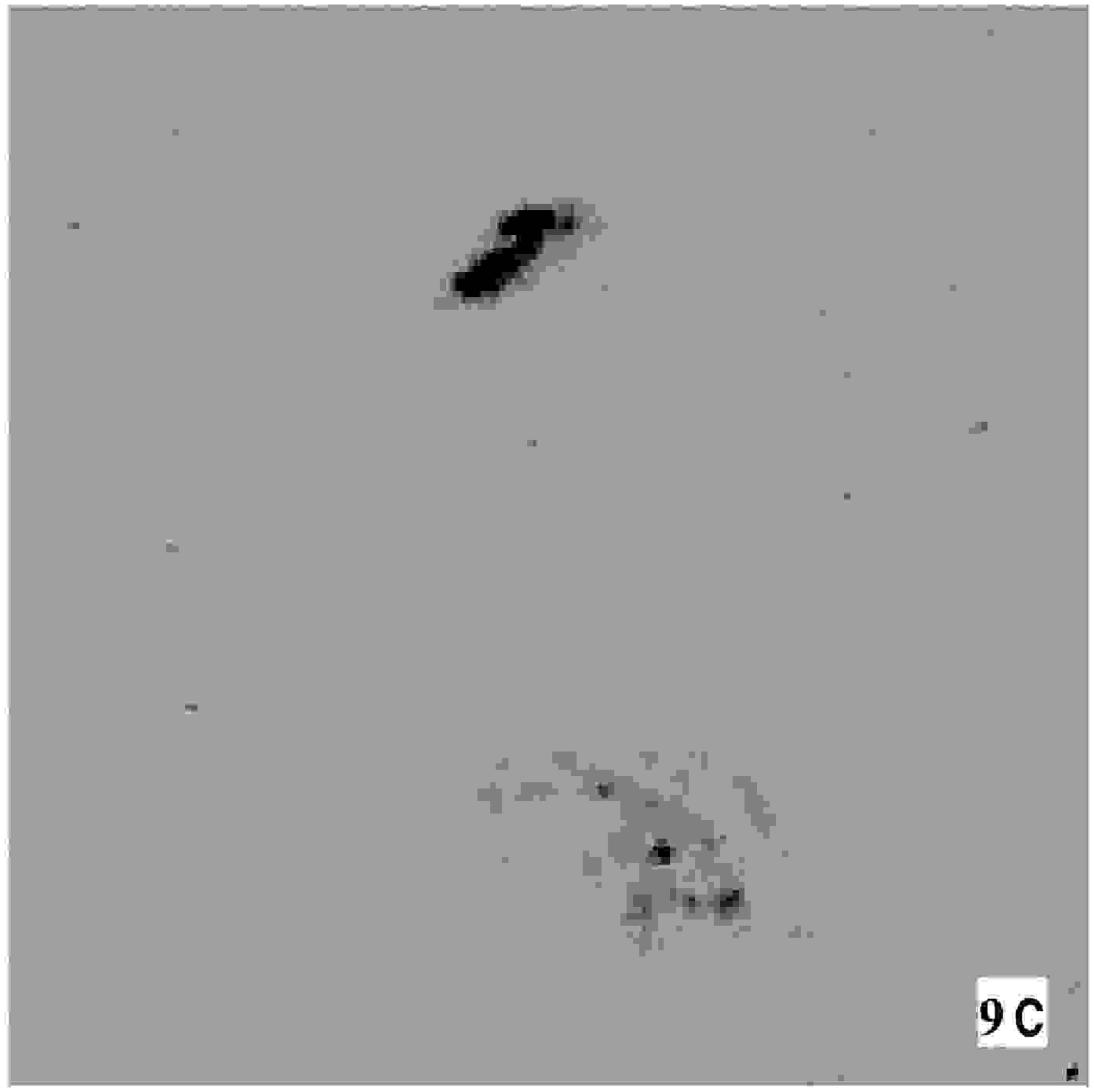}\includegraphics[width=7cm,height=7cm]{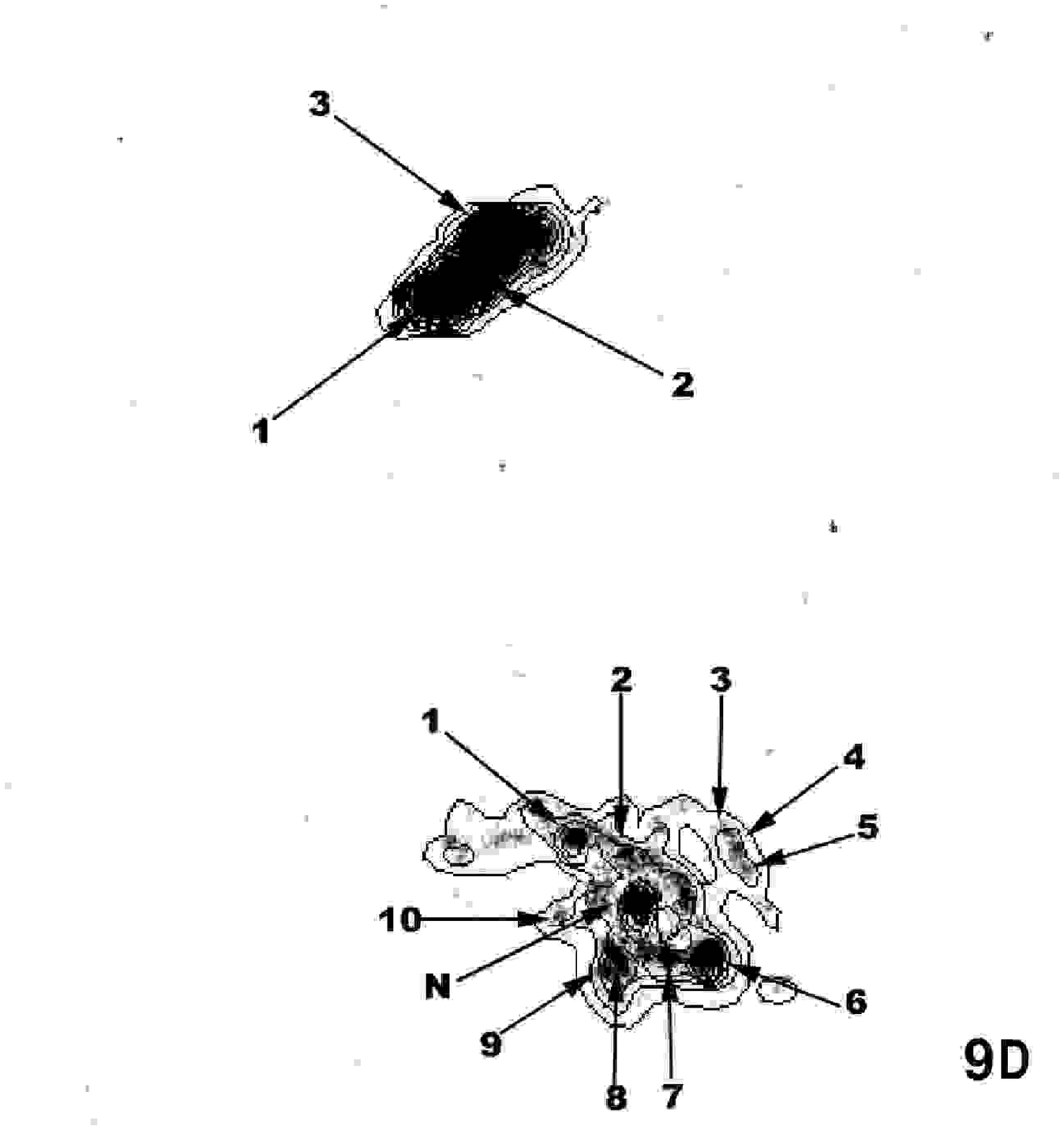}\\
\includegraphics[width=7cm,height=7cm]{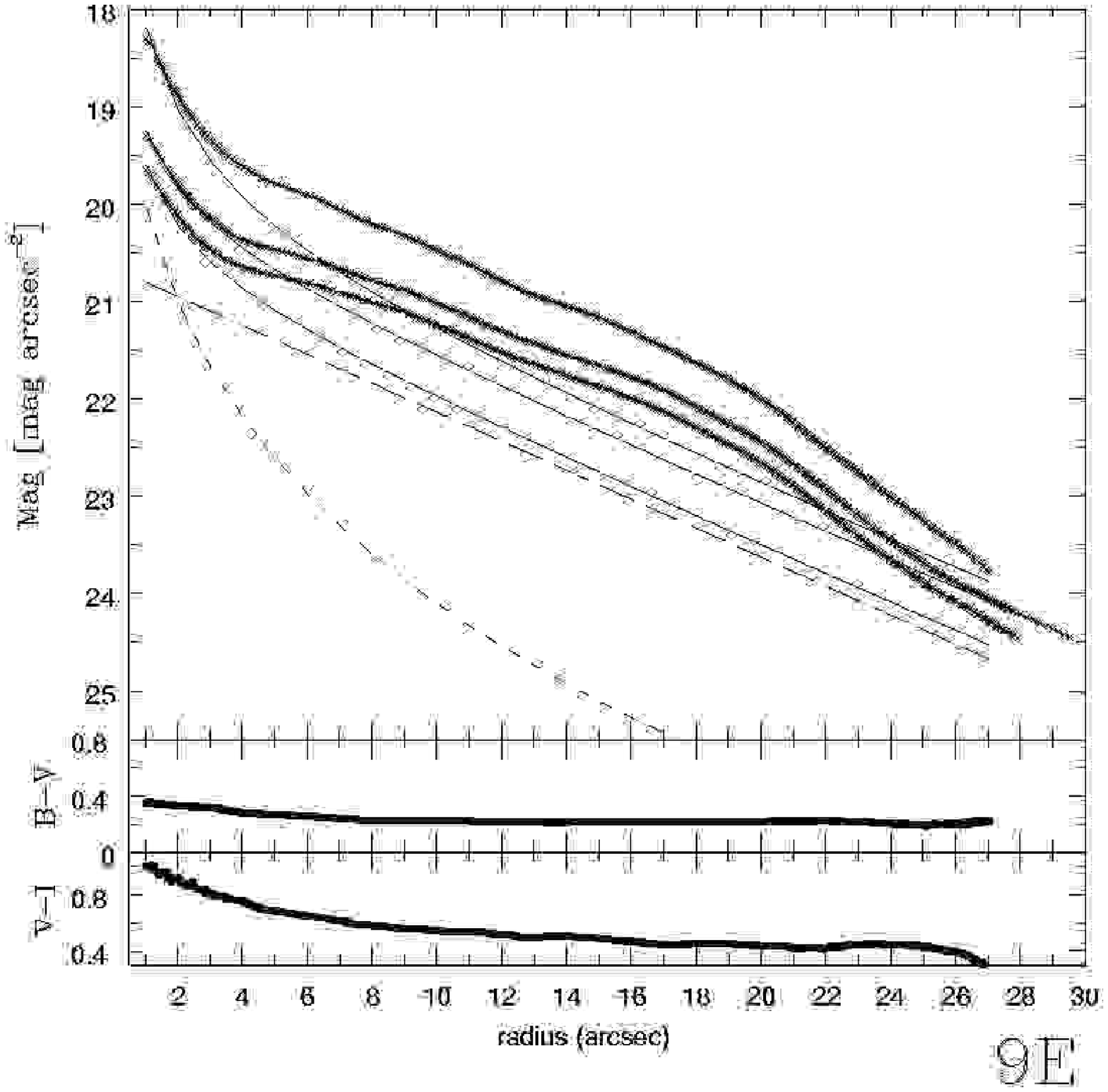}\includegraphics[width=7cm,height=7cm]{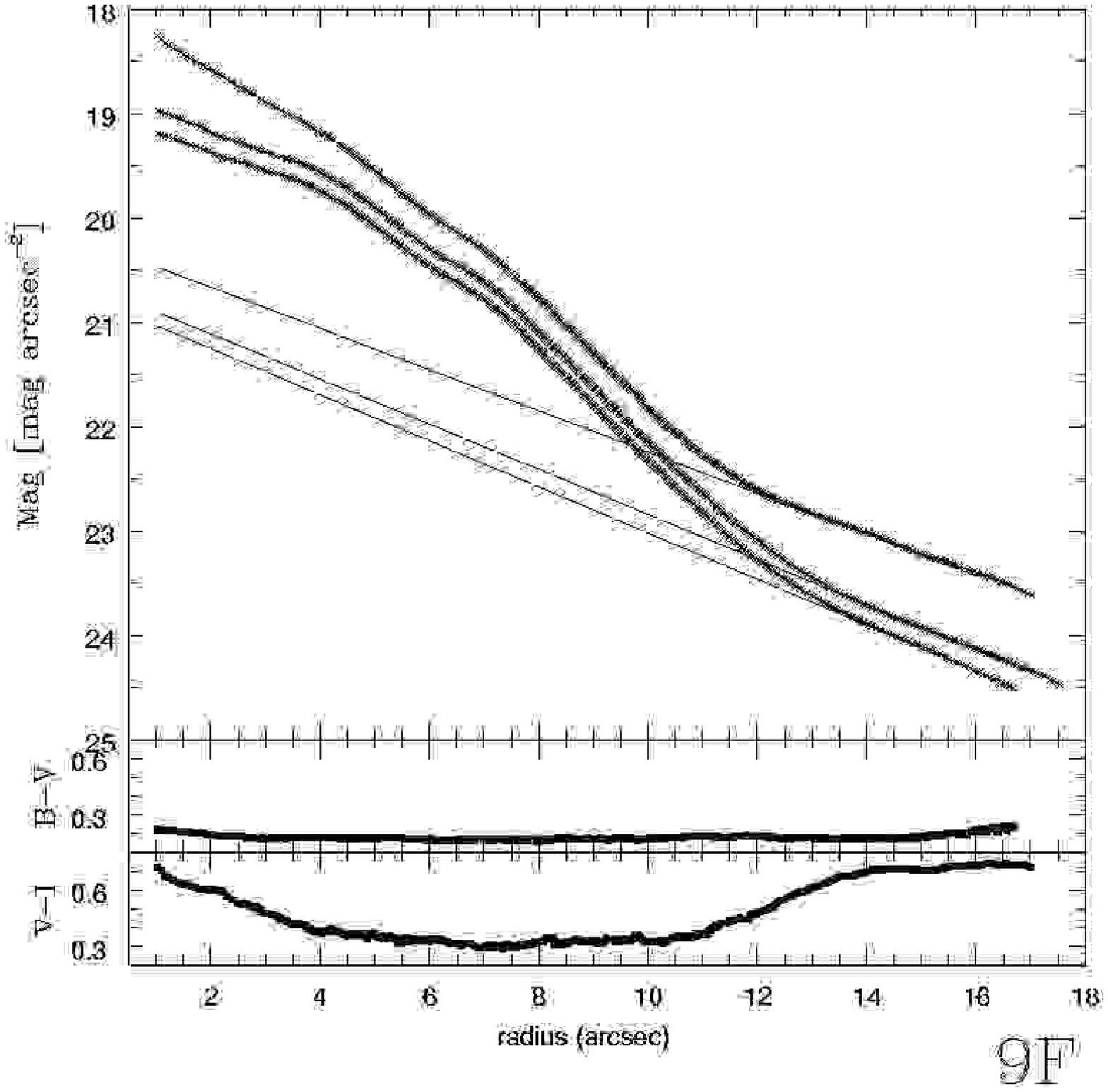}\\
\caption{AM2306-721. Same as Fig 1.}
\label{fig9}
\end{center}
\end{figure*}

AM2306-721: This pair at  $cz$ = 8916 $km$ $s^{-1}$ is formed by a peculiar spiral with disturbed arms interacting with an irregular galaxy, probably disrupted by tidal forces (Fig.~\ref{fig9}). The main component does not show evidence of enhanced star formation   but is blue, (B-V) = 0.24.

The main body of the secondary galaxy is resolved into several bright HII regions and has a blue colour, (B-V) = 0.18.\\

\begin{figure*}
\begin{center}
\includegraphics[width=7cm,height=7cm]{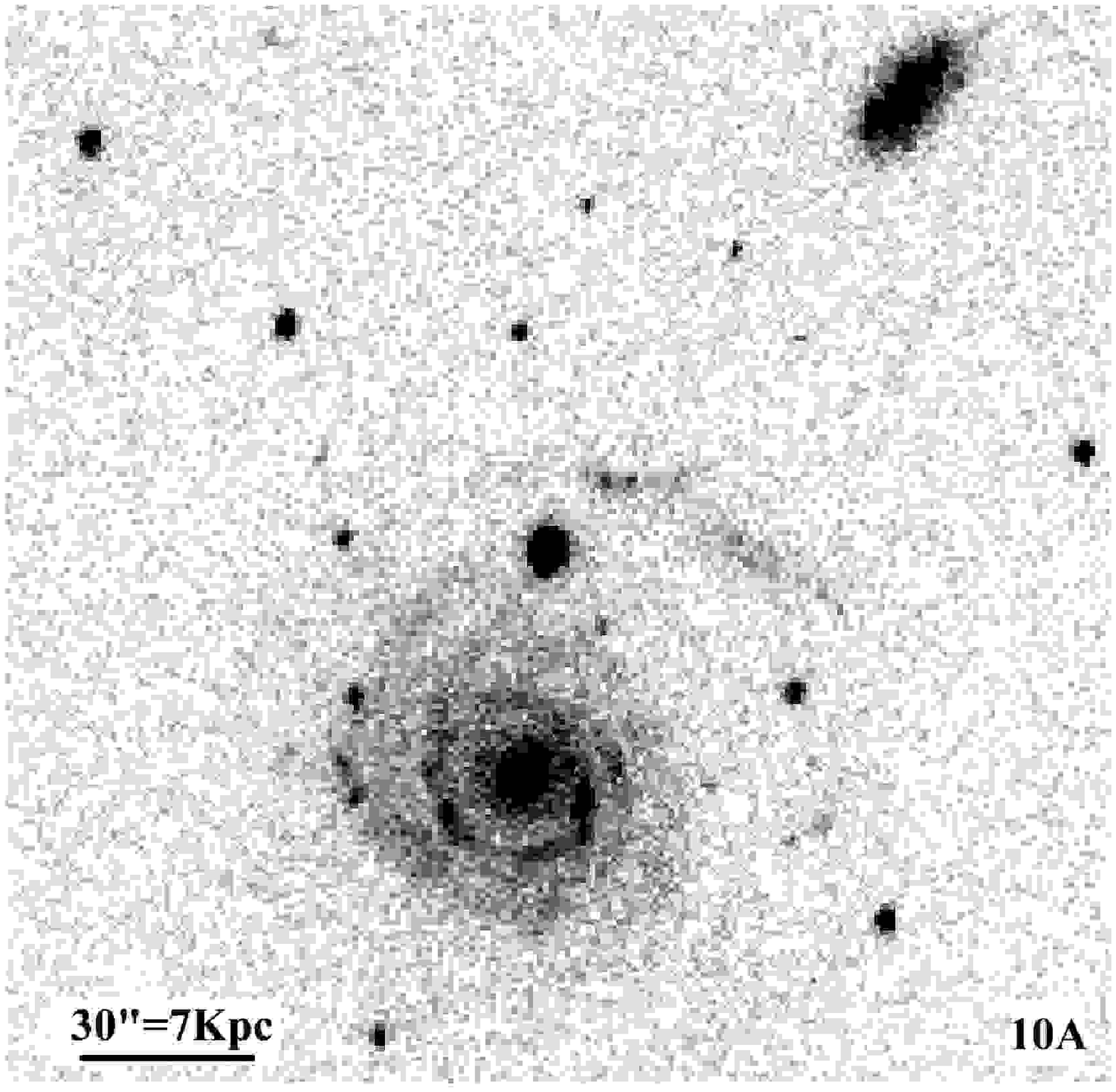}\includegraphics[width=7cm,height=7cm]{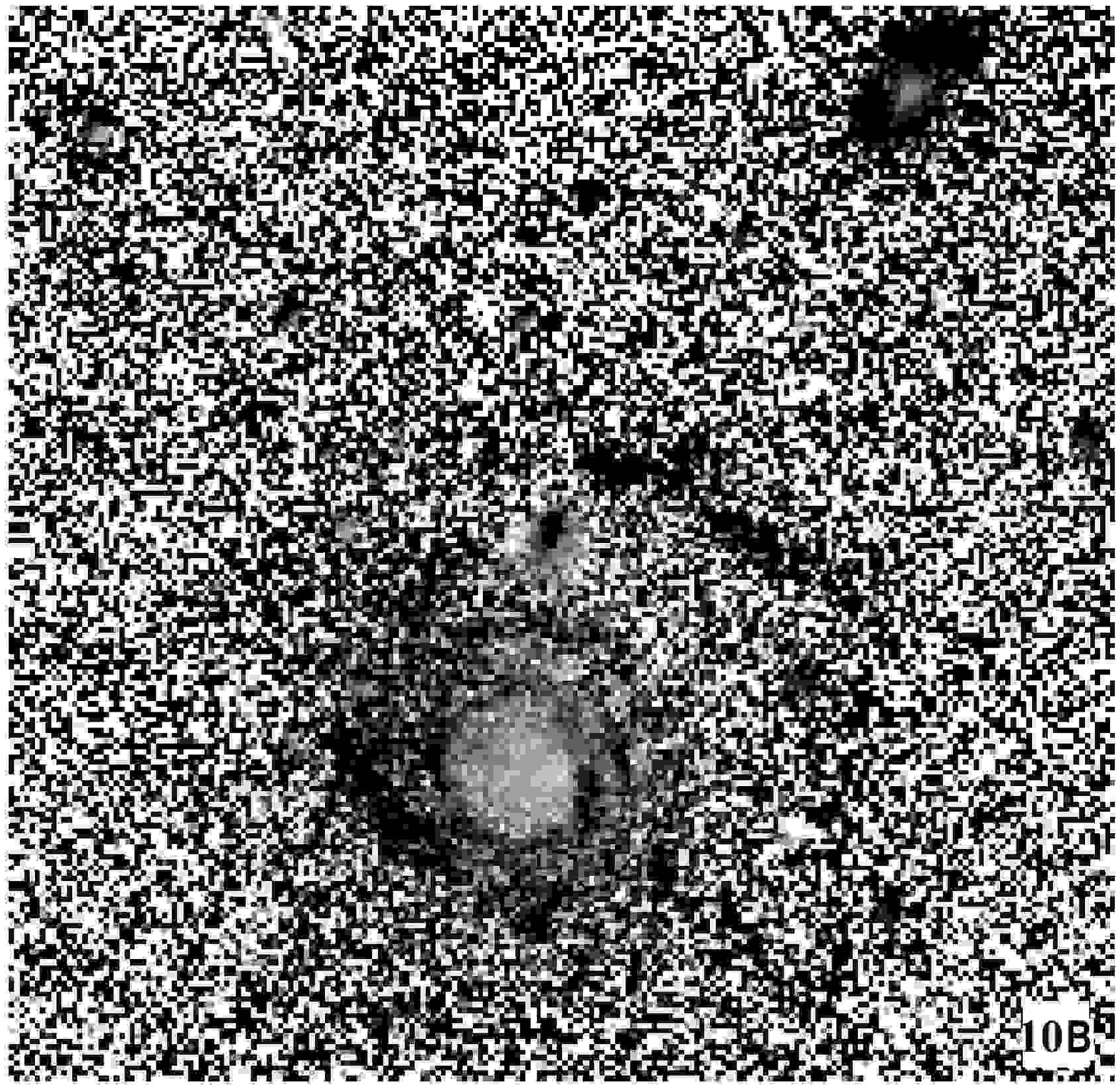}\\
\includegraphics[width=7cm,height=7cm]{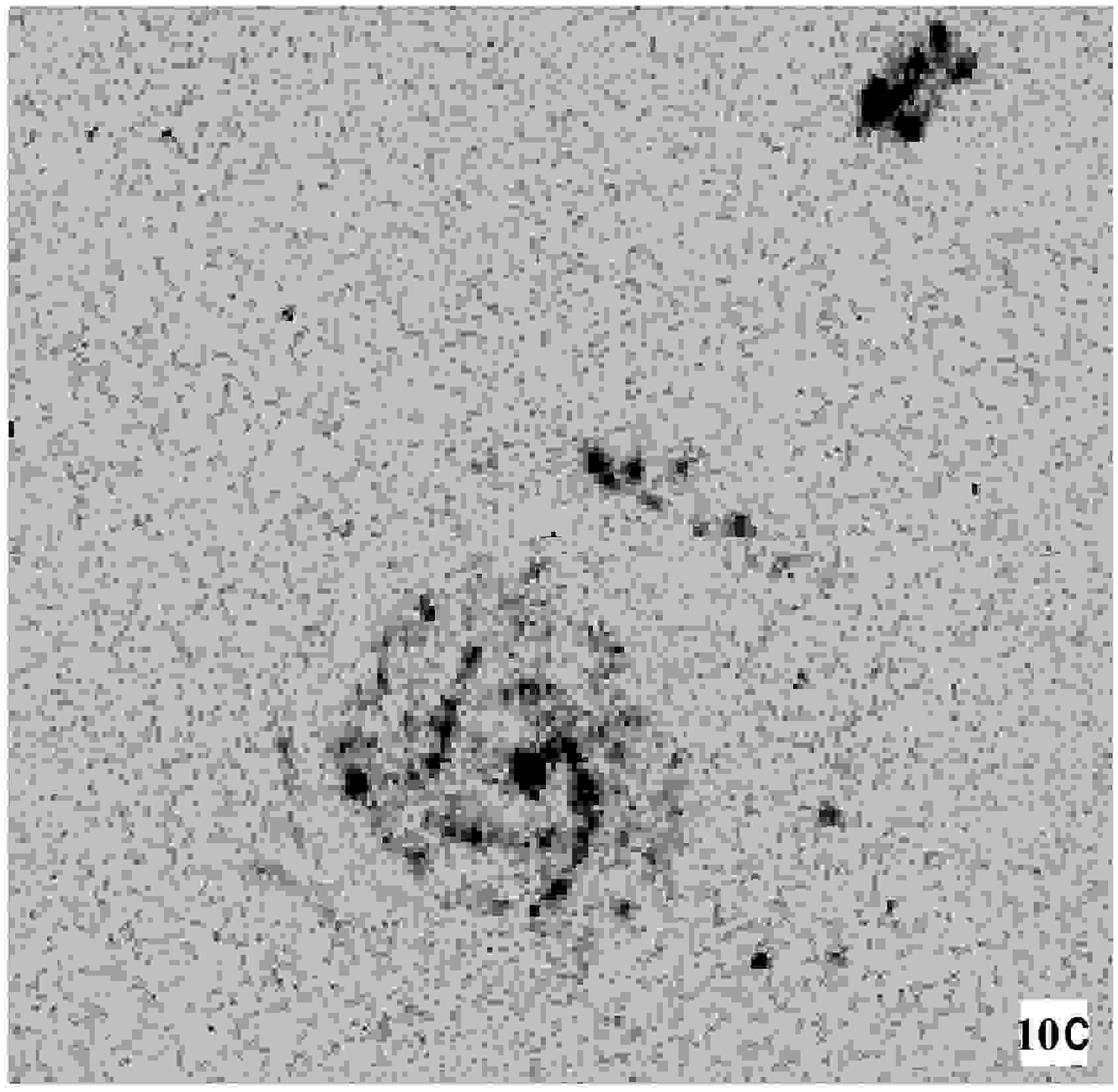}\includegraphics[width=7cm,height=7cm]{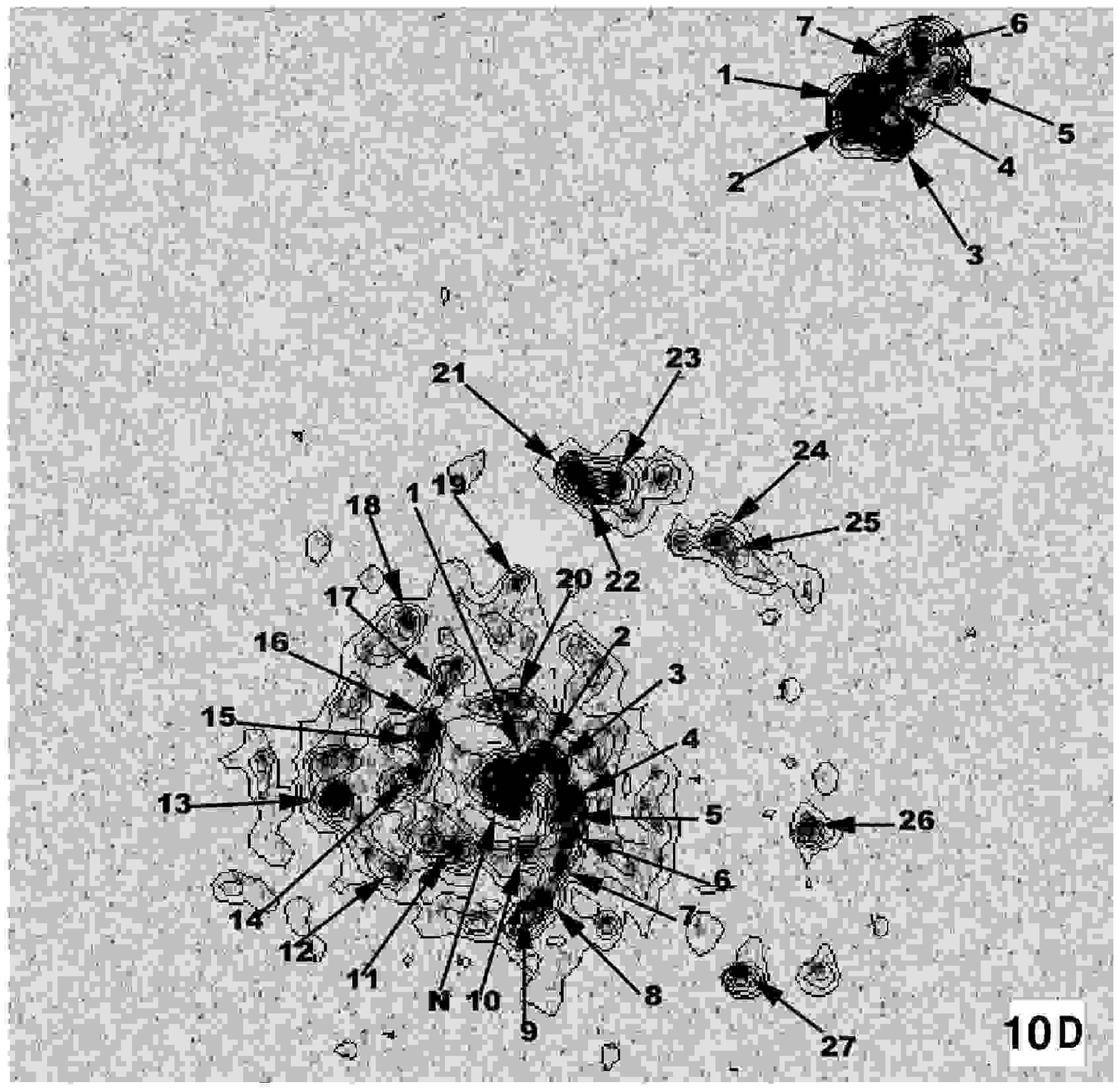}\\
\includegraphics[width=7cm,height=7cm]{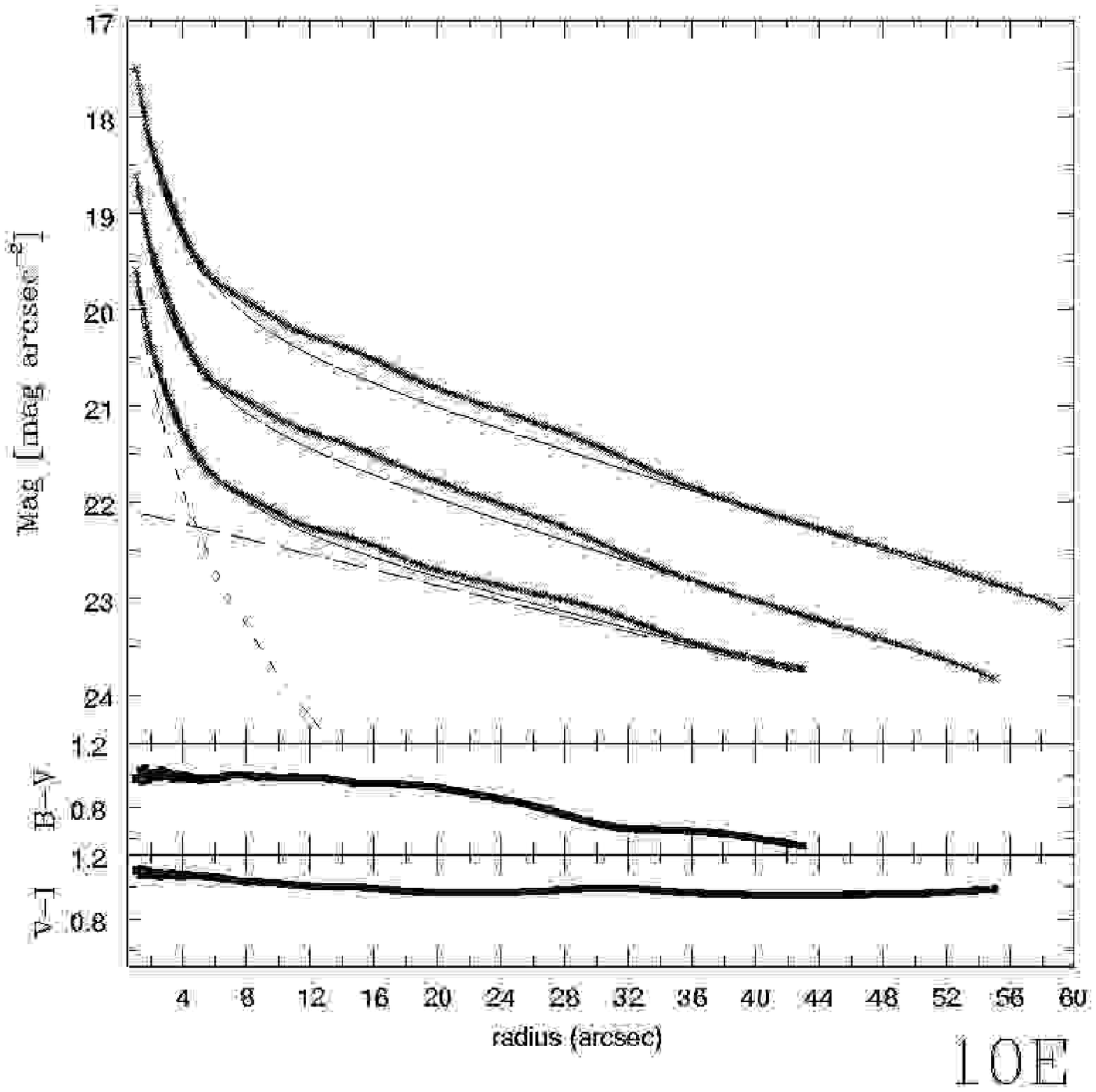}\includegraphics[width=7cm,height=7cm]{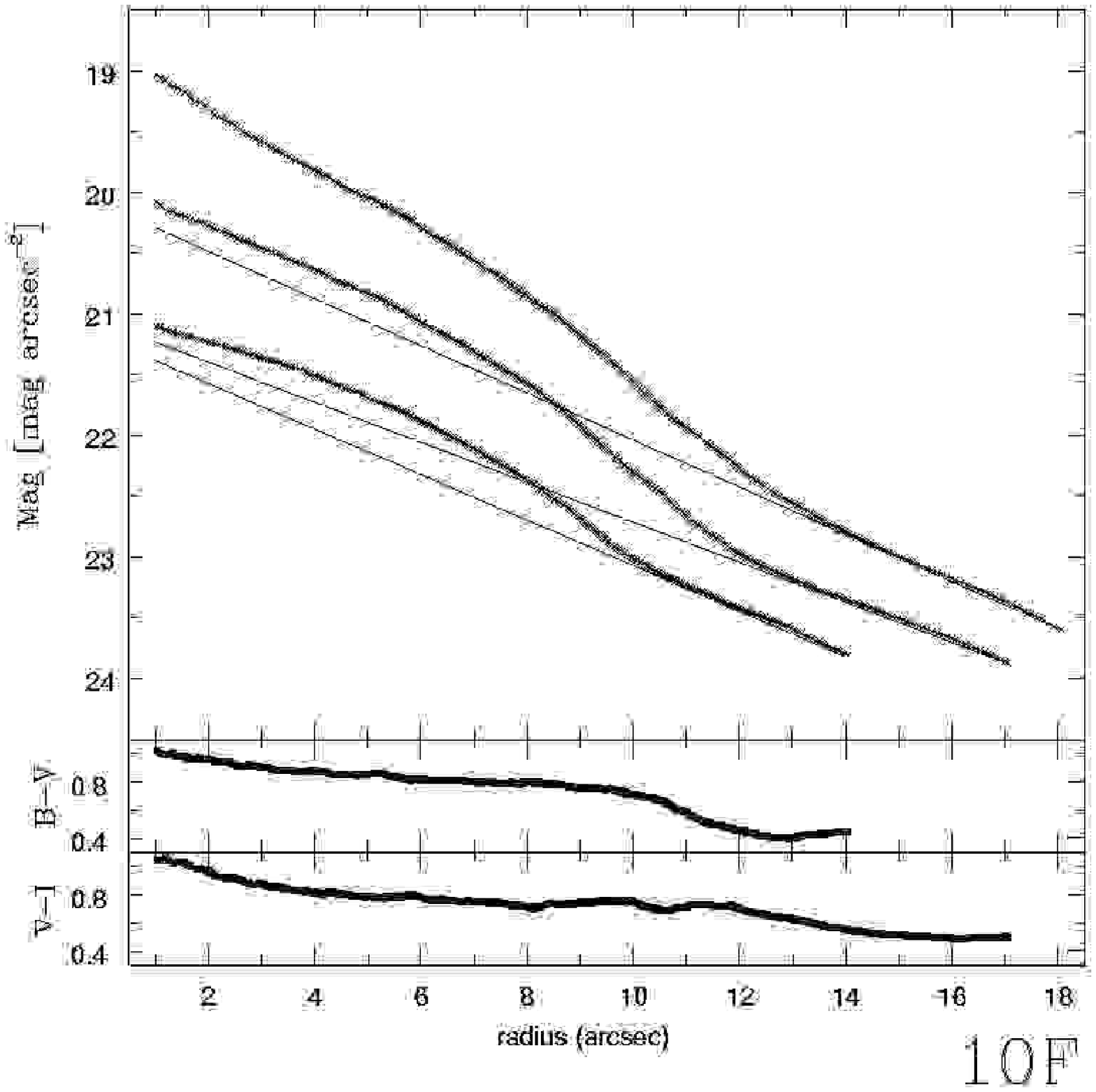}\\
\caption{AM2322-821. Same as Fig 1.}
\label{fig10}
\end{center}
\end{figure*}

AM2322-821: This pair at $cz$ = 3680 $km$ $s^{-1}$ is formed by a Sa/b galaxy and an irregular object. The main galaxy has a typical colour of Sb galaxy, (B-V) = 0.81 and shows the typical bulge plus disk profile. An off-center ring can be observed in the images (Fig.~\ref{fig10}).

The secondary galaxy is almost as faint as a dwarf, $M_B$ = --18.9 and its luminosity profile is exponential in the outer regions. Many HII regions are observed in both components.\\

AM2330-451: This pair is composed by an SO and a disk galaxy, at $cz$ = 3137 $km$ $s^{-1}$. The secondary galaxy exhibits signs of tidal interaction but has a pure exponential luminosity profile.

\begin{figure*}
\begin{center}
\includegraphics[width=7cm,height=7cm]{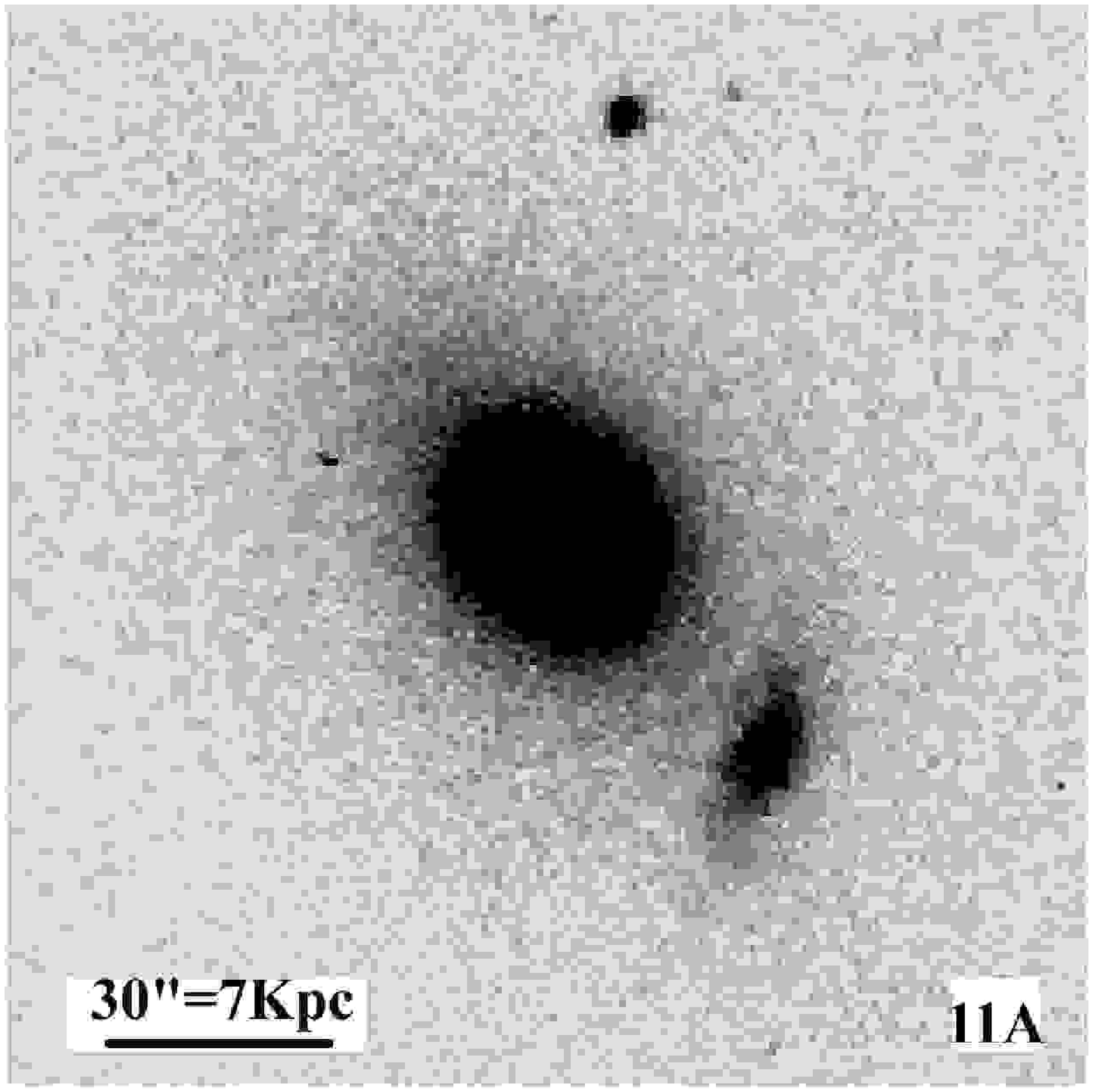}\includegraphics[width=7cm,height=7cm]{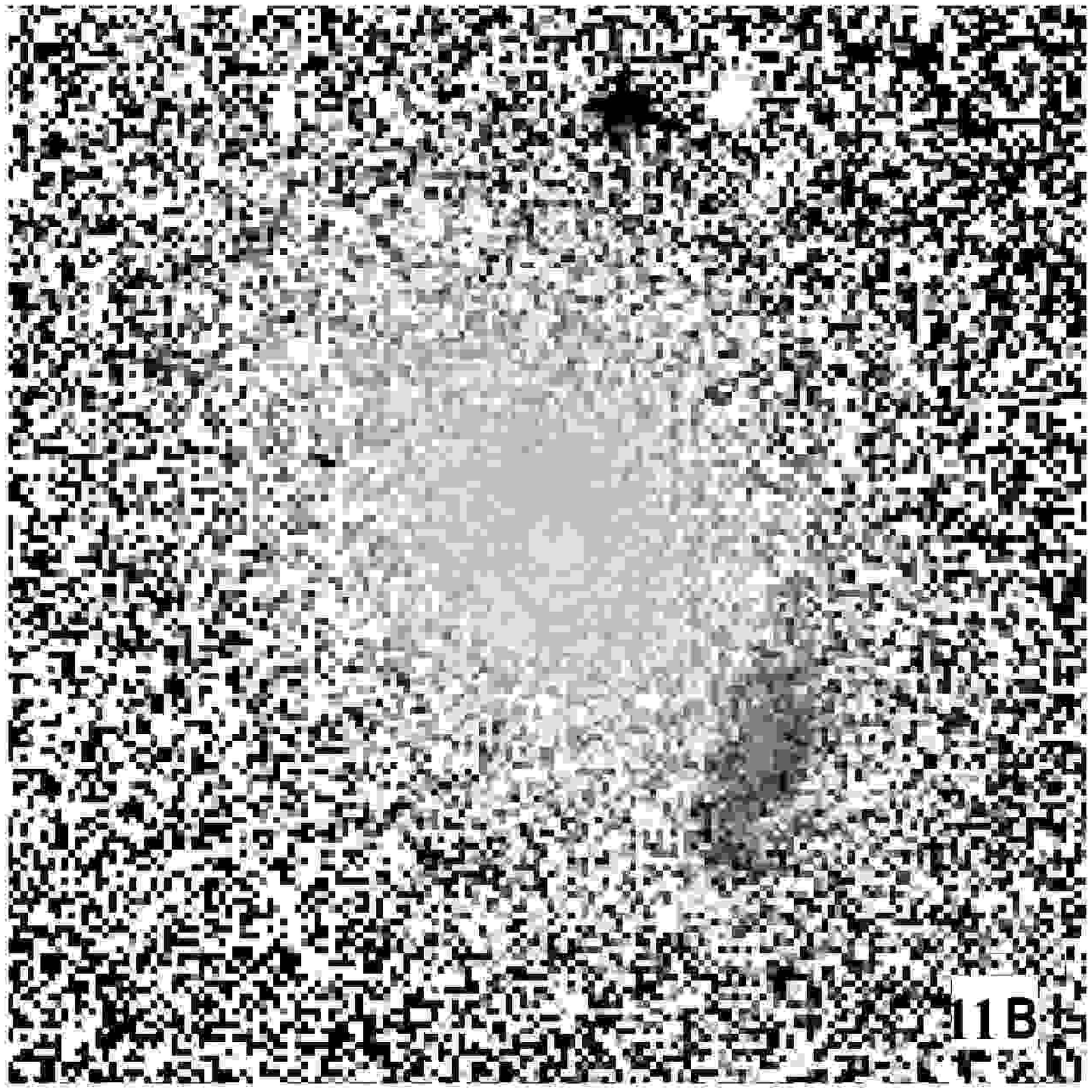}\\
\includegraphics[width=7cm,height=7cm]{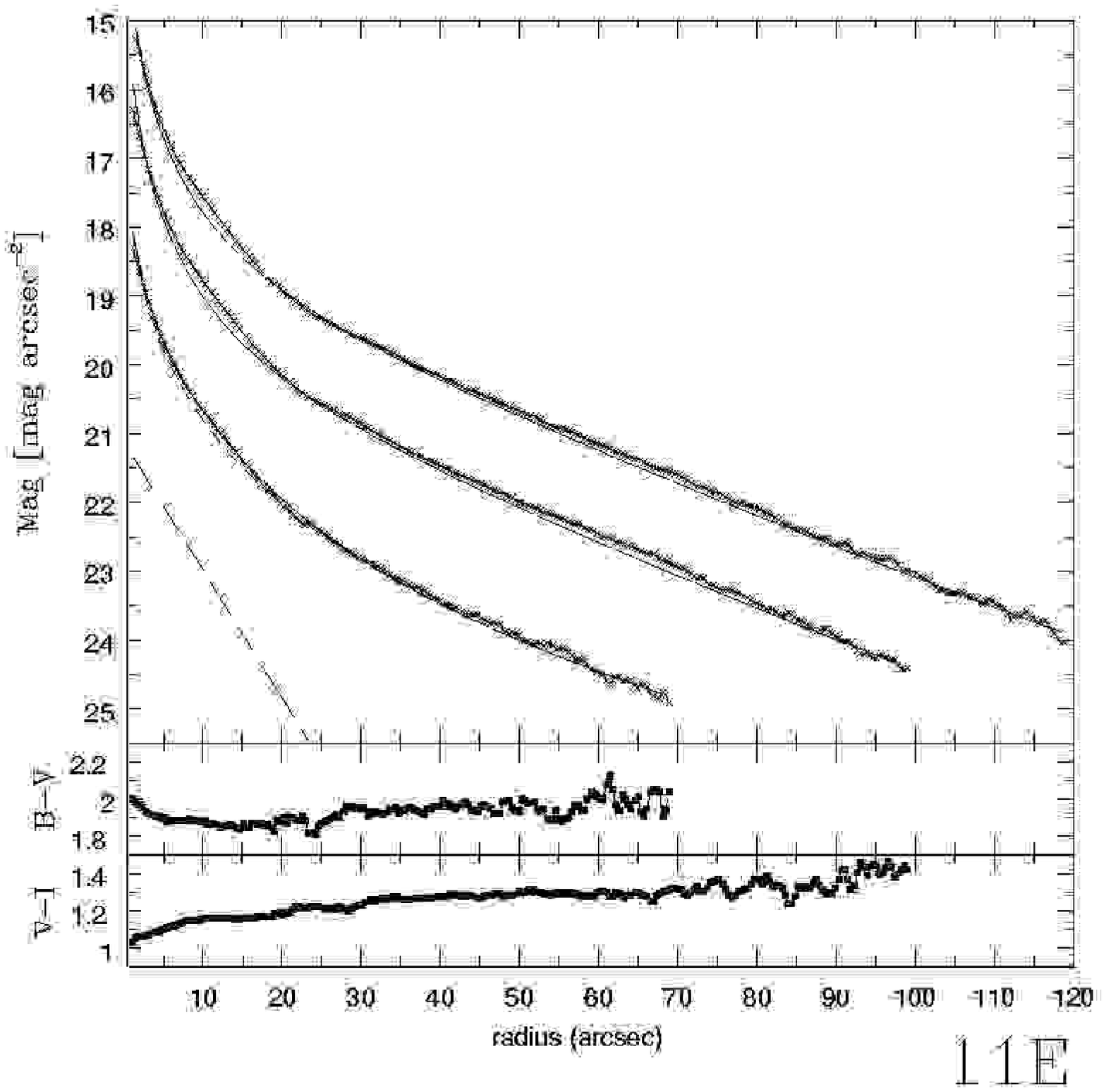}\includegraphics[width=7cm,height=7cm]{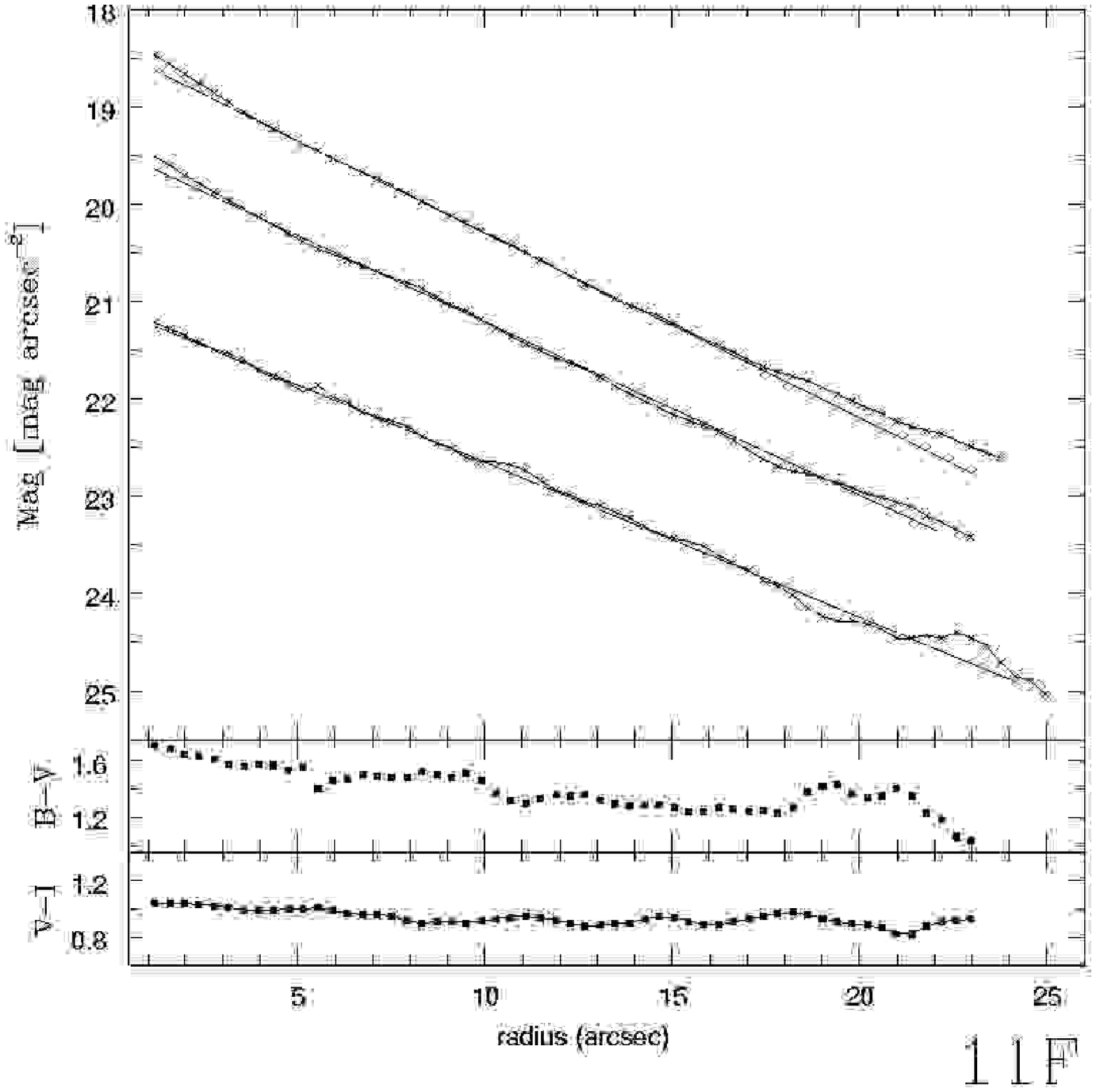}\\
\caption{AM2330-451. Same as Fig 1.}
\label{fig11}
\end{center}
\end{figure*}

The main galaxy has luminosity profile typical of an SO galaxy. A few faint HII regions are observed in the disk galaxy of this pair (Fig.~\ref{fig11}).\\

\section{Results and discussion}
\label{res}
\subsection{Optical integrated properties of the components}

\begin{figure*}
\begin{center}
\includegraphics[width=8cm,height=8cm]{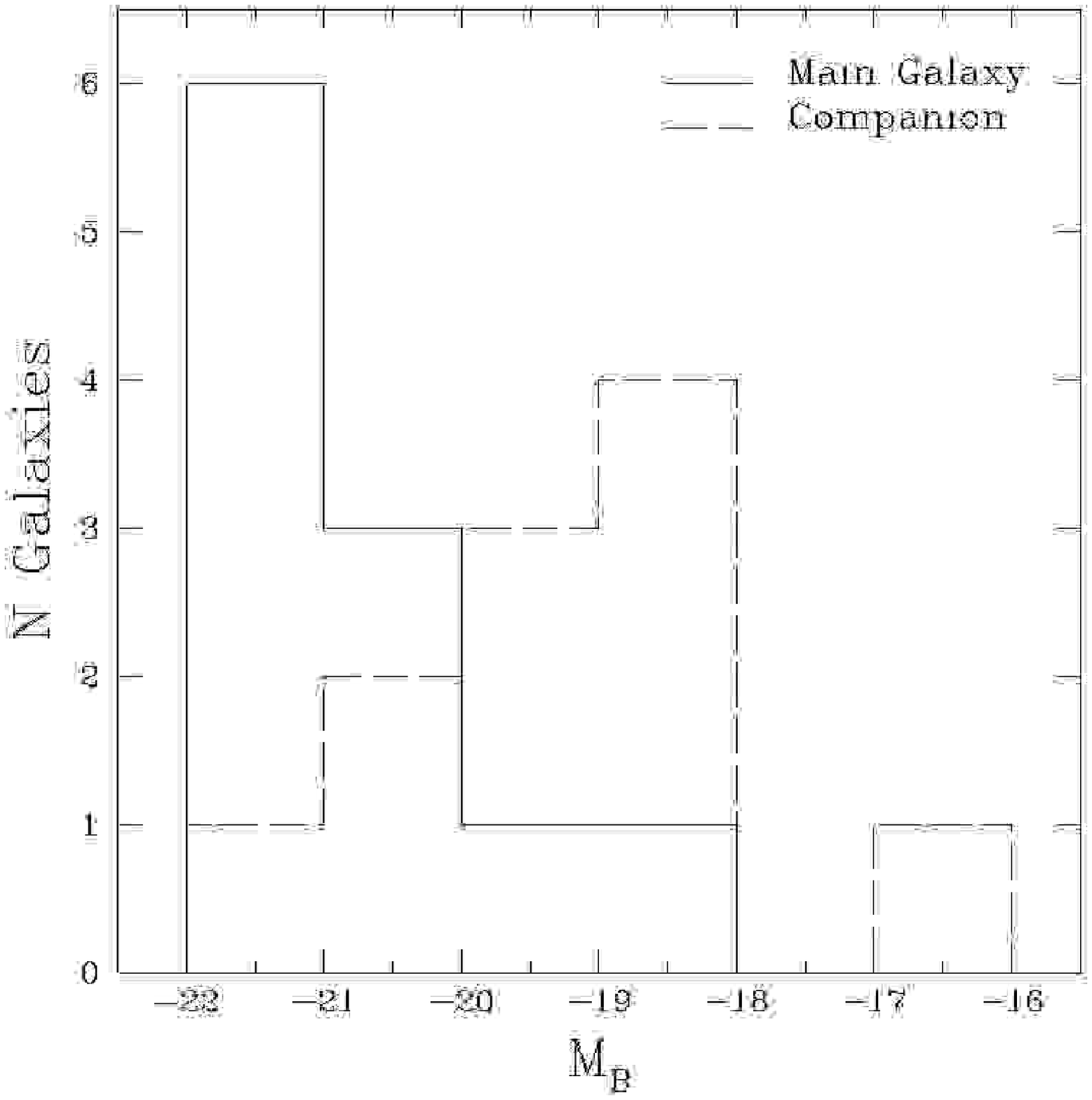}
\includegraphics[width=8cm,height=8cm]{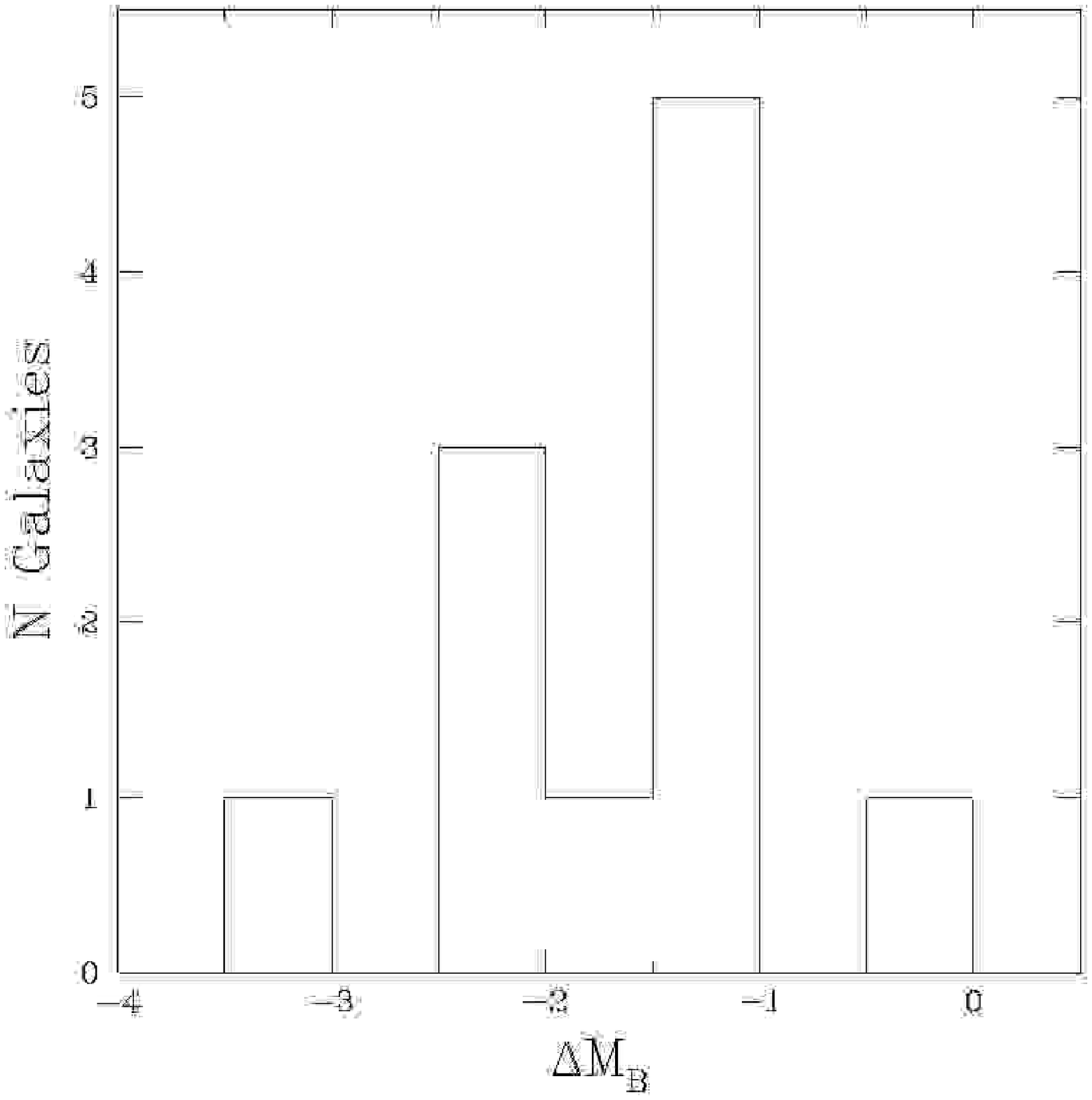}

\caption{Left panel: Luminosity distribution for each component, primary:  continuons line; secondary: short dashed line. Right panel: Luminosity difference distribution ($\Delta$$M_B$) of the pair components. }
\label{F lumi}
\end{center}
\end{figure*}

The above calculated integrated magnitudes, colours and diameters allowed us to
analyze the optical properties of the present sample. The luminosity distribution of the pair components is illustrated in Fig.~\ref{F lumi}A. The continuous line corresponds the primary components and a dashed line to the secondary ones. Primary components have blue absolute magnitudes in the range --22 $< M_B <$ --18, with a peak at $M_B$ = -- 22. The magnitudes of the secondary components lie in the range --22 $< M_B <$ --16 with the maximum at $M_B$ = -- 19. The distribution of the magnitude difference between the components is shown in Fig.~\ref{F lumi}B. Most pairs have $\Delta$$M_B$ $\sim$ 2, which means that in luminosity the primary galaxy is on average 6 times brighter than the secondary. In the sample there are exceptions like AM2306-721 where the primary   is 1.5 times more luminous than the secondary and AM1401-324 where the primary is 15 times brighter.
We have compared $M_B$ magnitudes of our sample with magnitudes of
ultraluminous infrared galaxies (ULIG, $L_{IR}$ $>$ 10$^{12}$ $L_0$) (Surace
et al. 1998, hereafter S1998) and very luminous infrared galaxies (VLIRGs:
$10^{11}$$L_0$ $<$ L(8-1000$\mu$m $<$ 10$^{12}$$L_0$) (Arribas et al. 2004, hereafter A2004) and found that the range in magnitudes of the primary components of our sample is similar to that found for 19 low redshift VLIRGs galaxies (--19 $< M_B <$ --22) (Arribas et al. 2004). In addition the ultraluminous infrared galaxy ULIG sample expands a similar range in blue magnitudes (S1998). Most ULIG and VLIRG galaxies are also interacting or merging. However the main difference between the galaxies of these last samples and the objects of this paper is that the magnitudes difference between the components is larger for the minor mergers ($\Delta$$M_B$ $\sim$ 2 mag) than for the VLIRG component ($\Delta$$M_B$ $\sim$ 0.5 mag) value estimated from Table 3 of A2004.

\begin{figure*}
\begin{center}
\includegraphics[width=8cm,height=8cm]{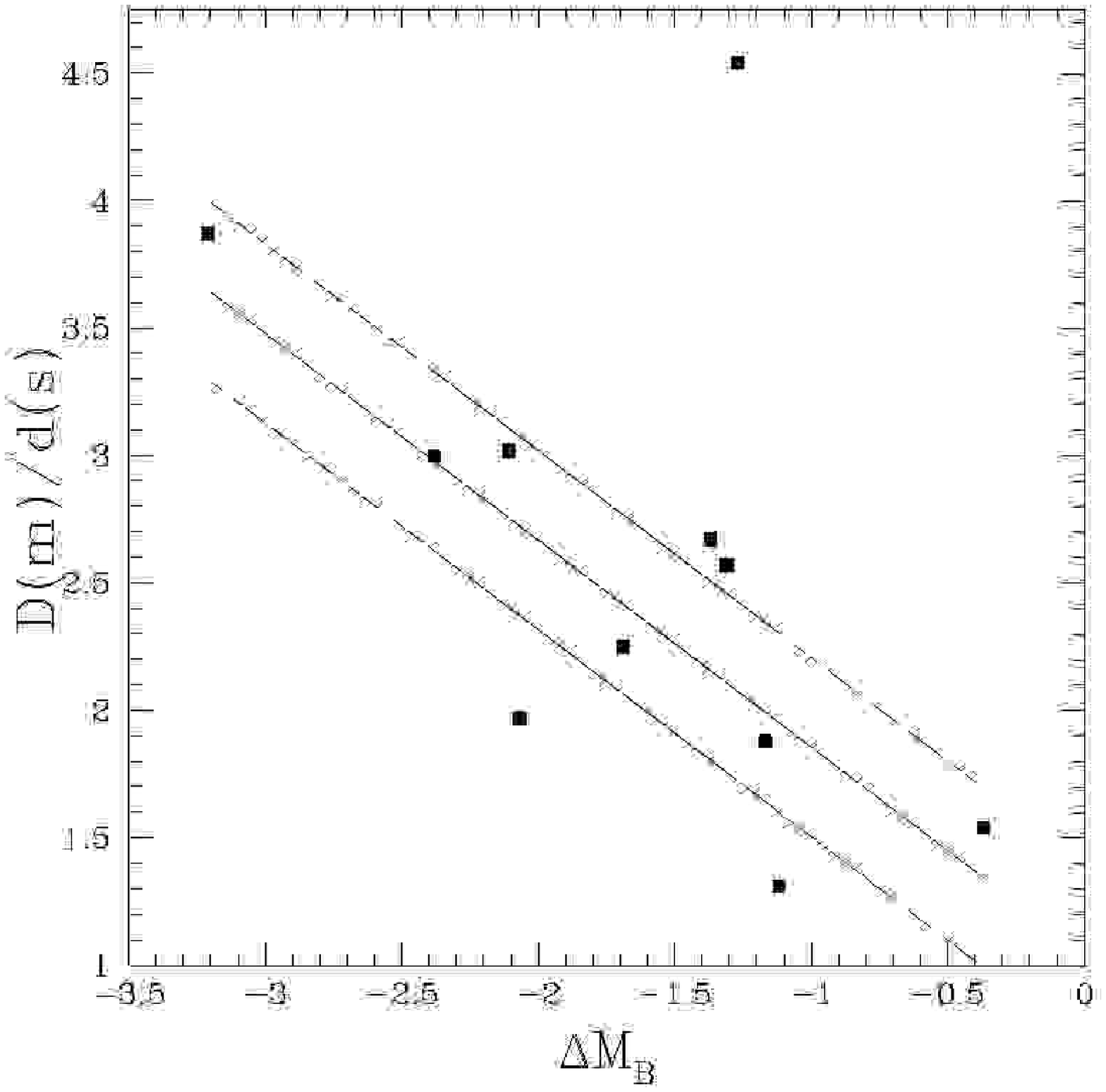}

\caption{Correlation between  ($\Delta$$M_B$) and the ratio of the major diameters (D(m))/(d(s). The solid line is the least square fit to the data and dashed lines are the standard dispersion. The upper point is (AM1256-433).}
\label{F diflum}
\end{center}
\end{figure*}

The major diameter of the galaxies corresponds to the diameter of the isophote of  $\mu$ = 25 mag arcsec$^{-2}$. A linear correlation between magnitude difference of the components with their ratio of major diameters is show in Fig.~\ref{F diflum}. A least square fit gives:

\noindent

\begin{equation}
\label{diam}
D(m)/d(s) = (-0.81 \pm 0.2) \Delta M_B  + (1.04 \pm 0.35)
\end{equation}

\noindent

We can interpret this correlation in terms of the mass ratio between the components. Since most of the galaxies are spirals we adopt a mass luminosity ratio $\cal M/L$ = 3. Therefore for the smallest observed difference of ($M_B$(primary)-$M_B$(secondary)) = --0.5 correspond a mass ratio ($\cal M$ $_{secondary}$/$\cal M$ $_{primary}$) of 0.2, and for the largest differences of --3.5 corresponds to a mass ratio of 0.04. Therefore the mass ratios of our sample are in the range of 0.04 $<$ $\cal M$ $_{secondary}$/$\cal M$  $_{primary}$ $<$ 0.2 indicating according to SD2000 that our sample is indeed formed by minor mergers. This together with accretion events of comparable mass ratios represents a frequent mechanism driving evolution of galaxies in the local universe \\

\subsection{Colour-Colour Diagram}
\label{col col}

Whether interaction plays an important role in driving star formation or not is not yet clear. Interaction between galaxies may lead to a high star-formation rate (SFR) (Combes 1993); however, Bergvall et al. (2003) conclude that interacting and merging galaxies do not substantially differ from similar normal isolated galaxies with respect to the global star forming events. They also found no significant scatter in the colours of Arp galaxies with respect to normal ones. Therefore it is important to determine if the colors of
the observed sample fit into the colours of normal galaxies.  The colour-colour diagram of the pair components (Fig.~\ref{col col}) shows, on average,  similar colours. 
The dereddening vector is  shown in the colour-colour diagram for A$_V$=1 mag, as well as the colors of an HII region of 10$^6$ Myr. The mean visual extinction observed in our galaxies is  A$_V$=1.48 mag (there are two galaxies in the sample with A$_V$ $>$ 2.5 mag, AM1256-433W (A$_V$=2.5 mag) and AM2105-332SE (A$_V$=3 mag)). Larger corrections would give, for several galaxies, dereddened colors similar to that of an HII region 10$^6$ years old, which is not the case, since we have estimated for the HII regions of the these systems ages at least older than 6 $\times 10^6$ Myr.

The secondary components (triangles) have on average (B-V) = 0.71  $\pm$ 0.4 and the main components (squares) have (B-V) = 0.69 $\pm$ 0.3. On the other hand the colour of our sample galaxies are rather bluer than the galaxies with same morphological type, for example, the S0 of our sample have on average (B-V) = 0.77, while for isolated galaxies (B-V) = 0.92. The Sa and irregular galaxies of our sample have average values of (B-V) = 0.71 and (B-V) = 0.53 respectively. For a sample of galaxies of the same morphological type taken from the RC3 we found (B-V) = 0.82 and (B-V) = 0.5 respectively.

The integrated (B-I) color of our galaxies, computed from columns 4 and 5 of Table~\ref{Tmag}, range from 0.62 to 3.3 with a mean value of $<(B-I)>$ = 1.92 for the main component, and $<(B-I)>$ = 1.91 for the secondary component. These values are in agreement with the range in (B-I) values found for VLIRGs (1.2 $< (B-I) <$ 2.8) (A2004) and ULIGs (S1998).\\ 

\begin{figure*}
\begin{center}
\includegraphics[width=9cm,height=9cm]{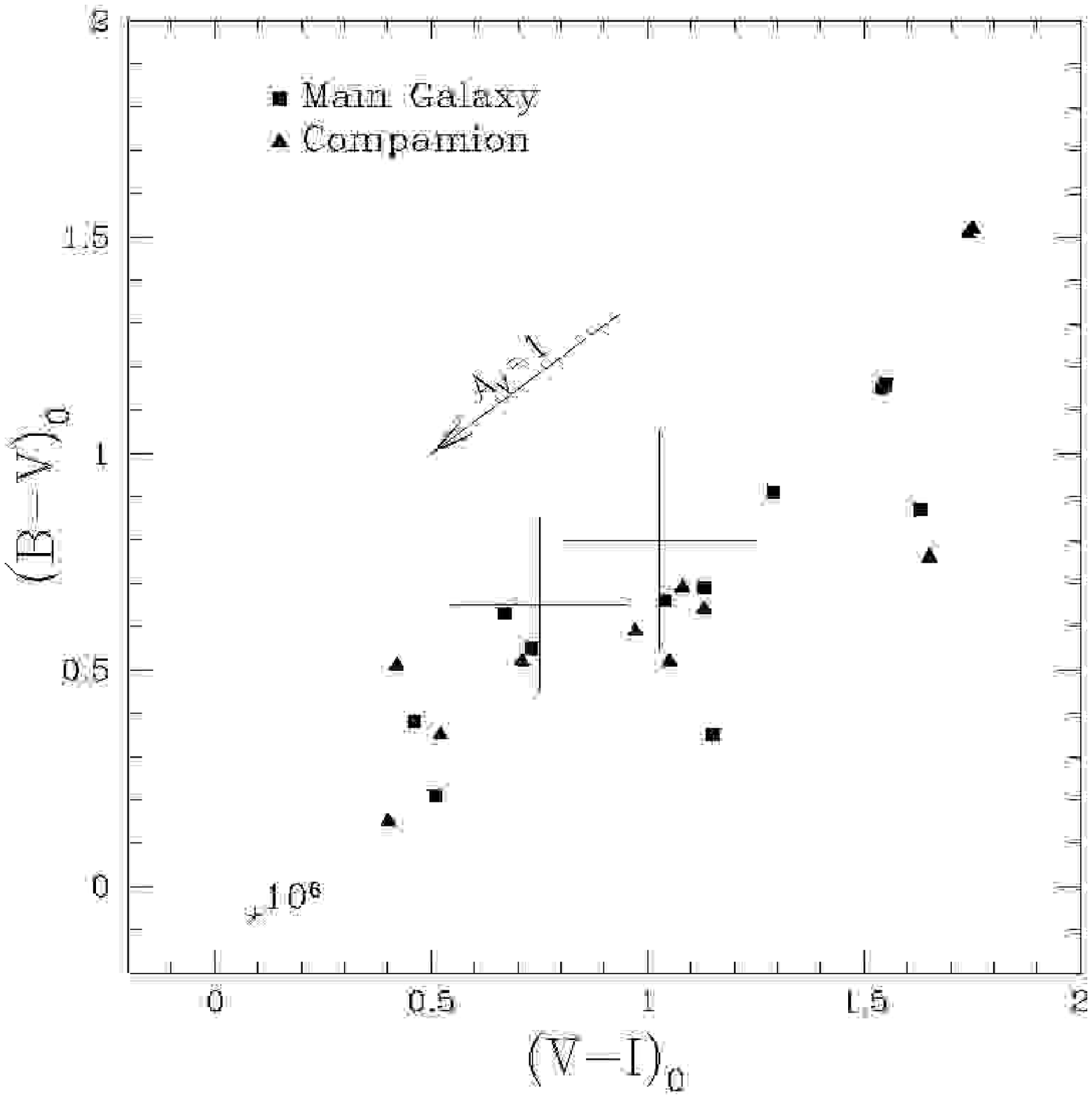}

\caption{Colour-colour diagram, squares are primary and triangle secondary components. }
\label{col col}
\end{center}
\end{figure*}

\subsection{Properties of the HII regions}
\label{sfr}

The equivalent width $EW(H\alpha+[NII])$, luminosities and colours of the HII regions allow us to estimate ages, using  the synthesis models of Leitherer et al. (1999). We found a good agreement between ages obtained from equivalent width or colours. The average age for the HII regions   are listed in Table~\ref{Tregions}. Most of the HII regions and evolved star forming regions of the sample have been formed between 3.6 to 13.7 Myr ago, with an average of (6.3$\pm$0.7) Myr.
 The average values of the luminosity, size and age of the HII regions of the primary and secondary component have been calculated and the values are the following:
\noindent

\begin{eqnarray}
\label{diam}
Age_{main} =  (6.44 \pm 1.5) Myr &;& Age_{sec.} =  (7.06 \pm 1.8) Myr\\
Log(Radius)_{main} =  2.03 \pm 0.3 &;& Log(Radius)_{sec.} =  2.14 \pm 0.3\\
Log(\cal{L} \rm(H\alpha+[NII])_{main} =  39.72 \pm 0.7 &;& Log(\cal{L} \rm(H\alpha+[NII])_{sec.} = 39.87 \pm 1.1
\end{eqnarray}

\noindent

\begin{figure*}
\begin{center}
\includegraphics[width=8cm,height=8cm]{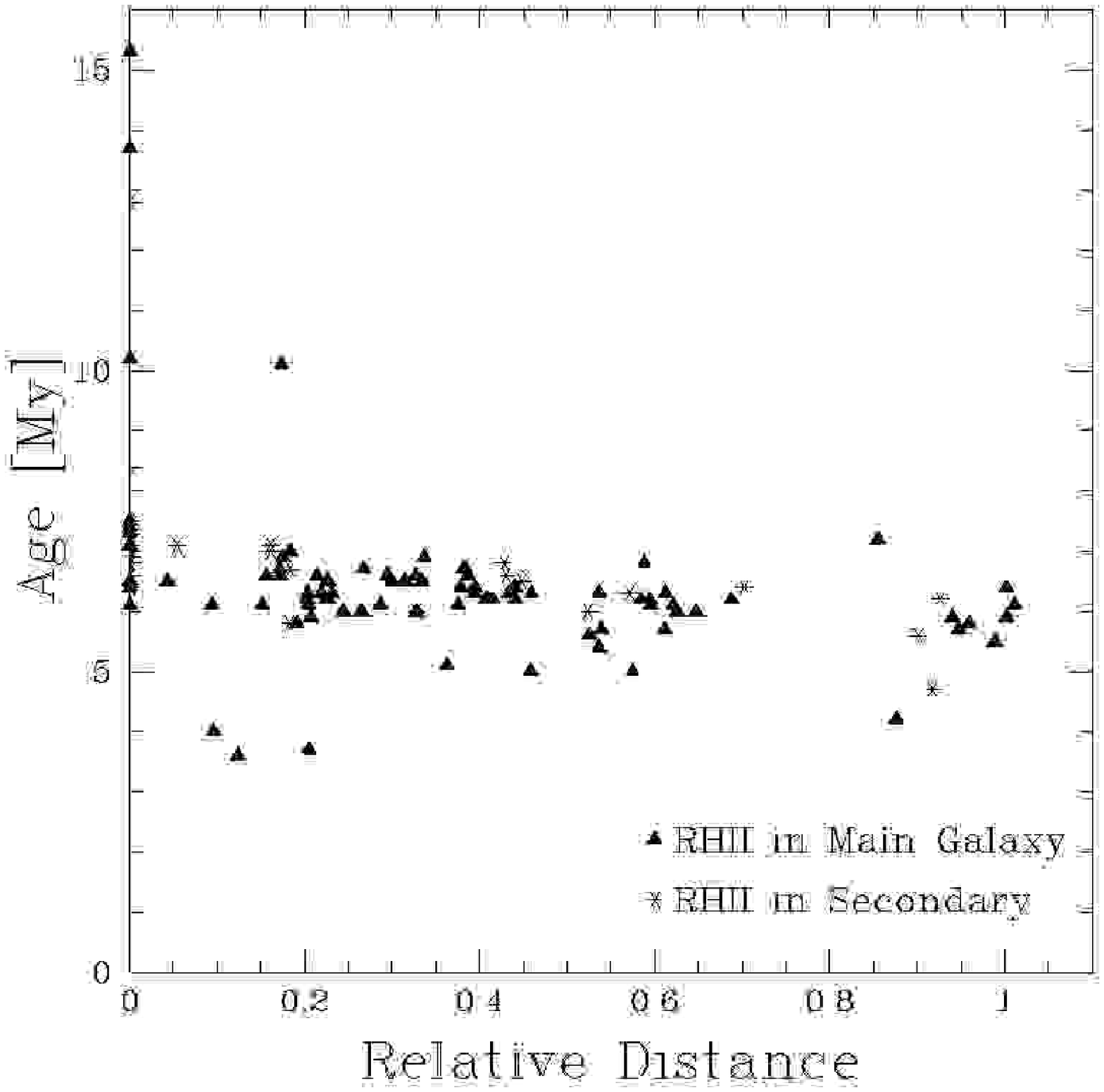}
\includegraphics[width=8cm,height=8cm]{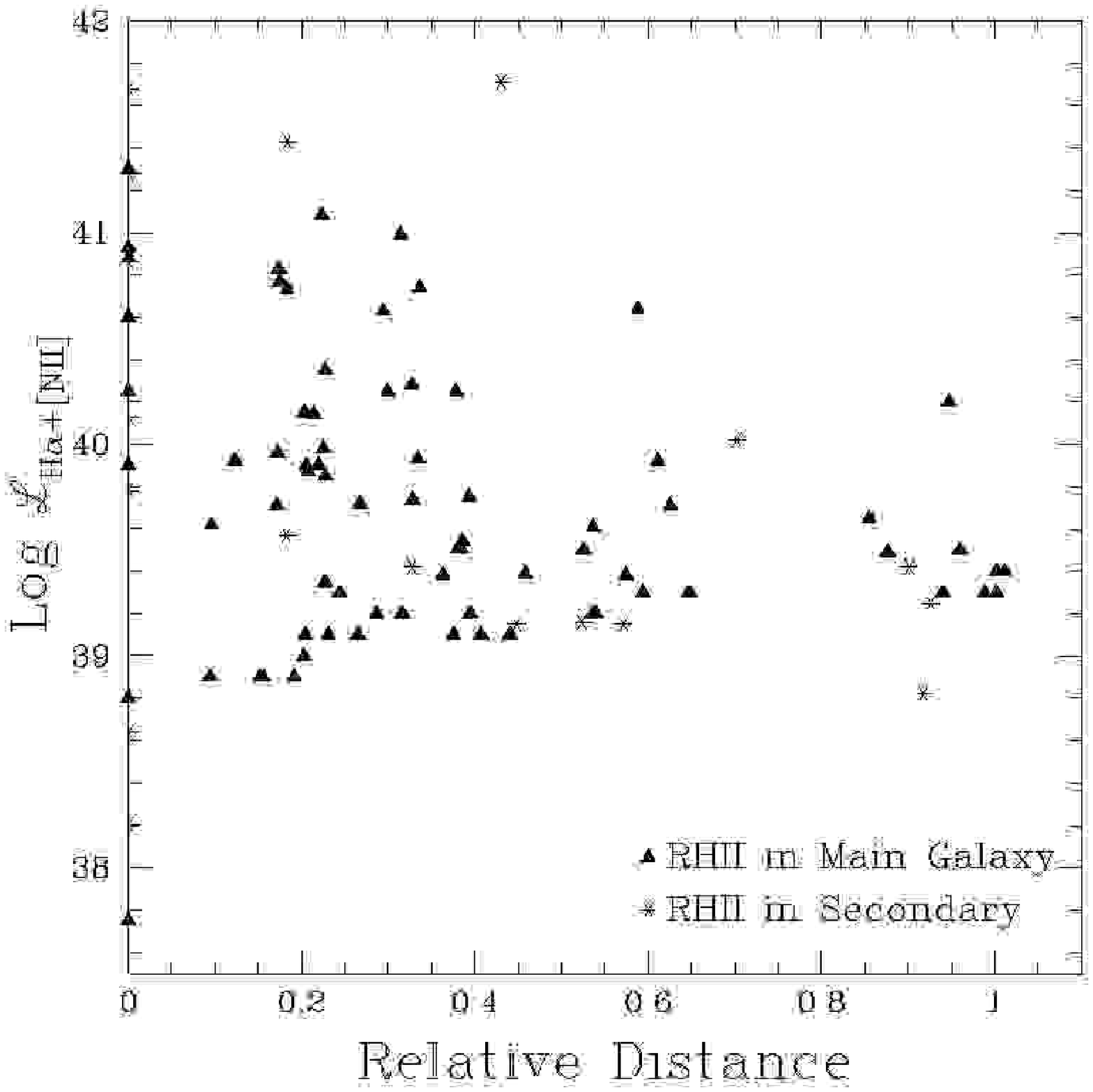}

\caption{Left panel: Age, in megayears, versus relative distance to the galaxy center. Right panel: Logarithm of the luminosity H$\alpha$+[NII] ($\cal L$(H$\alpha$+[NII]) in erg s$^{-1}$) versus relative distance to the galaxy center in Kpc. In both panels triangles  correspond to the HII regions  of the primary and stars to  the secondary components.}
\label{age lum}
\end{center}
\end{figure*}

The HII region properties are similar in both components. This result was also obtained by Hancock et al. (2003) for the interacting pair NGC 3395 and NGC 3396. These authors did not find a correlation between the age of the knots and their positions in the galaxies. Fig.~\ref{age lum} illustrates ages as a function of  the relative distance to the galaxy center for the present  HII region sample. No correlation was found between these parameters.

S1998, analyzing a sample of nine "warm" ultraluminous infrared galaxies
(ULIGs), found blue compacts knots of star formation with mean radius
($r_{eff}$) of 65 pc; the largest one has a radius of 244 pc. The ages for the
knots in individual galaxies are in the range of 6 to 370 Myr. A2004 found that the HII regions observed in the VLIRGs are
young bursts with ages between 5-10 Myr. The mean integrated B magnitude of the
observed knots in the individual warm ULIGs is $<M_B>$ = -- 16.4, ranging from
--13.16 to a maximum of --20.4. In addition for a sample of 18 cool ULIGs Surace et al. (2000) found $M_B$ magnitudes of the knots ranging from --14.4 to --18.5.mag. The range in age, size and B luminosity of the HII regions of ULIG, VLIRG and warm ULIG galaxies are similar to those found in this paper for the HII regions of the minor merger.\\

\subsection{Luminosity function}
\label{lum fun}

\begin{figure*}
\begin{center}
\includegraphics[width=9cm,height=9cm]{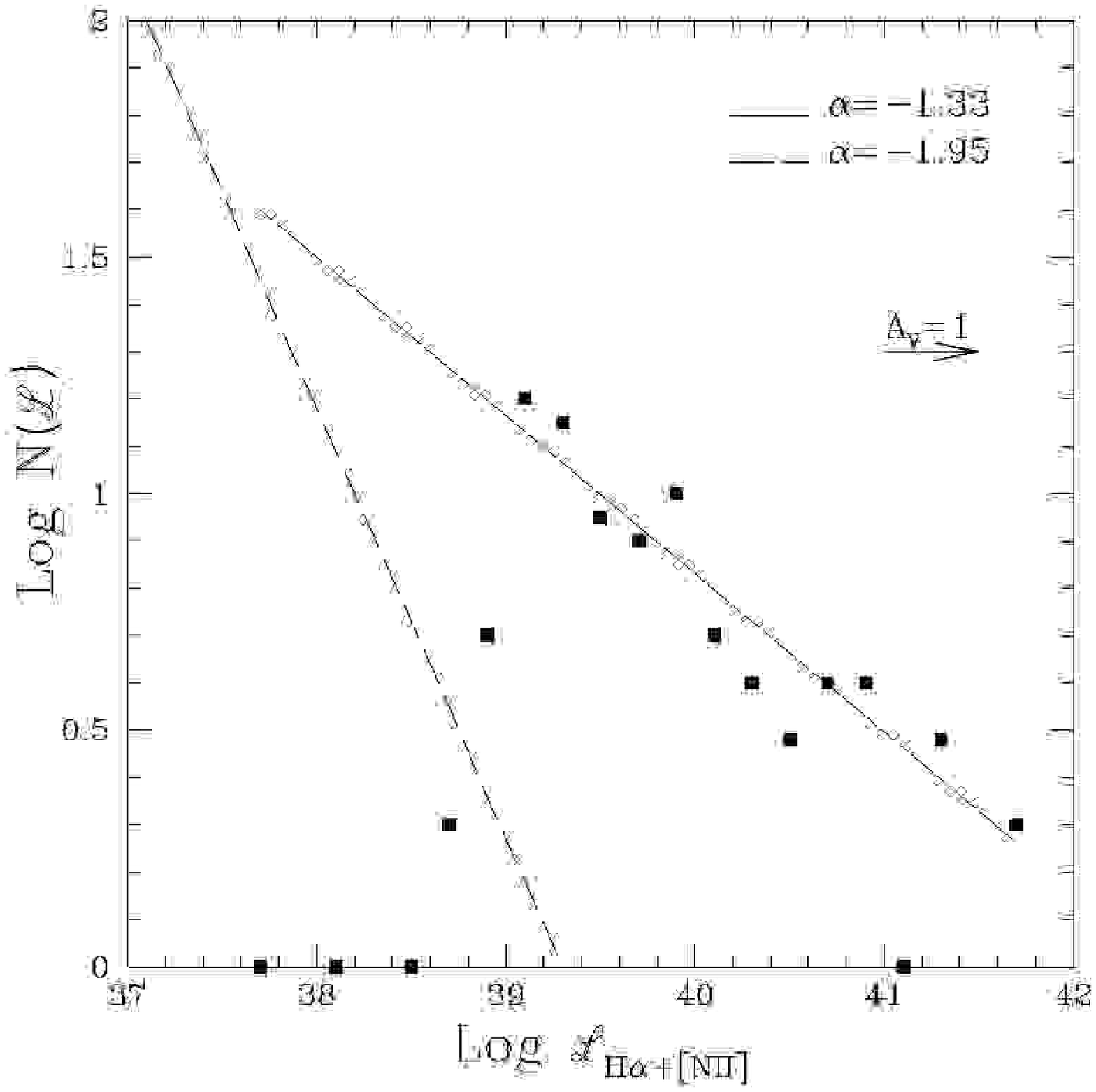}

\caption{HII region luminosity function for the whole sample. The  solid line is the power-law  with index $\alpha$ = -- 1.33 fitted to the data; dashed line is the fit obtained by Crocker from  HII regions of ringed galaxies  with index $\alpha$  =  -- 1.9.}
\label{F fun lum}
\end{center}
\end{figure*}

The luminosity function of the HII regions is often represented by a power law.
 Keniccutt \& Hodge (1980) claim that the power law is universal with average  index $\alpha$  = -- 2. However a broken power law better fits the luminosity function of several galaxies of the sample observed by Gonz\'alez Delgado \& P\'erez (1997). The H$\alpha$+[NII] luminosities of our HII regions are found in the interval 39 $<$ log(H$\alpha$+[NII]) $<$ 42 with an average value of  $\sim$ 40,  clearly more luminous  than  HII regions of  ringed galaxies (Crocker et al. (1996)). We have shown that the optical properties, luminosity, size and age of the HII regions of the main and secondary components are similar. Since we detected a few HII regions per galaxy, we constructed a luminosity function containing all observed HII regions, in order to compare the results obtained by Crocker et al. (1996). The luminosity function was constructed in bins of 0.2 in logarithm, and fitted by a power-law (solid line) of index $\alpha$ =  -- 1.3 Fig.~\ref{F fun lum}. The fit obtained by Crocker et al. (1996) (dashed line) from HII regions of ringed galaxies with index $\alpha$  =  -- 1.9 is also shown in the figure. Comparing these results we conclude that the present sample of HII regions has an excess of high luminosity  regions with respect to the ringed galaxies.
In addition the power law index, found for the luminosity function of our HII regions, is in close agreement with the index $\alpha$ = -- 1.39 $\pm$ 0.08 obtained combining the data for all the warm galaxies (S1998). This value is more positive than that found in other systems and is due to an overabundance of bright knots in ULIGs.\\

\subsection{Luminosity versus size}
\label{lum size}

\begin{figure*}
\begin{center}
\includegraphics[width=9cm,height=9cm]{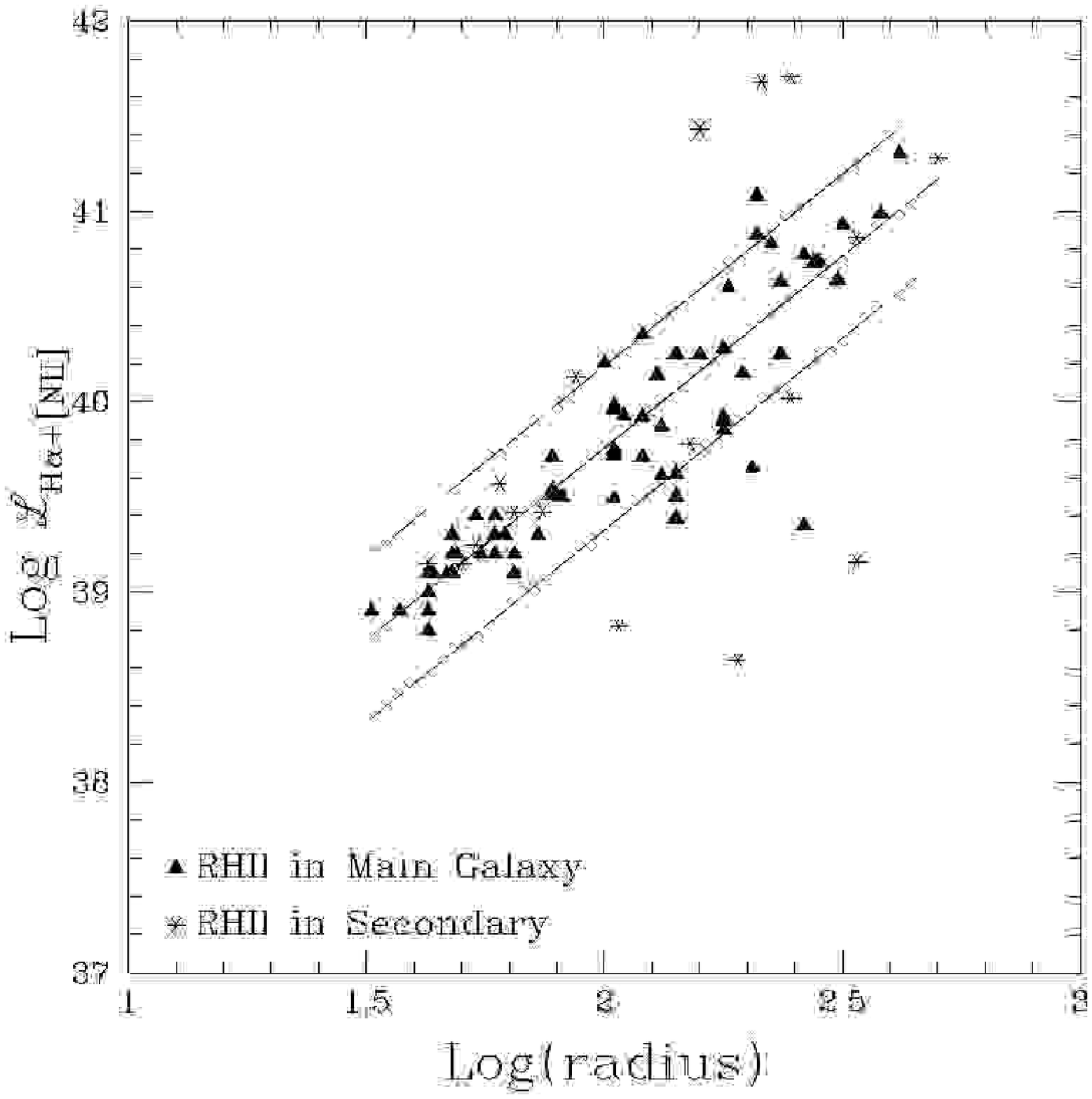}

\caption{ H$\alpha$+[NII] luminosities  as a function of the equivalent radius. The  HII regions of the   primary  and secondary  components are represented by triangles and stars respectively. The solid line is the least square fit to the data and dashed lines are the standard deviation. }
\label{F lum size}
\end{center}
\end{figure*}

Fig.~\ref{F lum size} illustrates the relationship between the H$\alpha$+[NII] luminosities as a function of the equivalent radius. The data fit a linear regression in the log-log plane given by:

\begin{equation}
Log(\cal{L} \rm(H\alpha+[NII])) = (2.12 \pm 0.06) Log(radius) + (35.5 \pm 0.1)\\
\end{equation}

Luminosities are given in erg sec$^{-1}$ and radius in
parsecs and were determined as described in Sect.\ref{halpha}.
A similar analysis of the HII regions of individual galaxies of our sample give a similar correlation between luminosity and size. However, slopes were different from galaxy to galaxy. Slope values range from 1.9 to 2.8. with mean value of 2.4 $\pm$ 0.6, in agreement with the result  obtained by Gonz\'alez Delgado \& P\'erez (1997) for a sample of HII regions of galaxies with nuclear activity.
For HII regions that have constant density and are ionization-bounded, the H$\alpha$ luminosity scales with the third power of the Stromgren radius (Kennicutt 1988). If the power is smaller than 3, which is the case for our HII region sample,  then we are dealing with radiation bounded regions. It is possible that slopes flatter than 3 indicate the effect of dust on the Lyman continuum photons. For that reason, larger HII regions would have more dust than smaller ones (Gonz\'alez Delgado \& P\'erez 1997).

\subsection{Interaction vs. galaxy disk parameters}
\label{int disk}

 The relevant parameters determined in Sect~\ref{prof} are the central surface magnitude $m_0$ and the scale length $d_l$. Prior to the analysis we transformed the length scale $d_l$, that is in angular length, to linear length according to the equation of Lu (1998):

\begin{equation}
log (d_s) = log(d_l) -0.2 M_B -4.416
\end{equation}

In this equation, d$_s$ is the disk scale length in Kpc, d$_l$ is the disk scale length in arcsec. The disk central surface magnitude was corrected
for the inclination according to the following relation:

\begin{eqnarray}
b_0^c = b_0-2.5 K\; log(1-e);\;\; 	e&\leq& 0.8\\
b_0^c = b_0-2.5 K\; log(1-0.8);\;\; 	e&>& 0.8
\end{eqnarray}

b$_0$$^c$ being the central surface magnitude for the disk component corrected for inclination and b$_0$ the central observed surface magnitude. The eccentricity of the galaxy is defined as e$^2$ = (D$^2$-d$^2$)/D$^2$,
 D and d being the major and minor diameters of the galaxy. K is the opacity of the galaxy disk, K = 0 corresponds to a totally opaque disk and K = 1 to a fully transparent disk (Lu 1998).

\begin{figure*}
\begin{center}
\includegraphics[width=8cm,height=8cm]{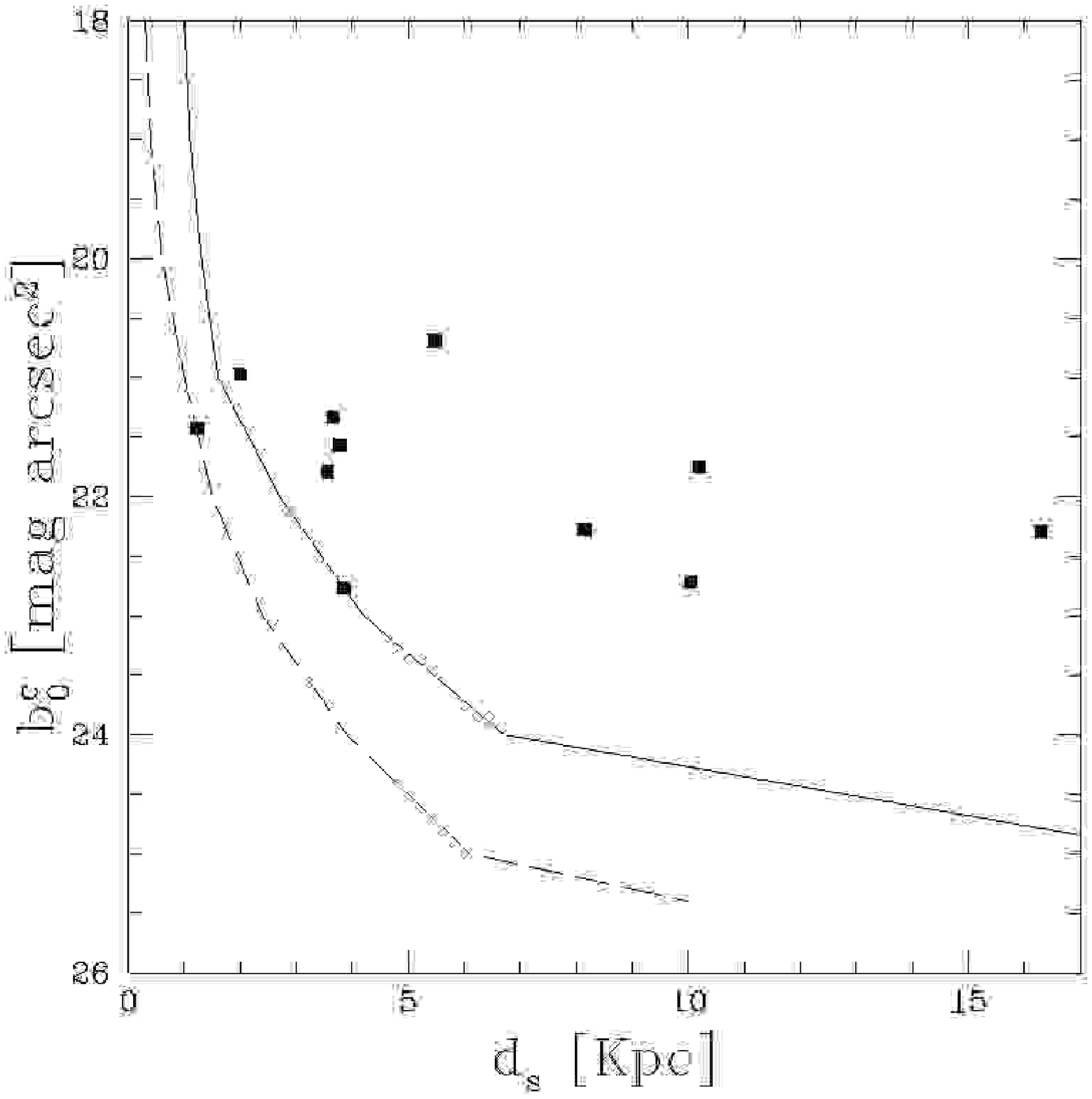}
\includegraphics[width=8cm,height=8cm]{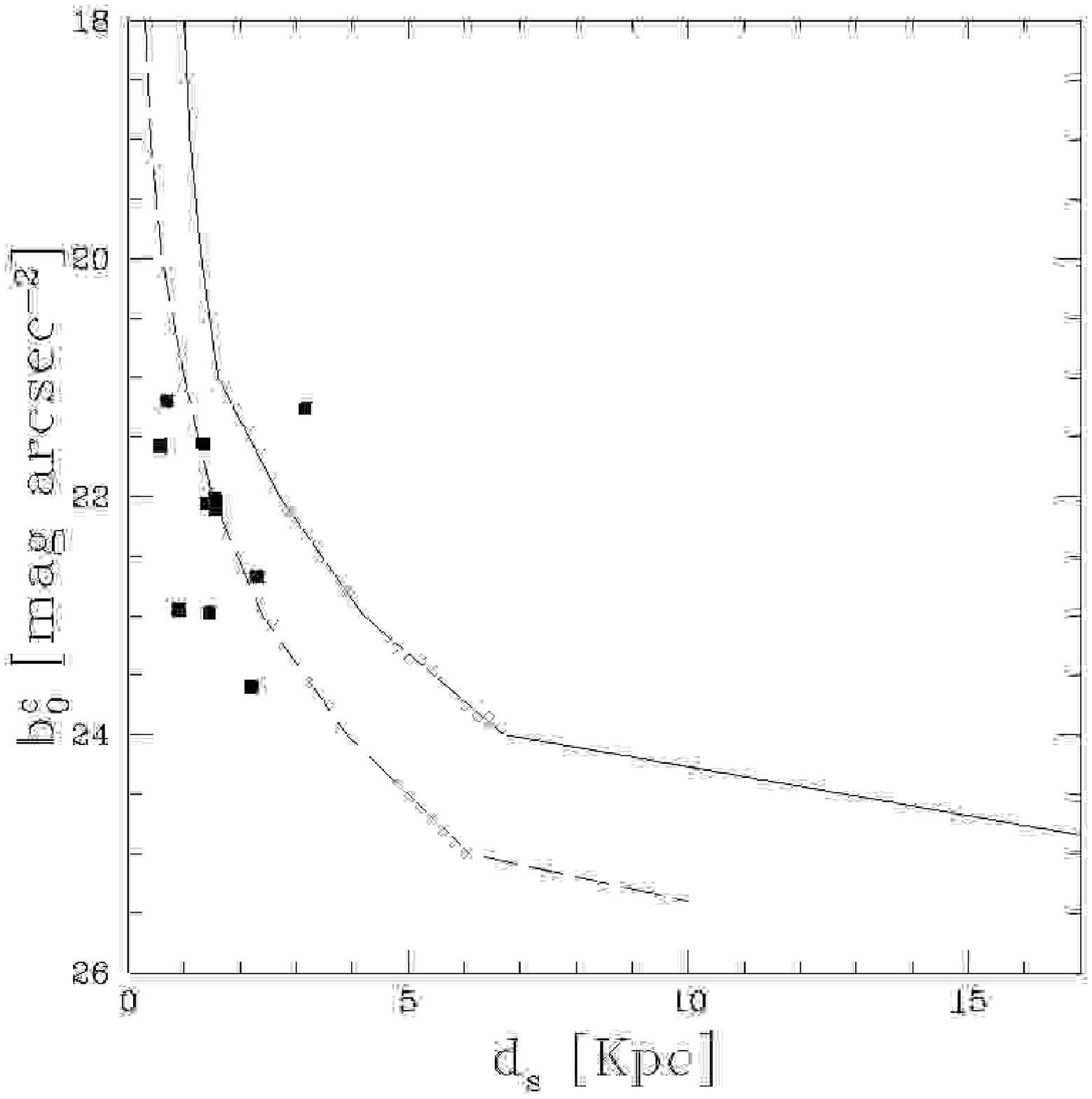}

\caption{Central disk brightness in mag arcsec$^{-2}$ as a
function of the linear disk scale length in Kpc. The
dashed line indicates the path of a face-one disk galaxy of $M_B$ = --17.5 mag
that is detectable up to half of the maximum of the UGC sample distance. The solid curve indicates the threshold above which a face-one galaxy will be detectable up the maximum sample distance. The right panel corresponds to the disk of the primary components and left panel to the secondary components.}
\label{F inter disk1}
\end{center}
\end{figure*}

Fig.~\ref{F inter disk1} shows the corrected central surface magnitude of the disk as a function of the linear scale length. Lu (1998) analyzed a sample of 76 disk galaxies (B$_T$ $<$ 14.5 mag) from the Uppsala General Catalogue (UGC) with  radial velocities cz $<$ 3000 Km s$^{-1}$ and found a correlation  between the face-on central surface brightness and the linear disk scale length up to 24.5 mag arcsec$^{-2}$.

\begin{figure*}
\begin{center}
\includegraphics[width=8cm,height=8cm]{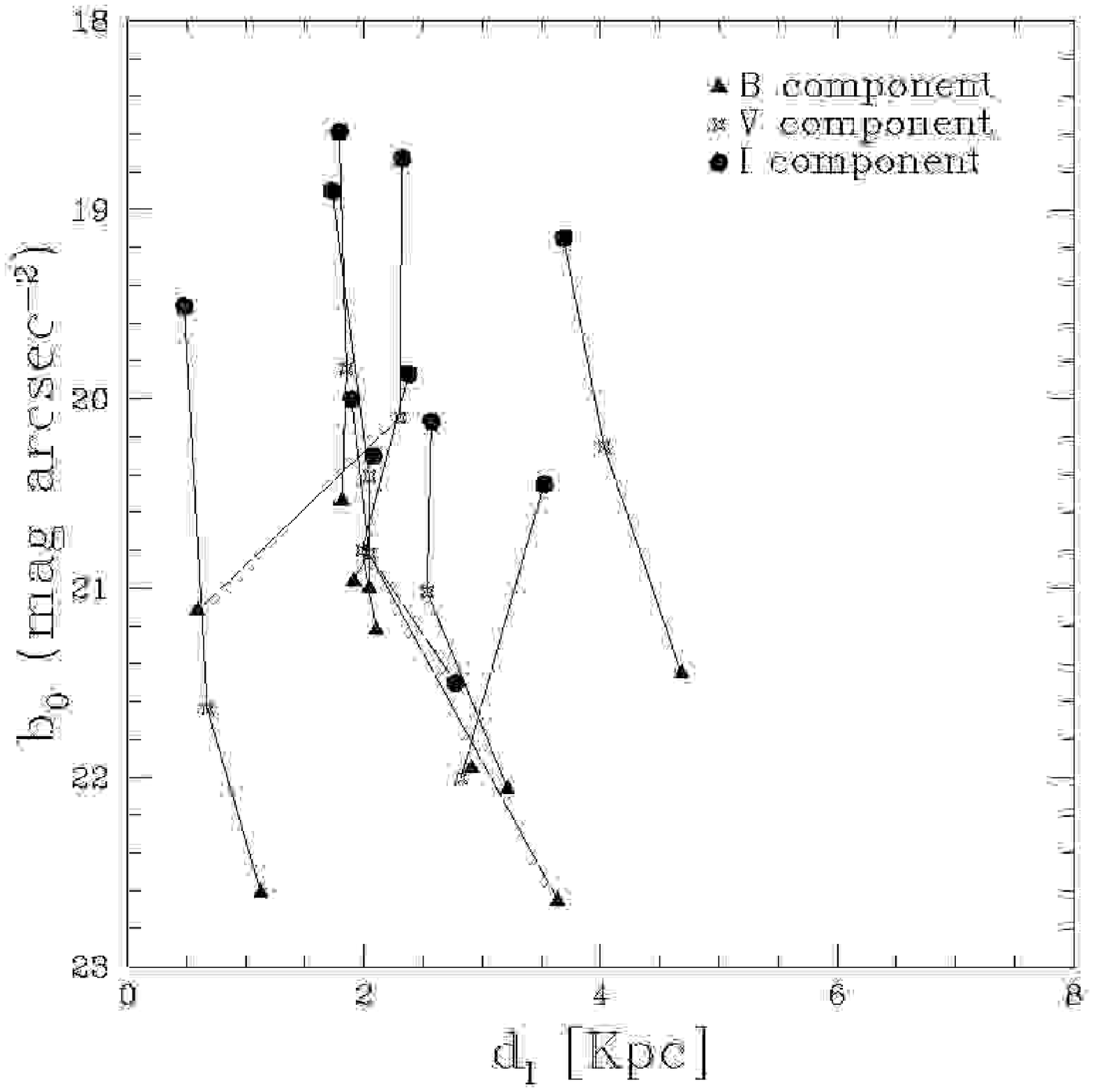}
\includegraphics[width=8cm,height=8cm]{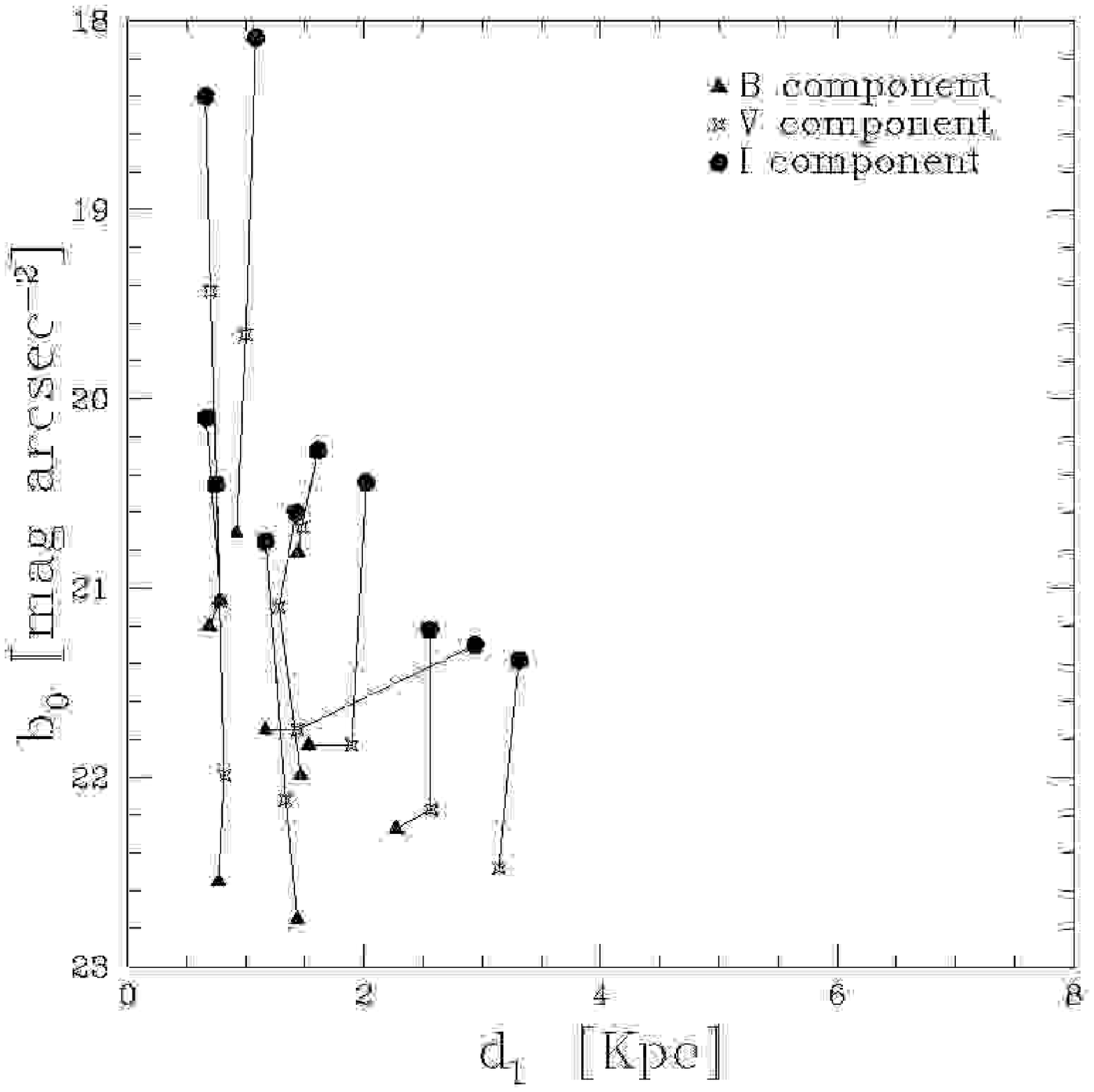}

\caption{Central brightness (mag arcsec$^{-2}$) as a function of the linear disk scale length given in Kpc. The lines link the disks of each galaxy in the B, V and I colours. The right panel corresponds to the disk of the primary component and the left panel to the secondary.}
\label{F inter disk2}
\end{center}
\end{figure*}

A comparison with Lu (1998) shows that, in general, both samples behave similarly. The disks of the main galaxy are more luminous than those in Lu's sample and the disks of the secondary are smaller and fainter  (Fig.~\ref{F inter disk1}a  and Fig.~\ref{F inter disk1}b).  A plot of the disk parameters obtained in each filter for the main and secondary galaxy is given (see Fig.~\ref{F inter disk2}). The scale length does not change, in general, with colour. This indicates that the different stellar populations in the disks were affected in the same way.

\section{Conclusions}
\label{con}

\begin{enumerate}

\item We have shown the results of B, V and I photometry of eleven southern minor mergers. The total apparent B magnitude, integrated B-V and V-I colours were measured. We built B, V, and I equivalent profiles for each galaxy  and  decomposed them into bulge and disk components when possible. All galaxies show light excess due to the contribution of star-forming regions, arms, bar and rings. From H$\alpha$+N[II] images we have estimated the  basic photometric parameters of the HII regions, such as position, size, B-V and V-I colours, H$\alpha$+[NII] luminosity and $EW(H\alpha+[NII])$ equivalent width.\\

\item Among the primary and secondary components we found a variety of morphological types: two elliptical, six SO, five SO/a-Sa, one Sb, one Sc, one Sd and six irregulars. Most of them present signs of tidal interactions.\\

\item Primary components have blue absolute magnitudes in the range --22 $< M_B <$ --18, with a peak at $M_B$ = -- 22. The magnitudes of secondary components are in the range  --22 $< M_B <$ --16 with maximum at $M_B$ = -- 19. Most pairs have $\Delta$$M_B$ $\sim$ 2, which means that in luminosity the primary galaxy is on average about 6 times brighter than the secondary.\\

\item We found a linear  correlation between the luminosity ratios of the components and their ratio of major diameters, leading to mass ratios
 between 0.04 $<$ $\cal M$ $_{secondary}$/$\cal M$ $_{primary}$ $<$ 0.2, suggesting indeed that our sample is  formed by  minor mergers.\\

\item We found a wide range of  integrated colours, 0.15 $\leq$ (B-V)
$\leq$ 1.52 and 0.40 $\leq$ (V-I) $\leq$ 1.75. In general, the primary and secondary components have similar colours.  Most of the galaxies have colours bluer than those of isolated galaxies with the same morphological type.\\

\item Most of the  HII regions and evolved star forming regions of the sample have been formed between 3.6 to 13.7 Myr ago, with an average of  (6.3$\pm$0.7) Myr. The HII region properties, luminosity, sizes and ages are similar in both components. No correlation between age and luminosity of the HII regions with the distance to the galaxy center was   found. The HII regions have log H$\alpha$+[NII] luminosity between 38.6 and 41.7. The HII region luminosity function for the whole sample   fits a power law of index  $\alpha$ = --1.33. The linear correlation between the luminosity $\cal L$(H$\alpha$+[NII]) and the size of the HII regions has slope of 2.12$\pm$0.06. For individual galaxies the luminosity-size correlations have a large dispersion, varying from slopes of 1.91 to 2.79.\\

\item The corrected central surface magnitude of the disk versus the linear scale length was compared with that of Lu's sample (1998). We found that the disk of the primary components is more luminous than those in Lu's sample, while the disk of the secondary is smaller and fainter.  A plot of the disk parameters does not change with colour. This indicates that the different stellar populations in the disks were affected in the same way.\\

\end{enumerate}

\begin{acknowledgements}
This work has been partially supported by the Secretar\'{\i}a de Ciencia y T\'ecnica de la Universidad Nacional de C\'ordoba (SeCyT) and The Brasilian Institution CNPq (PRONEX 66.2088/1997-2). We thank Dr. Carlos Donzelli and Dr. Eduardo Bica for helpful comments and suggestions.

\end{acknowledgements}

\end{document}